\documentclass{amsart}

\usepackage{amssymb}
\usepackage{amssymb,amsmath,amscd}
\usepackage{pstricks,pst-node}
\psset{unit=.5cm}

\usepackage{graphicx}
\usepackage{epstopdf}
\usepackage{amscd}
\usepackage{graphics}
\newtheorem{Theorem}{Theorem}[section]
\newtheorem{Proposition}{Proposition}[section]

\newtheorem{Lemma}{Lemma}[section]

\def\proof{\par{\it Proof}. \ignorespaces}
\def\endproof{{\ \vbox{\hrule\hbox{%
     \vrule height1.3ex\hskip0.8ex\vrule}\hrule }}\par}
\newenvironment{Proof}{\proof}{\endproof}

\theoremstyle{definition}
\newtheorem{Definition}[Theorem]{Definition}
\newtheorem{Example}[Theorem]{Example}

\theoremstyle{remark}

\numberwithin{equation}{section}

\numberwithin{figure}{section}

\let\trueint=\int
\let\truesum=\sum
\def\int{\mathop{\textstyle\trueint}\limits}
\def\sum{\mathop{\textstyle\truesum}\limits}

\let\<=\langle
\let\>=\rangle
\def\Real{{\mathbb{R}}}

\renewcommand\labelitemi{\ifmmode\circ\else$\circ$\fi}

\newlength{\myVSpace}
\newlength{\bigmyVSpace}
\newcommand\bigxstrut{\raisebox{\bigmyVSpace}
  {\rule{0pt}{\bigmyVSpace}}%
}

\begin{document}


\title{The Finite Non-periodic Toda Lattice: A Geometric and Topological Viewpoint}

\author{Yuji Kodama$^1$}
\address{Department of Mathematics, Ohio State University, Columbus,
OH 43210}
\email{kodama@math.ohio-state.edu}
\author{Barbara Shipman$^{2}$}
\thanks{$^1$Partially
supported by NSF grant DMS0404931}

\address{Department of Mathematics, The University of Texas at Arlington, Arlington TX}
\email{bshipman@uta.edu}

\begin{abstract} In 1967, Japanese physicist Morikazu Toda published the seminal papers \cite{Toda I} and \cite{Toda II}, 
exhibiting soliton solutions to a chain of particles with nonlinear interactions between nearest neighbors.
In the decades that followed, Toda's system of particles has been generalized in different directions, each with its own analytic, geometric, and topological characteristics that sets it apart from the others.  These are known collectively as the Toda lattice.  This survey describes and compares several versions of the finite non-periodic Toda lattice from the perspective of their geometry and topology. 

\end{abstract}

\maketitle

\thispagestyle{empty}
\pagenumbering{roman}\setcounter{page}{1}
\tableofcontents

\pagenumbering{arabic}
\setcounter{page}{1}

\section{Outline of the paper}

We organize the paper as follows:

Section \ref{Finite non-periodic real Toda lattice} introduces the real non-periodic Toda lattices.  It begins with two formulations of the Toda 
lattice in which the flow obeys a Lax equation on a set of real tridiagonal matrices whose subdiagonal entries are positive. 
The matrices are symmetric in one formulation and Hessenberg in the other.  The flows exist for all time and preserve the spectrum of the initial 
condition, and the system is completely integrable.  The references cited for these results are 
\cite{Arnold, Flaschka I, Flaschka II, Henon, Lax, manakov:74, Moser, Symes 1980}.
When the matrices of the Toda lattice system are extended to allow the subdiagonal entries to take on any real value, the two forms of the Toda lattice differ in the behavior of their solutions and in the topology of their isospectral sets.  In the symmetric case, the flows exist for all time and the isospectral manifolds are compact, while in the Hessenberg form, the flows blow up in finite time and the isospectral manifolds are not compact.  
For these results, the cited references are \cite{CK:02, Casian Kodama, Kodama Ye I, Kodama Ye II, RSTS, RS:94, Tomei}.
Section 1 concludes with the full symmetric real Toda lattice, in which the flows evolve on the set of all real symmetric matrices.  The symplectic structure comes from the Lie-Poisson structure on the dual of a Borel subalgebra of $sl(n,\Real)$.  In the full Toda lattice, more constants of motion are needed to establish complete integrability.  Cited references for this material are \cite{DLNT, Kodama McLaughlin, Symes 1982}.

Section \ref{Complex Toda lattices} describes the iso-level sets of the constants of motion in complex Toda lattices. 
The complex tridiagonal Toda lattice in Hessenberg form differs from the real 
case in that the flows no longer preserve "signs" of the subdiagonal entries.  The flows again blow up in finite (complex) time, and the isospectral manifolds are not compact.  These manifolds are compactified by embedding them into a flag manifold in two different ways.  In one compactification of isospectral sets with distinct eigenvalues, the flows enter lower-dimensional Bruhat cells at the blow-up times, where the singularity at a blow-up time is characterized by the Bruhat cell \cite{EFH, Kostant Whittaker}.
Under a different embedding, which works for arbitrary spectrum, the Toda flows generate a group action on the flag manifold, where the group depends on how eigenvalues are repeated.  The group is a product of a diagonal torus and a unipotent group, becoming the maximal diagonal torus when eigenvalues are distinct and a unipotent group when all eigenvalues coincide \cite{Nongeneric}.  
These group actions, together with the moment map \cite{Atiyah, GS, Kirwan}, are used in \cite{Complex Tridiagonal} to study the geometry of arbitrary isospectral sets of the complex tridiagonal Toda lattice in Hessenberg form.  These compactified iso-level sets are generalizations of toric varieties 
\cite{Oda}.  Further properties of the actions of these torus and unipotent components of these groups are found in 
\cite{FH G/P, Fixed Points, Torus Actions, Unipotent Actions, Unipotent Fixed Points, Connected}.
The survey then considers the full Kostant-Toda lattice, where the complex matrices in Hessenberg form are extended to have arbitrary complex entries below the diagonal.  The techniques for finding additonal constants of motion in the real symmetric case are adapted to the full Kostant-Toda lattice to obtain a complete family of integrals in involution on the generic symplectic leaves; the geometry of a generic iso-level set is then explained using flag manifolds \cite{EFS, GS:99, Ikeda, Symmetry}.
Nongeneric flows of the full Kostant-Toda lattice are described in terms of special faces and splittings of moment polytopes, and monodromy around nongeneric iso-level sets in special cases is determined \cite{PJM, Monodromy, Sl4, Nongeneric}.

Section \ref{Other Extensions} provides other extensions of the Toda lattice. In this paper, we only discuss those related to finite non-periodic Toda lattices. 
We first introduce the flow in the Lax form on an arbitrary diagonalizable matrix, which can be integrated by
the inverse scattering method (or equivalently by the factorization method) \cite{general matrix}.
We then show how the tridiagonal Hessenberg and symmetric Toda lattices, which are defined on the Lie algebra of type $A$ (that is, $sl(n)$),
are extended to semisimple Lie algebras using the Lie algebra splittings given by the Gauss decomposition and the QR decomposition, respectively
\cite{B:76, CK:02, CK:02a, Kostant, Guest, Perelomov}.
As an important related hierarchy of flows, we explain the Kac-van Moerbeke system, which can be considered as a 
square root of the Toda lattice \cite{GHSZ:93, KM:75}. 
We also show that the Pfaff lattice, which evolves on symplectic matrices, is related to the indefinite Toda lattice \cite{AM:02, KP:07, KP:08}.
As another aspect of the Toda lattice, we mention its gradient structure \cite{Bloch:90, BBR:90, BBR:92, BG:98, BG:07, dMP:99},
which explains the sorting property in the asymptotic behavior of the solution and can be used to solve problems in 
combinatorian optimization and linear programming \cite{Br:88, HM:94}.

Section \ref{KP connection} explains the connections with the KP equation \cite{IR:00, KP:70} in the sense that 
the $\tau$-functions as the solutions of the Toda lattices also provide a certain class of solutions of the KP equation
in a Wronskian determinant form \cite{BK:03, ChK:07, F:83, H:04, K:04, Kodama McLaughlin, MJD:00}.
This is based on the Sato theory of the KP hierarchy, which states that the solution of the
KP hierarchy is given by an orbit of the universal Grassmannian \cite{S:81}.
The main proposition in this section shows that there is a bijection between the set of $\tau$-functions 
that arise as a $k \times k$ Wronskian determiant and
the Grassmannian $Gr(k,n)$, the set of all $k$-dimensional subspaces of $\Real^n$. 
We present a method to obtain soliton solutions of the KP equation and give an elementary
introduction to the Sato theory in a finite-dimensional setting.
The moment polytopes of the fundamental representations of $SL(n,\mathbb{C})$ play a crucial role in describing the geometry of the 
soliton solutions of the KP equation.

Section \ref{Toda-cohomology connections} shows that the singular structure given by
the blow-ups in the solutions of the Toda lattice
contains information about the integral cohomology of real flag varieties \cite{CK:06, CK:07}. We begin with a detailed
study of the solutions of the indefinite Toda lattice hierarchy \cite{Kodama Ye II}.  The singular structure is determined by
the set of zeros of the $\tau$-functions of the Toda lattice.
First we note that the image of the moment map of the isospectral variety is a convex polytope whose vertices are the orbit of the Weyl group action
\cite{CK:02a, FH:91}.  
Each vertex of the polytope corresponds to
a fixed point of the Toda lattice, and it represents a unique cell of the flag variety.
Each edge of the polytope can be considered as an orbit of the $sl(2)$ Toda lattice (the smallest
nontrivial lattice), and it represents a simple reflection of the Weyl group.
There are two types of orbits, either regular (without blow-ups) or singular (with blow-ups).
Then one can define a graph, where the vertices are the fixed points of the Toda lattice and where two fixed points are connected
by an edge when the $sl(2)$ flow between them is regular.  If the flow is singular, then there is no edge between the two points.
  The graph defined in this way turns out to be the incidence
 graph that gives the integral cohomology of the real flag variety, where the incidence numbers associated to the edges are 
either $2$ or $-2$ \cite{K:95, CS:99}. We also show that the total number of blow-ups in the flow
 of the Toda lattice is related to the polynomial associated with the rational cohomology of a
 certain compact subgroup \cite{C:89, CK:06, CK:07}.

\section{Finite non-periodic real Toda lattice}\label{Finite non-periodic real Toda lattice}

Consider $n$ particles, each with mass 1, arranged along a line at positions $q_1, ..., q_n$.
Between each pair of adjacent particles, there is a force whose magnitude depends exponentially on the distance 
between them.  Letting $p_k$ denote the momentum of the $k$th particle, and noting that $\frac{d}{dt}q_k = p_k$ since 
each mass is 1, the total energy of the system is the Hamiltonian
\begin{equation}
H(p,q) = \frac{1}{2} \sum_{k = 1}^n\, p_k^2 + \sum_{k = 1}^{n-1} \,e^{-(q_{k+1} - q_k)} \ .
\label{Hamiltonian} \end{equation}
The equations of motion 
\begin{equation}\label{Hamiltonian DEs}\begin{array}{lllllll}
\displaystyle{\frac{dq_k}{dt} = \frac{\partial H}{\partial p_k} }\,, \\[2.0ex]
\displaystyle{\frac{dp_k}{dt} = -\frac{\partial H}{\partial q_k}}\,,
\end{array}\end{equation}
yield the system of equations for the finite non-periodic Toda lattice,
\begin{equation}\label{Toda Equations}\begin{array}{lllllll}
\displaystyle{\frac{dq_k}{dt}=p_k,} & k = 1, ..., n,  \\[2.0ex]
\displaystyle{\frac{dp_k}{dt} = -e^{-(q_{k+1} - q_k)} + e^{-(q_k - q_{k-1})},} \ \ \ & k = 1, ..., n.
\end{array}\end{equation}
Here we set $e^{-(q_1 - q_0)} = 0$ and $e^{-(q_{n+1} - q_n)} = 0$
with the formal boundary conditions
$$q_0 = -\infty, \hspace{1cm} q_{n+1} = \infty \ .$$

In the 1970's, the complete integrability of the Toda lattice was discovered by Henon \cite{Henon}
and Flaschka \cite{Flaschka I} in the context of the periodic form of the lattice, where the boundary condition
is taken as $q_0 = q_n$. 
Henon \cite{Henon} found analytical expressions for the constants of motion  
following indications by computer studies at the time that the Toda lattice should be completely 
integrable.  That same year, Flaschka \cite{Flaschka I, Flaschka II} (independently also by Manakov \cite{manakov:74}) showed that the periodic Toda lattice equations can be written in Lax form through an appropriate change of variables.  The complete integrability of the finite non-periodic Toda lattice
was established by Moser \cite{Moser} in 1980.

A system in Lax form \cite{Lax} gives the constants of motion as
eigenvalues of a linear operator. 
In the finite non-periodic case, there are two standard Lax forms of the Toda equations.

\subsection{Symmetric form} \label{Symmetric Toda}

Consider the change of variables (Flaschka \cite{Flaschka I}, Moser \cite{Moser})
\begin{equation}\label{Symmetric Variables}\begin{array}{llll}
\displaystyle{a_k = \frac{1}{2}e^{-\frac{1}{2}(q_{k+1} - q_k)}}, \ \ \ & k = 1, ..., n-1\\[2.0ex]
\displaystyle{b_k = -\frac{1}{2}\,p_k,} \ \ \ & k = 1, ..., n \ .
\end{array}\end{equation}
In these variables, the Toda system (\ref{Toda Equations}) becomes
\begin{equation}\label{Symmetric Toda Equations} \begin{array}{llll}
\displaystyle{\frac{da_k}{dt} = a_k(b_{k+1} - b_k),} \ \ \ & k = 1, ..., n-1 \\[2.0ex]
\displaystyle{\frac{db_k}{dt} = 2(a_k^2 - a_{k-1}^2),} \ \ \  & k = 1, ..., n  
\end{array}\end{equation}
with boundary conditions 
$$a_0 = 0, \ \ \ a_n = 0 \ .$$
This can be written in Lax form as 
\begin{equation}
\frac{d}{dt}L(t) = [\mbox{Skew}(L(t)), \ L(t)] \label{Symmetric Lax Equation}
\end{equation}
with
\begin{equation}
L = \left( \begin{array}{cccc} b_1 & a_1 &  & \\ 
a_1 & \ddots & \ddots &  \\
 & \ddots & \ddots & a_{n-1} \\
 &  & a_{n-1} & b_n
\end{array}  \right)\,, \label{Symmetric}
\end{equation} 
\begin{equation} 
\mbox{Skew}(L): = \left( \begin{array}{cccc} 0 & a_1 &  & \\ 
-a_1 & \ddots & \ddots &  \\
 & \ddots & \ddots & a_{n-1} \\
 &  & -a_{n-1} & 0
\end{array} \right) \label{Skew} \ .
\end{equation}

Any equation in the Lax form $\frac{d}{dt}L = [B, L]$ for matrices $L$ and $B$ has the 
immediate consequence that the flow
preserves the spectrum of $L$.  To check this, it suffices to show that the functions $\mbox{tr}(L^k)$ 
are constant for all $k$.
One shows first by induction that $\frac{d}{dt}L^k= [B, L^k]$ and then observes that 
$\frac{d}{dt}[\mbox{tr} (L^k)]  =  \mbox{tr}[\frac{d}{dt}(L^k)] =  \mbox{tr}[B, L^k] = 0$.
We now have $n-1$ independent invariant functions
$$H_k(L) = \frac{1}{ k + 1}  \ \mbox{tr} L^{k + 1} \,.$$
The Hamiltonian (\ref{Hamiltonian}) is related to $H_1(L)$ by 
$$H_1(L) = \frac{1}{4}H(p,q)$$
with the change of variables (\ref{Symmetric Variables}).

If we now fix the value of $H_0(L) = \ \mbox{tr} L$ (thus fixing the momentum of the system), the resulting phase space
has dimension $2(n-1)$.  With total momentum zero, this is
\begin{equation}
{\mathcal S} = \left\{ \left( \begin{array}{cccc} b_1 & a_1 &  & \\ 
a_1 & \ddots & \ddots &  \\
 & \ddots & \ddots & a_{n-1} \\
 &  & a_{n-1} & b_n
\end{array} \right) \, :~ a_i >0, \ \ b_i \in {\mathbb{R}}, \ \  \sum_{i=1}^n b_i = 0 \right\} \ . \label{S}
\end{equation}
A property of real tridiagonal symmetric matrices (\ref{Symmetric}) with $a_k \neq 0$ for all $k$ is that the eigenvalues $\lambda_k$ are real and distinct.  Let $\Lambda$ be a set of $n$ real distinct eigenvalues, and let ${\mathcal M} = \{L \in {\mathcal S} : \mbox{spec}(L) = \Lambda \ \}$.  Then
${\mathcal S} = \cup_{\Lambda} {\mathcal M}_{\Lambda}$.
${\mathcal S}$ is in fact a symplectic manifold.  Each invariant function $H_k(L)$ generates a Hamiltonian flow via the symplectic structure, and the flows are involutive with respect to the symplectic structure (see \cite{Arnold} for the general framework and \cite{Notes} for the Toda lattice specifically).  
We will describe the Lie-Poisson structure for the Toda lattice equations (\ref{Symmetric Lax Equation})
in Section \ref{Full Symmetric real Toda lattice}; however, 
we do not need the symplectic structure explicitly here.   

Moser \cite{Moser} analyzes the dynamics of the Toda particles, showing that  
for any initial configuration, $q_{k+1} - q_k$ tends to 
$\infty$ as $t \to \pm \infty$.  Thus,  
the off-diagonal entries of $L$ tend to zero as $t \to \pm \infty$ so that
$L$ tends to a diagonal matrix whose diagonal entries are the eigenvalues.
We will order them as $\lambda_1 < \lambda_2 < \cdots < \lambda_n$.
The analysis in \cite{Moser} shows that $L(\infty) = \mbox{diag}(\lambda_n, \lambda_{n-1}, \cdots, \lambda_1)$
and $L(-\infty) = \mbox{diag}(\lambda_1, \lambda_2, \cdots, \lambda_n)$.  The physical interpretation of this is
that as $t \to -\infty$, the particles $q_k$ approach the velocities $p_k(-\infty) = -2\lambda_k$, and as $t \to \infty$, 
the velocities are interchanged so that $p_k(\infty) = -2\lambda_{n-k}$.
Asymptotically, the trajectories behave as
\begin{eqnarray}
q_k(t) & \approx & \lambda_k^{\pm} t + c_k^{\pm} \nonumber \\
p_k(t) & \approx & \lambda_k^{\pm} \ , \nonumber
\end{eqnarray}
where $\lambda_k^+ = \lambda_k$ and $\lambda_k^- = \lambda_{n-k}$.

Symes solves the Toda lattice using matrix factorization, the QR-factorization;
his solution, which he verifies in \cite{Symes 1982} and proves in a more general context in \cite{Symes 1980}
is equivalent to the following.  To solve (\ref{Symmetric Toda Equations}) with initial matrix $L(0)$,
take the exponential $e^{tL(0)}$ and use Gram-Schmidt orthonormalization to factor it as
\begin{equation}
e^{tL(0)} = k(t)r(t) \ , \label{orthog x upper}
\end{equation}
where $k(t) \in SO(n)$ and $r(t)$ is upper-triangular.
Then the solution of (\ref{Symmetric Toda Equations}) is
\begin{equation}\label{TodaL}
L(t) = k^{-1}(t) L(0) k(t)=r(t)L(0)r^{-1}(t) \ .
\end{equation}
Since the Gram-Schmidt orthonormalization of $e^{tL(0)}$ can be done for all $t$, this shows that
the solution of the Toda lattice equations (\ref{Symmetric Toda Equations}) on the set (\ref{S})
is defined for all $t$.  

We also mention the notion of the $\tau$-functions which play a key role of the theory of
integrable systems (see for example \cite{H:04}). Let us first introduce the following symmetric
matrix called the moment matrix,
\begin{equation}\label{Moment Toda}
M(t):=e^{tL(0)^T}\cdot e^{tL(0)}=e^{2tL(0)}=r^T(t)k^T(t)k(t)r(t)=r^T(t)r(t)\,,
\end{equation}
where $r^T$ denotes the transpose of $r$, and note $k^T=k^{-1}$.
The decomposition of a symmetric matrix to an upper-triangular matrix times its transpose on the left is called the Cholesky factorization.
This factorization is used to find the matrix $r$, and then the matrix $k$ can be found
by $k=e^{tL(0)}r^{-1}$.  The $\tau$-functions, $\tau_j$ for $j=1,\ldots, n-1$, are defined by
\begin{equation}\label{Todatau}
\tau_j(t):={\rm det}\,(M_j(t))=\prod_{i=1}^j r_i(t)^2\,, 
\end{equation}
where $M_j$ is the $j\times j$ upper-left submatrix of $M$, and we denote ${\rm diag}(r)={\rm diag}(r_1\ldots,r_n)$. Also note from (\ref{TodaL}), i.e.
$L(t)r(t)=r(t)L(0)$, that
we have
\[
a_j(t)=a_j(0)\frac{r_{k+1}(t)}{r_{k}(t)}\,.
\]
With (\ref{Todatau}), we obtain
\begin{equation}\label{a}
a_j(t)=a_j(0)\frac{\sqrt{\tau_{j+1}(t)\tau_{j-1}(t)}}{\tau_j(t)}\,,
\end{equation}
and from this we can also find the formulae for $b_j(t)$ of $L(t)$ as
\[
b_j(t)=\frac{1}{2}\frac{d}{dt}\ln\left(\frac{\tau_j(t)}{\tau_{j-1}(t)}\right)\,.
\]
One should note that the $\tau$-functions are just defined from the moment matrix $M=e^{2tL(0)}$,
and the solutions $(a_j(t), b_j(t))$ are explicitly given by those $\tau$-functions {\it without}
the factorization.

\subsection{Hessenberg form}

The symmetric matrix $L$ in (\ref{Symmetric}), when conjugated by the diagonal matrix \linebreak
$D = \mbox{diag}(1, a_1, \ldots, a_{n-1})$, yields a matrix $Y = D L D^{-1}$ in Hessenberg form:
\begin{equation}
Y = \left( \begin{array}{cccc} b_1 & 1 &  & \\ 
a_1^2 & \ddots & \ddots &  \\
 & \ddots & \ddots & 1 \\
 &  & a_{n-1}^2 & b_n
\end{array} \right) \ , \label{Hessenberg Y}
\end{equation}
The Toda equations now take the Lax form
\begin{equation}
\frac{d}{dt}Y = 2[Y, F]  \label{Hessenberg equation I}
\end{equation}
with
$$F = \left( \begin{array}{cccc} 0 & 0 &  & \\ 
a_1^2 & \ddots & \ddots &  \\
 & \ddots & \ddots & 0 \\
 &  & a_{n-1}^2 & 0
\end{array} \right) \ .$$ 
Notice that $Y(t)$ satisfies (\ref{Hessenberg equation I}) if and only if $X(t) = 2Y(t)$ satisfies
\begin{equation}
\frac{d}{dt}X = [X, \Pi_{{\mathcal N}_-}X] \ , \label{Hessenberg equation}
\end{equation}
where $\Pi_{{\mathcal N}_-}A$ is the strictly lower-triangular 
part of $A$ obtained by setting all entries on and above the diagonal equal to zero.
Equation (\ref{Hessenberg equation}) with 
\begin{equation}
X = \left( \begin{array}{cccc} f_1 & 1 &  & \\ 
g_1 & \ddots & \ddots &  \\
 & \ddots & \ddots & 1 \\
 &  & g_{n-1} & f_n
\end{array} \right)  \label{Hessenberg}
\end{equation}
is called the asymmetric, or Hessenberg, form of the non-periodic Toda lattice.
Again, since the equations are in Lax form, the functions 
$H_k(X) = \frac{1}{ k + 1}  \ \mbox{tr} X^{k + 1} \,.$ are constant in $t$.  

Notice that the Hessenberg and symmetric Lax formulations of (\ref{Toda Equations})
are simply different
ways of expressing the same system.  The solutions exist for all time and exibit the same behavior as $t \to \pm \infty$ in both cases.  However, when we generalize the Toda lattice to allow the subdiagonal entries to take on any real value, the
symmetric and Hessenberg forms differ in their geometry and topology and in the character of their solutions.  

\subsection{Extended real tridiagonal symmetric form}

Consider again the Lax equation
\begin{equation}
\frac{d}{dt}L = [\mbox{Skew}(L), \ L] \ , \label{Symmetric Lax}
\end{equation}
where we now extend the set of matrices $L$ of the form (\ref{Symmetric}) by allowing $a_k$ to take on any real value.
(Recall that in our original definition of the matrix $L$, each $a_k$ was an exponential and was therefore strictly positive.)
As before, $b_k$ may be any real value, and   
$\mbox{Skew}L$ is defined as in (\ref{Skew}).

Given any initial matrix in this extended form, the factorization method of Symes, described in Section \ref{Symmetric Toda}, still works.
Indeed, for any such initial matrix $L(0)$, $e^{tL(0)}$ can be factored into 
(orthogonal)$\times$(upper-triangular) via the Gram-Schmidt procedure, 
i.e. the QR-factorization, and one can verify as before that
$L(t) = k^{-1}(t) L(0) k(t)$, where $k(t)$ is the orthogonal factor. 
Thus, in the extended symmetric form, solutions are still defined for all $t$. 
Given the initial eigenmatrix of $L$, 
an explicit solution of (\ref{Symmetric Lax})
can be found in terms of the eigenmatrix of $L(0)$ using the method of inverse scattering. 

For a general Lax equation $\frac{d}{dt}Y = [B, Y]$, if $Y(0)$ has distinct eigenvalues $\lambda_1, ..., \lambda_n$, then the inverse scattering scheme works as follows.  Let $\Lambda$ be the diagonal matrix of eigenvalues,
$\Lambda = \mbox{diag}(\lambda_1, ..., \lambda_n)$, and let $\Phi(t)$ be a matrix of normalized eigenvectors, 
varying smoothly in $t$,
where the $k$th column is a normalized eigenvector of $Y(t)$ with eigenvalue $\lambda_k$.  
Then the Lax equation $\frac{d}{dt}Y = [B, Y]$ is the compatibility condition for the equations
\begin{align}
Y(t)\Phi(t)& =  \Phi(t) \Lambda \label{A} \\
\displaystyle{\frac{d}{dt}\Phi(t) }& =  B(t) \Phi(t) \ . \label{B} 
\end{align}
We see this as follows: Denoting $(\cdot)'=\frac{d}{dt}(\cdot)$,
\[\begin{array}{llllll}
 \displaystyle{(Y \Phi)' = (\Phi \Lambda)'} & 
 \Rightarrow & \displaystyle{(Y)' \Phi +  Y(\Phi)'  =( \Phi)'\Lambda } & \\[0.5ex]
& \Rightarrow &(Y )'\Phi +  Y B \phi  =  B \Phi \Lambda  & \quad\mbox{by} \ (\ref{B}) \\[0.5ex]
& \Rightarrow &(Y )'\Phi  =  -Y  B \Phi + B Y \Phi &\quad \mbox{by} \ (\ref{A}) \\[0.5ex]
& \Rightarrow &(Y)' \Phi = [B, Y] \Phi   & \\[0.5ex]
& \Rightarrow &(Y)' = [B, Y] \ . &
\end{array}\]
The inverse scattering method solves the system (\ref{A}) and (\ref{B}) for $\phi(t)$ and then recovers $Y(t)$
from (\ref{B}).  Since $B(t)$ is defined as a projection of $Y(t)$, which can be written in terms of $\Phi(t)$ and 
$\Lambda$, we obtain a differential equation for $\Phi(t)$  by replacing $B(t)$ in (\ref{B}) by its expression
in terms of $\Phi(t)$ and $\Lambda$.  
Given $Y(0)$, we then obtain $\Phi(0)$ from (\ref{A}) with $t = 0$.
This solution is in fact equivalent to the QR factorization given above (see Section \ref{Symmetric Toda}).

For real matrices of the form (\ref{Symmetric}), the inverse scattering method can be used on the 
open dense subset where all $a_k$ are nonzero.  This is because
a real matrix $L$ of the form (\ref{Symmetric}) has distinct real eigenvalues if $a_k \neq 0$ for all $k$.
The eigenvalues are real because $L$ is a real symmetric matrix; the fact that they are distinct follows 
from the tridiagonal form with nonzero $a_k$, which forces there to be one eigenvector (up to a scalar) for each 
eigenvalue.

Let $\mathcal{M}_{\Lambda}$ denote the set of $n \times n$ matrices of the form (\ref{Symmetric}) with fixed eigenvalues 
$\lambda_1 < \lambda_2 < \cdots < \lambda_n$.  $\mathcal{M}_{\Lambda}$ contains $2^{n-1}$ components of dimension $n-1$, 
where each component consists of
all matrices in $\mathcal{M}_{\Lambda}$ with a fixed choice of sign for each $a_k$.   The solution of (\ref{Symmetric Lax}) with initial condition in a given 
component remains in that component for all $t$; that is, the solutions preserve the sign of each $a_k$. 
Each lower-dimensional component, where one or more $a_k$ is zero and the signs of the other $a_k$ are fixed,
is also preserved by the Toda flow. Adding those lower dimensional components gives a
compactification of each component of $\mathcal{M}_{\Lambda}$ with fixed signs in $a_k$'s.
Tomei \cite{Tomei} shows that $\mathcal{M}_{\Lambda}$ is a compact smooth manifold of dimension $n-1$.  In the proof of this, he uses the Toda flow to construct coordinate charts around the fixed points.  Tomei shows that
$\mathcal{M}_{\Lambda}$ is orientable with universal covering ${\mathbb{R}}^{n-1}$ and calculates its Euler characteristic. 

\begin{figure}[t!]
\centering
\includegraphics[scale=0.9]{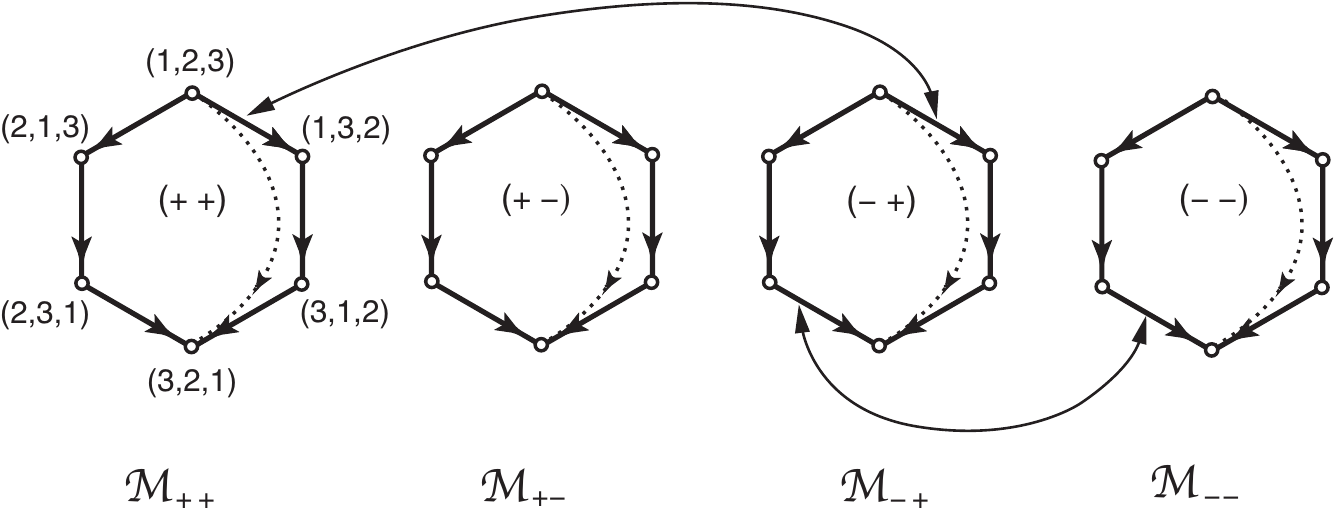}
\caption{The Tomei manifold $\mathcal{M}_{\Lambda}$ for the $sl(3,\Real)$ Toda lattice.
The 3-tuples $(i,j,k)$ on the vertices indicate the diagonal matrices $L=\mbox{diag}(\lambda_i,\lambda_j,\lambda_k)$.
Each hexagon $\mathcal{M}_{\epsilon_1,\epsilon_2}$ corresponds to the moment polytope
(see Section \ref{moment}) for the Toda lattice with the signs $(\epsilon_1,\epsilon_2)=({\rm sgn}(a_1),{\rm sgn}(a_2))$. 
The boundaries correspond to the $sl(2,\Real)$ Toda lattices associated with either $a_1=0$ or $a_2=0$.
The Tomei manifold $\mathcal{M}_{\Lambda}$ is given by gluing edges of the hexagons. For example,
the edge between $(1,2,3)$ and $(1,3,2)$ in $\mathcal{M}_{++}$ is glued with the same edge in
$\mathcal{M}_{-+}$, since this edge indicates $a_1=0$. The other gluing shown in the figure is for the edges
corresponding to $a_2=0$.}
\label{fig:tomei}
\end{figure}
In his analysis, Tomei shows that $\mathcal{M}_{\Lambda}$ contains $2^{n-1}$ open components diffeomorphic to ${\mathbb{R}}^{n-1}$.  On each of these components, $a_k \neq 0$ for all $k$, and the sign of each $a_k$ is fixed.   They are glued together along the 
lower-dimensional sets where one or more $a_k$ is zero.  For example, in the case $n = 3$, there are four 2-dimensional
components, denoted as $\mathcal{M}_{++}, \mathcal{M}_{+-}, M_{-+}$, and $\mathcal{M}_{--}$, according to the signs of $a_1$ and $a_2$. 
The closure of each component is obtained by adding six diagonal matrices where all the $a_k$ vanish (these are the fixed points of the Toda flow)
and six 1-dimensional sets where exactly one $a_k$ is zero.  Denote the closure of $\mathcal{M}_{++}$
by $\overline{\mathcal{M}_{++}}$, and so on.
The boundary of $\overline{\mathcal{M}_{++}}$, for example, contains three 
1-dimensional sets with $a_1 = 0$ and $a_2 > 0$.  Each is characterized by having a
fixed eigenvalue as its first diagonal entry.  The other three 1-dimensional components in the boundary
of $\overline{\mathcal{M}_{++}}$, have $a_1 > 0$ and $a_2 = 0$, with a fixed eigenvalue in the third diagonal entry. Adding those boundaries with the 6 vertices corresponding to the diagonal matrices
$L={\rm diag}(\lambda_i,\lambda_j,\lambda_k)$ gives the compactified set $\overline{\mathcal{M}}_{++}$.
Notice that the three 1-dimensional components
with $a_1 = 0, a_2 > 0$, and a fixed $\lambda_k$ in the first diagonal entry also lie along the boundary of 
$\overline{\mathcal{M}_{-+}}$; the other three components, with $a_1 > 0$ and $a_2 = 0$ are shared by the boundary of $\overline{\mathcal{M}_{++}}$. 
In this manner, the four principal components are glued together along the subset of $\mathcal{M}_{\Lambda}$ where one or more
$a_k$ vanish.   
In Figure \ref{fig:tomei}, we illustrate the compactification of the Tomei manifold 
$\mathcal{M}_{\Lambda}$ for the $sl(3,\Real)$ symmetric Toda lattice,
\[
\mathcal{M}_{\Lambda}=\overline{\mathcal{M}}_{++}\cup\overline{\mathcal{M}}_{+-}\cup\overline{\mathcal{M}}_{-+}\cup\overline{\mathcal{M}}_{--}\,,
\]
where the cups include the specific gluing according to the signs of the $a_k$ as explained above.
The resulting compactified manifold $\mathcal{M}_{\Lambda}$ is a connected sum of two tori, the compact Riemann surface of genus two. This can be easily seen from Figure \ref{fig:tomei} as follows:
Gluing those four hexagons, $\mathcal{M}_{\Lambda}$ consists of 6 vertices, 12 edges and 4 faces. Hence the Euler characteristic
is given by $\chi(\mathcal{M}_{\Lambda})=6-12+4=-2$, which implies that the manifold has genus $g=2$ (recall $\chi=2-2g$).
It is also easy to see that $\mathcal{M}_{\Lambda}$ is orientable (this can be shown by giving
an orientation for each hexagon so that the directions of two edges in the gluing cancel each other).
Since the compact two dimensional surfaces are completely characterized by
their orientability and the Euler characters, we conclude that
the manifold $\mathcal{M}_{\Lambda}$ is a connected sum of two tori, i.e. $g=2$.

The Euler characteristic of $\mathcal{M}_{\Lambda}$ (for general $n$) is determined in  \cite{Tomei} as follows.  Let 
$L = \mbox{diag}(\lambda_{\sigma(1)}, ..., \lambda_{\sigma(n)})$ be a diagonal matrix in $\mathcal{M}_{\Lambda}$, where $\sigma$ is
a permutation of the numbers $\{1, ..., n\}$, and let $r(L)$ be the number of times that $\sigma(k)$ is less
than $\sigma(k+1)$.  Denote by $E(n,k)$ the number of diagonal matrices in $\mathcal{M}_{\Lambda}$ with $r(L) = k$.
Then the Euler characteristic of $\mathcal{M}_{\Lambda}$ is the alternating sum of the $E(n,k)$:
$$\chi(\mathcal{M}_{\Lambda}) = \sum_{k=0}^{n} (-1)^k E(n,k) \ .$$
[In M. Davis et al extends Tomei's result....]

\begin{figure}[t!]
\centering
\includegraphics[scale=0.9]{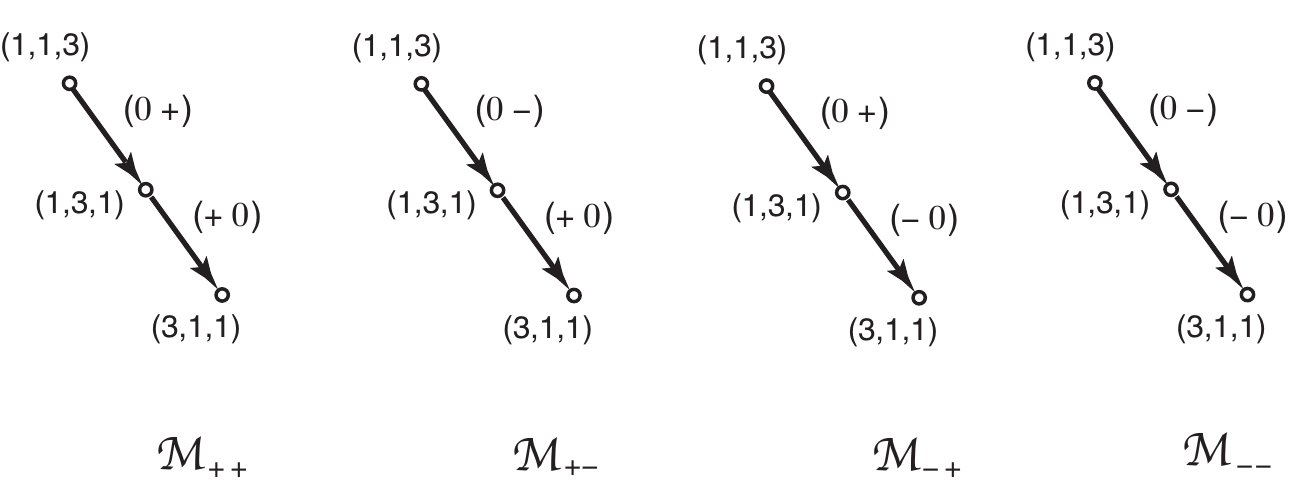}
\caption{The isospectral set of the matrices having the eigenvalues $\{1,1,3\}$.
Each polytope $\mathcal{M}_{\epsilon_1,\epsilon_2}$ contains the set of matrices with
either $a_1=0$ or $a_2=0$ and the signs $\epsilon_i={\rm sgn}(a_i)$.
These polytopes are obtained by squeezing the polytopes for the semisimple $sl(3)$ Toda
lattice in Figure \ref{fig:tomei} according to the degeneration of the eigenvalues $2\to 1$.
The gluing pattern is the same as in Figure \ref{fig:tomei}, that is,
identify, for example, the edge $(0\,+)$ in $\mathcal{M}_{++}$ with the same one in $\mathcal{M}_{-+}$. The resulting variety is singular 
and has a shape of figure eight.}
\label{fig:singular}
\end{figure}

If the eigenvalues of the tridiagonal real matrix $L$ are not distinct, then one or more $a_k$ must be zero.
The set of such matrices with fixed spectrum where the eigenvalues are not distinct is not a manifold.
For example, when $n = 3$ and the spectrum is $(1,1,3)$, the isospectral set is one-dimensional since one $a_i$ is zero.  It contains
three diagonal matrices $D_1 =  \mbox{diag}(3,1,1)$, $D_2 =  \mbox{diag}(1,3,1)$, and $D_3 =  \mbox{diag}(1,1,3)$,
and four 1-dimensional components.  Each 1-dimensional component has a 1 in either the first or last diagonal entry and a $2 \times 2$ block 
on the diagonal with eigenvalues 1 and 3, where the off-diagonal entry is
either positive or negative.  The two components with the $1 \times 1$ block in the last diagonal entry connect 
$D_1$ and $D_2$, and the two components with the $1 \times 1$ block in the first diagonal entry
connect $D_2$ and $D_3$. In Figure \ref{fig:singular}, we illustrate the isospectral set of those matrices which is singular
with a shape of figure eight.

\subsection{Extended real tridiagonal Hessenberg form}\label{Extended Hessenberg Toda}

We now return to the Hessenberg form of the Toda equations with
\begin{equation}\label{X indefinite}
X = \left( \begin{array}{cccc} f_1 & 1 &  & \\ 
g_1 & \ddots & \ddots &  \\
 & \ddots & \ddots & 1 \\
 &  & g_{n-1} & f_n
\end{array} \right)  
\end{equation}
as in (\ref{Hessenberg}), and allow the $g_k$ to take on arbitrary real values.
On the set of tridiagonal Hessenberg matrices $X$ with $f_k$ and $g_k$ real, the Toda flow is defined by
\begin{equation}
\frac{d}{dt}X(t) = [X(t), \Pi_{{\mathcal N}_-}X(t)] \ , \label{Hessenberg equation II}
\end{equation}
as in (\ref{Hessenberg equation}).

Recall that in the formulation of the original Toda equations, 
all the $g_k$ were positive, so that the eigenvalues were real and distinct.  When $g_k \neq 0$ for some $k$, the eigenvalues may now be complex or may coincide.  We will see that this causes blow-ups in the flows so that the topology of the isospectral manifolds are very different from the topology of the Tomei manifolds described in the previous section.

The matrices of the form (\ref{X indefinite}) with $g_k \neq 0$ for all $k$ are partitioned into $2^{n-1}$ different Hamiltonian systems,
each determined by a choice of signs of the $g_k$.   Letting $s_k = \pm 1$ for $k = 1, ..., n$
and taking the sign of $g_k$ to be $s_k s_{k+1}$, Kodama and Ye \cite{Kodama Ye II} give the Hamiltonian
for the system with this choice of signs as 
\begin{equation}
H = \frac{1}{2} \sum_{k = 1}^n\, y_k^2 + \sum_{k = 1}^{n-1} s_k s_{k+1} e^{-(x_{k+1} - x_k)} \ ,
\label{Indefinite Hamiltonian} \end{equation}
where
\[\begin{array}{llllll}
\displaystyle{f_k = -\frac{1}{2}\, y_k }\ , & k = 1, ..., n \\[2.0ex]
\displaystyle{g_k = \frac{1}{4}\,s_k s_{k+1}\, e^{-(x_{k+1} - x_k)},} \ \ \ & k = 1, ..., n-1 \ . 
\end{array}\]
The system (\ref{Hessenberg equation II}) is then called the {\it  indefinite} Toda lattice.
The negative signs in (\ref{Indefinite Hamiltonian}) correspond to attractive forces between adjacent particles,
which causes the system to become undefined at finite values of $t$, as is seen in the solutions obtained by
Kodama and Ye in \cite{Kodama Ye I} and \cite{Kodama Ye II} by inverse scattering.

The blow-ups in the solutions are also apparent in the factorization solution of the Hessenberg form.
To solve (\ref{Hessenberg equation II}) with initial condition $X(0)$,
factor the exponential $e^{tX(0)}$ as
\begin{equation}
e^{tX(0)} = n(t)b(t) \ , \label{lower x upper}
\end{equation}
where $n(t)$ is lower unipotent and $b(t)$ is upper-triangular.
Then, as shown by \cite{RSTS} and \cite{Reiman}, 
\begin{equation}
X(t) = n^{-1}(t) X(0) n(t)=b(t) X(0) b^{-1}(t) \label{factorization solution}
\end{equation}
solves (\ref{Hessenberg equation II}).
Notice that the factorization (\ref{lower x upper}) is obtained by Gaussian elimination, which multiplies 
$e^{tX(0)}$ on the left by elementary row operations to put it in upper-triangular form.  This process works only
when all principal determinants (the determinants of upper left $k \times k$ blocks, which are the 
$\tau$-functions defined below) are nonzero.
At particular values of $t \in {\mathbb{R}}$, this factorization can fail, and the solution (\ref{factorization solution})
becomes undefined.

The solutions $(f_k,g_k)$ can be expressed in terms of the $\tau$-functions which are defined by
\begin{equation}\label{indefinite tauk}
\tau_k(t):= {\rm det}\, \left[(e^{tX(0)})_k\right]=\prod_{j=1}^k d_{j}(t)\,,
\end{equation}
where $(e^{tX(0)})_k$ is the $k\times k$ upper-left submatrix of $e^{tL(0)}$, and
${\rm diag}(b)={\rm diag}(d_1,\ldots,d_n)$. With (\ref{factorization solution}), we have
\begin{equation}\label{g indefinite}
g_k(t)=g_k(0)\frac{d_{k+1}(t)}{d_k(t)}=g_k(0)\frac{\tau_{k+1}(t)\tau_{k-1}(t)}{\tau_k(t)^2}\,.
\end{equation}
The function $f_k(t)$ are given by
\begin{equation}\label{f indefinite}
f_k(t)=\frac{d}{dt}\ln\left(\frac{\tau_k(t)}{\tau_{k-1}(t)}\right)\,.
\end{equation}
Now it it clear that the factorization (\ref{lower x upper}) fails if and only if $\tau_k(t)=0$ for some $k$.
Then a blow-up (singularity) of the system (\ref{Hessenberg equation II}) can be characterized by
the zero sets of the $\tau$-functions.

\begin{Example}\label{example 1.4.1} To see how blow-ups occur in the factorization solution, consider the initial matrix
$$X_0 =  \left( \begin{array}{rr} 1 & 1 \\ 
 -1 & -1 \end{array} \right) \ .$$
When $t \neq -1$,
$$e^{t X_0} = \left( \begin{array}{cc} 1+t & t \\ 
 -t & 1-t \end{array} \right) 
= \left( \begin{array}{cc} 1 & 0 \\ 
 \frac{-t}{1 + t} & 1 \end{array} \right) \ 
\left( \begin{array}{cc} 1+t & t \\ 
 0 & \frac{1}{1+t} \end{array} \right) \ , $$
and the solution evolves as in (\ref{factorization solution}).
The $\tau$-function is given by $\tau_1(t)=1+t$, and
when $t = -1$, this factorization does not work.  However,
we can multiply $e^{-1 X_0}$ on the left by a lower unipotent matrix $n^{-1}$ (in this case the identity)
to put it in
the form $w b$, where $w$ is a permutation matrix:
$$e^{-1 X_0} = \left( \begin{array}{cc} 0 & -1 \\ 
 1 & 2 \end{array} \right) \ = \ 
\left( \begin{array}{cc} 1 & 0 \\ 
 0 & 1 \end{array} \right) 
\left( \begin{array}{cc} 0 & -1 \\ 
 1 & 0 \end{array} \right) \ 
\left( \begin{array}{cc} 1 & 2 \\ 
 0 & 1 \end{array} \right)  \ .$$
This example will be taken up again in Section \ref{Complex tridiagonal Hessenberg}, where it is shown how the factorization using a permutation matrix
leads to a compactification of the flows.  
\end{Example}

In general, when the factorization (\ref{lower x upper}) is not possible at time $t = \bar{t}$, 
$e^{\bar{t}X(0)}$ can be factored as
$e^{\bar{t}X(0)} = n(\bar{t}) \ w \ b(\bar{t})$, where  $w$ is  a
permutation matrix.  Ercolani, Flaschka, and Haine \cite{EFH} use this factorization to complete the flows 
(\ref{factorization solution}) 
through the blow-up times by embedding them into a flag manifold.  
The details will be discussed in the next section, where we consider the
complex tridiagonal Hessenberg form of the Toda lattice. 

Kodama and Ye find explicit solutions of the indefinite Toda lattices by inverse scattering.
Their method is used to solve a generalization of the full symmetric Toda lattice in \cite{Kodama Ye I} and is
specialized to the indefinite tridiagonal Hessenberg Toda lattice in \cite{Kodama Ye II}.
For the Hamiltonian (\ref{Indefinite Hamiltonian}), Kodama and Ye make the change of variables
\begin{equation} \label{Indefinite Variables s}\begin{array}{lllll}
\displaystyle{a_k = \frac{1}{2} e^{-(x_{k+1} - x_k)/2}, }\  \ \  & k = 1, ..., n-1\,, \\[2.0ex]
\displaystyle{ s_k b_k = -\frac{1}{2}y_k }\ , \ \ \  & k = 1, ..., n\,, \ 
\end{array}\end{equation}
together with $t \to \frac{1}{2}t$ so that Hamilton's equations take the form
\begin{equation}\label{Indefinite Toda Equation 1}\begin{array}{llllll}
\displaystyle{\frac{da_k}{dt}= \frac{1}{2}\, a_k(s_{k+1} b_{k+1} - s_k b_k) } \,, \\[2.0ex]
\displaystyle{\frac{db_k}{dt} = s_{k+1}a_k^2 - s_{k-1}a_{k-1}^2 }\,,
\end{array}\end{equation}
with $a_0 = a_n = 0$. Here we switched the notation $a_k$ and $b_k$ from the original one in \cite{Kodama Ye I, Kodama Ye II}. This system is equivalent to (\ref{Hessenberg equation II}) with
$f_k=s_kb_k$ and $g_k=s_ks_{k+1}a_k^2$.
The system (\ref{Indefinite Toda Equation 1})  
can then be written in Lax form as
\begin{equation}
\frac{d}{dt} \tilde L = [\tilde B, \tilde L] \ , \label{indefinite Lax equation}
\end{equation}
where $\tilde L$ is the real tridiagonal matrix
\begin{equation}
\tilde L = \left( \begin{array}{ccccc} s_1 b_1 & s_2 a_1 & 0 & \cdots & 0 \\ 
s_1 a_1 & s_2 b_2 & s_3 a_2 & \cdots & 0  \\
\vdots & \ddots & \ddots & \ddots & \vdots \\
0 & \cdots & \ddots & s_{n-1} b_{n-1} & s_n a_{n-1} \\
0 & \cdots & \cdots & s_{n-1} a_{n-1} & s_n b_n
\end{array}  \right) \label{Indefinite Symmetric1}
\end{equation} 
and $\tilde B$ is the projection 
\begin{equation} 
\tilde B = \frac{1}{2}[(\tilde L)_{>0} -(\tilde L)_{<0}] \ . \label{Indefinite Lax Equation}
\end{equation}
The inverse scattering scheme for (\ref{Indefinite Lax Equation}) is
\begin{equation}
\tilde L \Phi = \Phi \Lambda, \ \ \ \ \ \frac{d}{dt}\Phi = \tilde B \Phi  \label{Inverse Scattering Equations}
\end{equation}
where $\Lambda = \mbox{diag}(\lambda_1, ..., \lambda_n)$ and $\Phi$ is the eigenmatrix of $\tilde L$,
normalized so that
\begin{equation}
\Phi S^{-1} \Phi^T = S^{-1}, \ \ \ \ \Phi^T S \Phi = S  \label{Normalization}
\end{equation}
with $S = \mbox{diag}(s_1, ..., s_n)$. Note that the matrix $\tilde L$ of (\ref{Indefinite Symmetric1})
is expressed as $\tilde L=LS$ with the symmetric matrix $L$ given by
(\ref{Symmetric}) for the original Toda lattice.
When $S$ is the identity, (\ref{Normalization}) implies that $\Phi$ is orthogonal, and for
$S = \mbox{diag}(1, ..., 1,-1, ..., -1)$, $\Phi$ is a pseudo-orthogonal matrix in $O(p, q)$
with $p + q = n$.  This then defines an inner product for functions $f$ and $g$ on a set $C$,
\[
\langle f,g\rangle:=\int_{C}f(\lambda)g(\lambda)\,d\mu(\lambda)=\sum_{k=1}^nf(\lambda_k)g(\lambda_k)s_k^{-1}\,,
\]
with the indefinite metric $d\mu:=
\sum_{k=1}^n s_k^{-1}\delta(\lambda-\lambda_k)\, d\lambda$ and the eigenvalues $\lambda_k$ of $\tilde L$.
Then the entries of $\tilde L$ can be expressed in terms of the eigenvector $\phi(\lambda)=(\phi_1(\lambda),\ldots,\phi_n(\lambda))^T$, i.e. $\tilde L \phi(\lambda)=\lambda\phi(\lambda)$,
\begin{equation}
(\tilde L)_{ij} = s_j\langle\lambda \phi_i, \phi_j\rangle=s_j\sum_{k=1}^n \lambda_k\phi_i(\lambda_k)\phi_j(\lambda_k)s_k^{-1} \ . \label{Lij}
\end{equation}
The explicit time evolution of $\Phi$ can be obtained using an orthonormalization procedure on functions of the 
eigenvectors that generalizes the method used in \cite{Kodama McLaughlin} to solve the full symmetric Toda hierarchy. 
A brief summery of the procedure is as follows:
First consider the factorization (called the HR-factorization),
\begin{equation}
e^{tX(0)}=r(t)h(t)\,, \label{HR}
\end{equation}
where $r(t)$ is a lower triangular matrix and $h(t)$ satisfies $h^TSh=S$ (if $S=I$, then $h\in SO(n)$, i.e.
the factorization is the QR-type). Then the eigenmatrix $\Phi(t)=(\phi_i(\lambda_j))_{1\le i,j\le n}$ is given by $\Phi(t)=h(t)\Phi(0)$.
Now one can write $\Phi(t)$ as
\begin{align*}
\Phi(t) &= h(t)\Phi(0)=r^{-1}(t)r(t)h(t)\Phi(0)\\
&=r^{-1}(t)e^{tX(0)}\Phi(0)= r^{-1}(t)\Phi(0)e^{t\Lambda}\,.
\end{align*}
Since $r(t)$ is lower triangular, this implies
\[
\phi_i(\lambda,t)={\rm Span}_{\Real}\left\{\phi_1^0(\lambda)e^{\lambda t},\ldots,\phi^0_i(\lambda)e^{\lambda t}\right\},\qquad i=1,\ldots,n\,.
\]
Then using the Gram-Schmidt orthogonalization method, 
the functions $\phi_i(\lambda,t)$ can be found as \cite{Kodama Ye II},
\begin{equation}
\phi_i(\lambda,t) = \frac{e^{\lambda t}}{\sqrt{D_i(t)D_{i-1}(t)}} 
\left| \begin{array}{ccccc} 
s_1 c_{11} & \cdots & s_{i-1}c_{1,i-1} & \phi_1^0(\lambda) \\ 
s_1 c_{21} & \cdots & s_{i-1} c_{2,i-1} & \phi_2^0(\lambda)\\
\vdots &  \ddots & \vdots & \vdots \\
s_1 c_{i1} & \cdots & s_{i-1}c_{i,i-1} & \phi_i^0(\lambda) \end{array}  \right| \ ,
\label{Inverse Scattering Solution}
\end{equation} 
where $\phi_i^0(\lambda) = \phi_i^0(\lambda, 0)$, $c_{ij}(t) = \langle\phi_i^0, \phi_j^0 e^{\lambda t}\rangle$,
and $D_k(t) = |(s_i c_{ij}(t))_{1 \leq i, j \leq k}|$.
The solution of the inverse scattering problem (\ref{Inverse Scattering Equations}) is then obtained from
(\ref{Inverse Scattering Solution}) using (\ref{Lij}).
The matrix $\tilde M:=(c_{ij})_{1\le i,j \le n}$ is the moment matrix for the indefinite Toda lattice which 
is defined in the similar way as (\ref{Moment Toda}), i.e.
\[
\tilde{M}(t):=e^{\frac{t}{2}\tilde{L}(0)}S^{-1}e^{\frac{t}{2}\tilde{L(0)}^T}=\Phi_0e^{t\Lambda}S^{-1}\Phi_0^T\,,
\]
where we have used $\tilde{L}(0)\Phi_0=\Phi_0\Lambda$ and $\Phi_0^TS\Phi_0=S$. Then the $\tau$-functions are
defined by 
\begin{equation}\label{indefinite tau}
\tilde\tau_k(t)={\rm det}\,(\tilde{M}_k(t))=\left|(c_{ij}(t))_{1\le i,j\le k}\right|=\frac{1}{s_1\cdots s_k}D_k(t)\,.
\end{equation}

 From (\ref{Inverse Scattering Solution}) it follows that when $\tilde\tau_k(\bar{t}) = 0$ for some $k$ and time $\bar{t}$,
$\tilde L(t)$ blows up to infinity as $t \to \bar{t}$. In \cite{Kodama Ye II},
Kodama and Ye characterize the blow-ups with the zeros of $\tau$-functions and  study the topology of a generic isospectral set $\mathcal{M}_{\Lambda}$ 
of the extended real tridiagonal Toda lattice in Hessenberg form.

It is first shown, using the Toda flows, that because of the blow-ups in $\tilde L$,
$\mathcal{M}_{\Lambda}$ is a noncompact manifold of dimension $n-1$.
The manifold is then compactified by completing the flows through the blow-up times.  The $2 \times 2$ case is basic to the compactification for general $n$.
The set of $2\times 2$ matrices with fixed eigenvalues $\lambda_1$ and $\lambda_2$,
\begin{equation}
\mathcal{M}_{\Lambda} = \Bigg\{ \left( \begin{array}{cc} f_1 & 1 \\ 
 g_1 & f_2 \end{array} \right) :  \lambda_1 < \lambda_2 \  \Bigg\} \label{2x2}
\end{equation}
consists of two components, $\mathcal{M}_{+}$ with $g_1 > 0$ and $\mathcal{M}_{-}$ with $g_1 < 0$, together with two
fixed points, 
$$\tilde L_1 = \left( \begin{array}{cc} \lambda_1 & 1 \\ 
 0 & \lambda_2 \end{array} \right) \hspace{.4cm} \mbox{and} \ \ \ 
\tilde L_2 = \left( \begin{array}{cc} \lambda_2 & 1 \\ 
 0 & \lambda_1 \end{array} \right) \ .$$
Writing $f_2 = \lambda_1 + \lambda_2 - f_1$ and substituting this into the equation for the determinant,
$f_1 f_2 - g_1 = \lambda_1 \lambda_2$, shows that $M_{\lambda}$ is the parabola
\begin{equation}
g_1 = -(f_1 - \lambda_1)(f_1 - \lambda_2) \label{parabola} \ .
\end{equation}
This parabola opens down, crossing the axis $g_1 = 0$ at $f_1 = \lambda_1$ and $f_1 = \lambda_2$, corresponding
to the fixed points $\tilde L_1$ and $\tilde L_2$.
For an initial condition with $g_1 > 0$, the solution is defined for all $t$; it flows away from $p_2$ toward
$p_1$.  This illustrates what is known as the sorting property, which says that as $t \to \infty$, 
the flow tends toward the fixed point with the eigenvalues in decreasing order along the diagonal.
The component with $g_1 < 0$ is separated into disjoint parts, one with $f_1 < \lambda_1$ and the other
with $f_1 > \lambda_2$.   The solution starting at an initial matrix with $f_1 > \lambda_2$ flows toward the fixed
point $\tilde L_2$ as $t \to \infty$.  For an initial matrix with $f_1 < \lambda_1$, the solution flows away from
$\tilde L_1$, blowing up at a finite value of $t$.  By adding a point at infinity to connect these two branches of the 
parabola, the flow is completed through the blow-up time and the resulting manifold is the circle, $S^1$.

\begin{figure}[t!]
\centering
\includegraphics[scale=0.9]{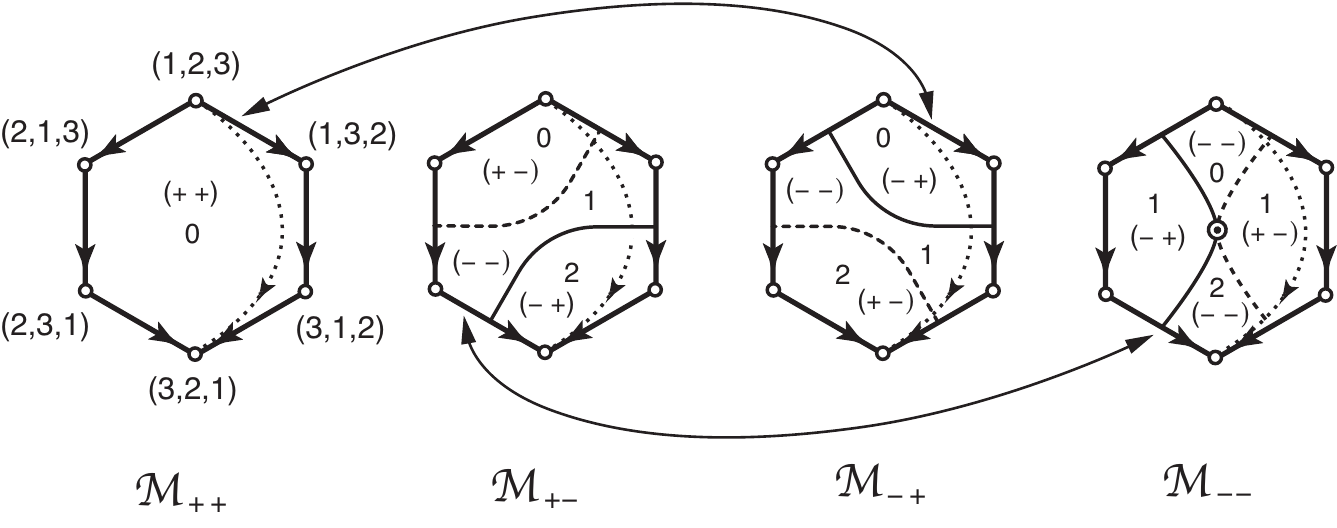}
\caption{The compactification of the isospectral manifold $\mathcal{M}_{\Lambda}$
for the indefinite $sl(3,\Real)$ Toda lattice. As in Figure \ref{fig:tomei}, each hexagon
indicates the moment polytope associated with the indefinite Toda lattice.
The signs $(\epsilon_1,\epsilon_2)$ in $\mathcal{M}_{\epsilon_1,\epsilon_2}$ are those of $(g_1,g_2)$ for $t\to-\infty$, and each sign in the hexagons indicates the signs of $(g_1,g_2)$.
The gluing rule according to the sign changes of $g_i$ is the same as that in the Tomei manifold, but the pattern is now different.
For example, the edge between $(2,3,1)$ and $(3,2,1)$ in $\mathcal{M}_{--}$ is now glued with
that in $\mathcal{M}_{+-}$. The solid and dashed lines in the hexagons
show the points where the solutions blow-up, i.e. $\tau_k=0$ for $k=1$ (solid) and for $k=2$ (dashed). 
Also the numbers in the hexagons indicate the number of blow-ups along the flow from
$t=-\infty$ to $+\infty$ (see Section \ref{blow-ups}).}
\label{fig:A2indefinite}
\end{figure}

For general $n$, the manifold $\mathcal{M}_{\Lambda}$ with spectrum $\Lambda$
contains $n!$ fixed points of the flow, where the eigenvalues are arranged along the diagonal.  
These vertices are connected to each other by incoming and outgoing edges analogous to the flows connecting the two
vertices in the case $n = 2$.  On each edge there is one $g_k$ that is not zero.
As in the case $n = 2$, edges in which the blow-ups occur are compactified by adding a 
point at infinity.   Kodama and Ye then show how to glue on the higher-dimensional components where more than one
$g_k$ is nonzero and compactify the flows through the blow-ups to produce a compact $n$-dimensional manifold.
The result is nonorientable for $n > 2$. In the case $n=3$, it is a connected sum of two Klein bottles. 
In Figure \ref{fig:A2indefinite}, we illustrate the compactification of $\mathcal{M}_{\Lambda}$ for
the $sl(3,\Real)$ indefinite Toda lattice. With the gluing, the compactified manifold $\overline{\mathcal{M}}_{\Lambda}$ has the Euler characteristic $\chi(\overline{\mathcal{M}}_{\Lambda})=-2$
as in the case of the Tomei manifold (see Figure \ref{fig:tomei}). The non-orientability can be shown by
non-cancellation of the given orientations of the hexagons with this gluing.

The compactification was further studied by Casian and Kodama \cite{CK:02} (also see
\cite{Casian Kodama}), where they show that the compactified isospectral manifold is identified as a connected completion of the disconnected Cartan subgroup of $G=Ad(SL(n,\mathbb{R})^{\pm})$. The manifold is diffeomorphic to a toric variety in the flag manifold associated with $G$.
They also give a cellular decomposition of the compactified manifold for computing the homology of
the manifold. We will show more details in Section \ref{blow-ups}, where
the gluing rules are given by the Weyl group action on the signs of the entries $g_j$ of the matrix $X$.

\subsection{Full Symmetric real Toda lattice}  \label{Full Symmetric real Toda lattice}

We now return to the symmetric Toda equation
\begin{equation}
\frac{d}{dt}L = [\, \mbox{Skew}(L), \ L] \label{full symmetric Toda}
\end{equation}
as in (\ref{Symmetric Lax Equation}), where $L$ is now a full symmetric matrix with distinct eigenvalues.  
As in the tridiagonal case, $\mbox{Skew}(L)$ is the skew-symmetric summand in the decomposition of 
$L$ into skew-symmetric
plus lower-triangular.
Deift, Li, Nanda, and Tomei \cite{DLNT} show that (\ref{full symmetric Toda}) remains completely integrable even in this case.
They present a sufficient number of constants of motion in involution and construct the associated angle variables.

The phase space of the full symmetric real Toda lattice is the set of symmetric matrices, which
we denote
\[
{\rm Sym}(n):=\left\{ L\in sl(n,\Real): L^T=L\right\}\,.
\]
One can then define the Lie-Poisson structure (the Kostant-Killilov 2-form) on this phase space as follows. First we define a nondegenerate
inner product (the Killing form), $\langle A,B\rangle={\rm tr}(AB)$, which identifies the dual space
$sl^*(n,\Real)$ with $sl(n,\Real)$. Then we consider the Lie algebra splitting,
\[
sl(n,\Real)=\mathcal{B}_-\oplus so(n)=\mathcal{B}_-^*\oplus so^*(n)\,,
\]
where $\mathcal{B}_-$ is the set of lower triangular matrices.  With the inner product, we identify
\[
\mathcal{B}_-^*\cong so^{\perp}(n)={\rm Sym}(n),\qquad
so^*(n)=\mathcal{B}_-^{\perp}=\mathcal{N}_-\,,
\]
where $\mathcal{N}_-$ is the set of strictly lower triangular matrices, and
$\mathcal{A}^{\perp}$ indicates the orthogonal complement of $\mathcal{A}$ with respect to
the inner product.
The Lie-Poisson structure is then defined as follows: For any functions $f$ and $h$ on $\mathcal{B}_-^*={\rm Sym}(n)$,
define
\[
\{f,h\}(L)=\langle\, L,[\Pi_{\mathcal{B}_-}\nabla f,\Pi_{\mathcal{B}_-}\nabla h]\,\rangle\,,
\]
where $\langle X, \nabla f\rangle=\lim_{\epsilon\to 0}\frac{d}{d\epsilon}f(L+\epsilon X)$,
and $\Pi_{\mathcal{B}_-}X$ is the projection of $X$ onto $\mathcal{B}_-$.
The Toda lattice (\ref{full symmetric Toda}) can now be expressed in Hamiltonian form as
\[
\frac{d}{dt}L=\{H_1,L\}(L) \qquad {\rm with}\quad H_1(L)=\frac{1}{2}{\rm tr}(L^2)\,.
\]
Using the Poisson structure, we can now extend equation (\ref{full symmetric Toda}) to define the Toda lattice hierarchy 
generated by the Hamiltonians $H_k(L)=\frac{1}{k+1}{\rm tr}(L^{k+1})$:
\begin{equation}
\frac{\partial}{\partial t_k}L=\{H_k,L\}(L)=[{\rm Skew}(L^k),L]\qquad {\rm with}\quad H_k=\frac{1}{k+1}{\rm tr}(L^{k+1})\,, \label{Symmetric Toda Hierarchy}
\end{equation}
where ${\rm Skew}(L^k)=\Pi_{so(n)}\nabla H_k$.
The flow stays on a co-adjoint orbit in Sym$(n)\cong \mathcal{B}_-^*$.
The Lie-Poisson structure is nondegenerate when restricted to the co-adjoint orbit,
and the level sets of the integrals found in \cite{DLNT}
are the generic co-adjoint orbits.

Deift and colleagues find the constants of motion by taking the matrices $(L)_k$ obtained by removing the first $k$ rows and last $k$ columns of $L$.
A co-adjoint orbit is obtained by fixing the trace of each $(L)_k$.  The remaining coefficients 
of the characteristic polynomials of
$(L)_k$ for $0 \leq k \leq [n/2]$ (that is, all coefficients except for the traces) provide a family of $[n^2/4]$ constants of motion in involution
on the orbit.   Generically, $(L)_k$ has $n-2k$ distinct eigenvalues $\lambda_{1,k}, ..., \lambda_{n-2k,k}$.  The constants of motion may be
taken equivalently as the eigenvalues $\lambda_{r,k}$ for $0 \leq k \leq [\frac{1}{2}(n-1)]$ and $1 \leq r \leq n-2k$.
In this case the associated angle variables are essentially the last components of the suitably normalized eigenvectors of the $(L)_k$.

In \cite{Kodama McLaughlin}, Kodama and McLaughlin give the explicit solution of the Toda lattice hierarchy (\ref{Symmetric Toda Hierarchy}) 
on full symmetric matrices with distinct eigenvalues by solving the inverse scattering problem of the system
\begin{align*}
L \,\Phi & =  \Phi\, \Lambda\,, \\
\frac{\partial}{\partial t_k}\Phi & =  \mbox{Skew}(L^k)\, \Phi \ .
\end{align*}
with $\Lambda={\rm diag}(\lambda_1,\ldots,\lambda_n)$.
Since $L$ is symmetric, the matrix $\Phi$ of eigenvectors is taken to be orthogonal:
$$
L = \Phi\, \Lambda\, \Phi^T
$$
with $\Phi = [\phi(\lambda_1), ..., \phi(\lambda_n)]$, where the $\phi(\lambda_k)$ is the normalized eigenvector of $L$ with eigenvalue $\lambda_k$.

The indefinite extension of the full symmetric Toda lattice (where $\tilde{L} = LS$ as in (\ref{Indefinite Symmetric1}) is studied in 
\cite{Kodama McLaughlin}, where explicit solutions of $\phi(\lambda_k,t)$ are obtained by inverse scattering.
The authors also give an alternative derivation of the solution using the factorization method of Symes \cite{Symes 1982}, where
$e^{tL(0)}$ is factored into a product of a pseudo-orthogonal matrix times an upper triangular matrix as in (\ref{HR}) (the HR-factorization).

\section{Complex Toda lattices}\label{Complex Toda lattices}
Here we consider the iso-spectral varieties of the complex Toda lattices. In order to describe the geometry of the iso-spectral variety, we first give a summary of the moment map on the flag manifold.
The general description of the moment map discussed here can be found in \cite{Kirwan,GS}.

\subsection{The moment map}\label{moment}
Let $G$ be a complex
semisimple Lie group, $H$ a Cartan subgroup of $G$, and $B$ a Borel
subgroup containing $H$.  If $P$ is a parabolic subgroup of $G$ that
contains $B$, then $G/P$ can be realized as the orbit of $G$ through
the projectivized highest weight vector in the projectivization, ${\mathbb P}(V)$, of an
irreducible representation $V$ of $G$. 
Let ${\mathcal A}$ be the set of weights of $V$, counted with multiplicity;
the weights belong to ${\mathcal H}^*_{\mathbb{R}}$, the real part of the dual of the Lie algebra ${\mathcal H}$ of $H$.
Let $\{v_{\alpha}:\alpha \in {\mathcal A}\}$ be a basis of $V$ consisting of weight vectors.
A point [X] in $G/P$, represented by $X \in V$, has homogeneous
coordinates $\pi_{\alpha}(X)$, 
where $X = \sum_{\alpha \in {\mathcal A}} \pi_{\alpha}(X) v_{\alpha}$.
The moment map as defined in \cite{Kirwan} sends $G/P$ into ${\mathcal H}^*_{\mathbb{R}}$:
\begin{equation}\label{moment map}\begin{array}{cccccc}
\mu &: &G/P & \longrightarrow & {\mathcal H}^*_{\mathbb{R}}  \\[1.5ex]
& & \mbox{}[X] & \longmapsto & 
\displaystyle{\frac{\sum_{\alpha \in {\mathcal A}} |\pi_{\alpha}(X)|^2 \alpha}{\sum_{\alpha \in {\mathcal A}} |\pi_{\alpha}(X)|^2} } 
\end{array}\end{equation}
Its image is the weight polytope of $V$, also referred to as
the moment polytope of $G/P$.   

The fixed points of $H$ in $G/P$ are the points in the orbit of
the Weyl group $W$ through the projectivized highest weight vector
of $V$;
they correspond to the vertices of the polytope under the moment map.
Let $\overline{H \cdot [X]}$ be the closure of the orbit of $H$ 
through $[X]$.  Its image under $\mu$ is the convex hull of the 
vertices corresponding to the fixed points contained in $\overline{H \cdot [X]}$; 
these vertices are the weights $\{\alpha \in W \cdot \alpha^{V} :
\pi_{\alpha}(X) \neq 0 \}$, where $\alpha^{V}$ is the highest weight
of $V$ \cite{Atiyah}.
In particular, the image of a generic orbit, where no $\pi_{\alpha}$ vanishes, is the full polytope.
The real dimension of the image is equal to the complex dimension of
the orbit.

In the case that $P = B$, $V$ is the representation whose highest weight is the sum of the fundamental weights of $G$,
which we denote as $\delta$.   Let $v_{\delta}$ be a weight
vector with weight $\delta$.  Then the action of $G$ through $[v_{\delta}]$ in ${\mathbb P}(V)$ has stabilizer $B$ 
so that the orbit $G \cdot [v_{\delta}]$ is identified with the flag manifold $G/B$.
The projectivized weight vectors that belong to $G/B$ are those in the
orbit of the Weyl group, $W = N(H)/H$, through $[v_{\delta}]$, where $N(H)$ is the
normalizer of $H$ in $G$. 
These are the fixed points of $H$ in $G/B$.  
The stabilizer in $W$ of $[v_{\delta}]$ is trivial 
so that in $G/B$, the fixed points of $H$ are in bijection
with the elements of the Weyl group.

\begin{figure}[t!]
\centering
\includegraphics[scale=0.31]{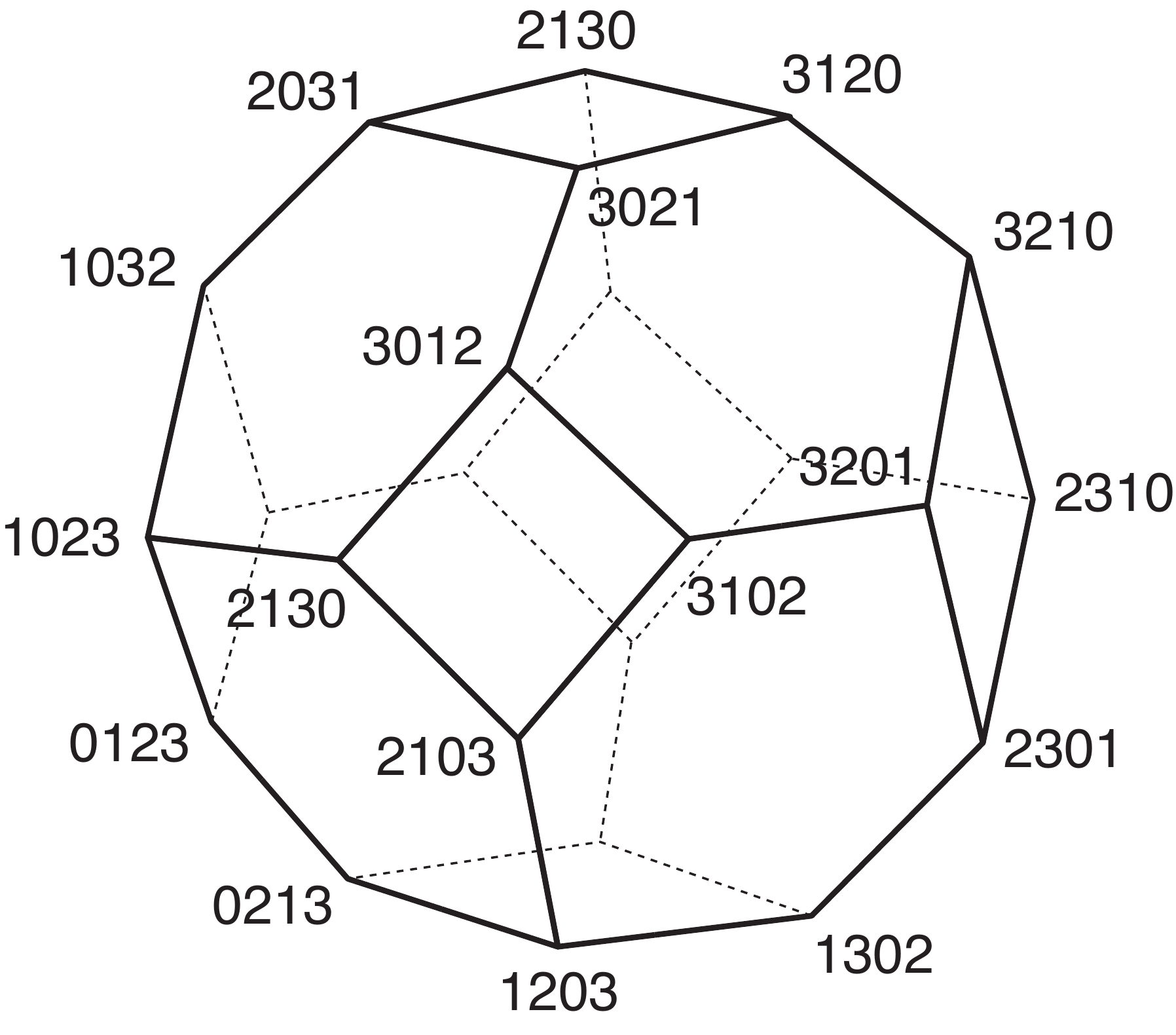}
\caption{The moment polytope of $G/B$ for $G=SL(4,\mathbb{C})$.
The number set $i_1i_2i_3i_4$ indicates the weight $L=i_1L_1+i_2L_2+
i_3L_3+i_4L_4$. The highest weight corresponds to the set $3210$.}
\label{fig:A3momentpolytope}
\end{figure}

Now take $G = SL(n,{\mathbb C})$, $B$ the upper triangular
subgroup, and $H$ the diagonal torus.  The choice of $B$ determines a 
splitting of the root system into positive and negative roots and a system
$\Delta$ of simple roots.  The simple roots are $L_i - L_{i+1}$, where $i = 1, ..., n-1$ and $L_i$ is
the linear function on $\mathcal H$ that gives the $i$th diagonal entry;
the Weyl group is the permutation group $\Sigma_n$, which acts by permuting the $L_i$. 
${\mathcal H}^*_{\mathbb{R}}$ is the quotient of the real span of the $L_i$
by the relation $L_1 + \cdots + L_n = 0$.
${\mathcal H}^*_{\mathbb{R}}$ may be viewed as the hyperplane in ${\mathbb{R}}^n$  
where the sum of the coefficients of the $L_i$ is equal to $1+2+\cdots+ (n-1)$.
Let $\{i_1, ..., i_n\} = \{0, ..., n\}$.
The moment polytope is the convex hull of the weights
$L=i_1L_{1} + i_2L_{2} + \cdots + i_nL_{n}$, where 
 $(i_1,i_2,\ldots,i_{n-1},i_{n})=(n-1, n-2, \cdots,1,0)$ 
corresponds to the highest weight.  In Figure \ref{fig:A3momentpolytope},
we illustrate the moment polytope for the flag manifold $G/B$ of $G=SL(4,\mathbb{C})$.

Let $S$ be the set of reflections
of ${\mathcal H}^*_{\mathbb{R}}$ in the hyperplanes perpendicular, 
with respect to the Killing form, to the simple roots (these are the simple reflections).
The group of motions of ${\mathcal H}^*_{\mathbb{R}}$ generated by $S$ is 
isomorphic to $W$; it is also denoted as $W$ and referred to as the Weyl group of $G$.
The vertices of the moment polytope are the orbit of $W$ through $\delta$.   
For $w \in W$, the moment map $\mu$ sends $w [v_{\delta}]$ to the 
vertex  $w(\delta)$. 
An arbitrary $w \in W$ can be written as a composition $w = s_r \ldots s_1$ of
simple reflections $s_i$.  The 
{\it length}, $l(w)$, of $w$ with respect to the simple system $\Delta$ 
is the smallest $r$ for which such an expression exists.

\subsection{Complex tridiagonal Hessenberg form} \label{Complex tridiagonal Hessenberg}
Here we consider the set ${\mathcal M}$ of complex tridiagonal Hessenberg matrices
\begin{equation}
X = \left( \begin{array}{cccc} f_1 & 1 &  & \\ 
g_1 & \ddots & \ddots &  \\
 & \ddots & \ddots & 1 \\
 &  & g_{n-1} & f_n
\end{array} \right) \ , 
\end{equation}
where the $f_k$ and $g_k$ are allowed to be arbitrary complex numbers.
As before, the Toda flow is defined by (\ref{Hessenberg equation II}) and the eigenvalues (equivalently, the traces of the powers of $X$)
are constants of motion.  The Hamiltonian
$$H_k(X) = \frac{1}{k + 1}  \ \mbox{tr} \left(X^{k + 1}\right)$$
generates the flow
\begin{equation}
\frac{\partial X}{\partial t_k} = [X(t_k),  \ \Pi_{{\mathcal N}_-}(X^k(t_k))] \ , \label{Hessenberg hierarchy}
\end{equation}
by the Poisson structure on ${\mathcal M}$ that we will define in Section (\ref{full Kostant-Toda}). 
This gives a hierarchy of commuting independent flows for $k = 1, ..., n-1$. From this we can see that the trace of $X$ is a Casimir with trivial flow.
The solution of (\ref{Hessenberg hierarchy}) can be found by factorization as in (\ref{lower x upper}):
Factor  $e^{t_kX^k(0)}$ as
\begin{equation}
e^{t_kX^k(0)} = n(t_k)b(t_k) \ , \label{general factorization}
\end{equation}
where $n(t_k)$ is lower unipotent and $b(t_k)$ is upper-triangular.
Then \begin{equation}
X(t_k) = n^{-1}(t_k) X(0) n(t_k) \label{general factorization solution}
\end{equation}
solves (\ref{Hessenberg hierarchy}).

\subsubsection{Characterization of blow-ups via Bruhat decomposition of $G/B$}

Fix the eigenvalues $\lambda_j$, and consider the level set 
${\mathcal M}_{\Lambda}$ consisting of all matrices in ${\mathcal M}$ with spectrum $(\lambda_1, ..., \lambda_n)$.
In the case of distinct eigenvalues, 
Ercolani, Flaschka, and Haine \cite{EFH} construct a minimal nonsingular compactification of ${\mathcal M}_{\Lambda}$ on which the
flows (\ref{Hessenberg hierarchy}) extend to global holomorphic flows.  
The compactification is induced by an embedding of ${\mathcal M}_{\Lambda}$ into the flag manifold $G/B$, where $G = SL(n,{\mathbb C})$ and
$B$ is the upper triangular subgroup of $G$.  

\bigskip

\begin{Proposition} (Kostant, \cite{Kostant}) \label{pro embedding}
Consider the matrix
\begin{equation}
\epsilon_{\Lambda} = 
\left( \begin{array}{ccccc} \lambda_1 & 1 & 0 & \cdots & 0 \\ 
0 & \lambda_2 & 1 & \cdots & 0  \\
\vdots & \ddots & \ddots & \ddots & \vdots \\
0 & \cdots & \ddots & \lambda_{n-1} & 1 \\
0 & \cdots & \cdots & 0 & \lambda_n
\end{array}  \right)  \label{Indefinite Symmetric}
\end{equation} 
in $\epsilon_{\Lambda}$.
Every $X \in {\mathcal M}_{\Lambda}$ can be conjugated to $\epsilon_{\Lambda}$ by a unique lower-triangular
unipotent matrix $L$:
\begin{equation}
X = L \epsilon_{\Lambda} L^{-1} \ .
\end{equation}
\end{Proposition}

This defines a map of ${\mathcal M}_{\Lambda}$ into $G/B$:
\begin{equation}\label{embedding}\begin{array}{cccc}
j_{\Lambda} : & {\mathcal M}_{\Lambda} &\rightarrow &G/B\\[0.4ex]
&X & \mapsto & L^{-1} \ \mbox{mod} \ B \ . 
\end{array}\end{equation}
This mapping is an embedding \cite{Kostant Whittaker}, and the closure, $\overline{j_{\Lambda}({\mathcal M}_{\Lambda})}$, of its image is a nonsingular and minimal compactification
of ${\mathcal M}_{\Lambda}$ \cite{EFH}.   Let $L_0$ be the unique lower unipotent matrix such that $X(0) = L_0 \epsilon_{\Lambda} L_0^{-1}$.  Then
the solution (\ref{general factorization solution}) is conjugate to $\epsilon_{\Lambda}$ as 
$X(t_k) = n^{-1}(t_k) L_0 \epsilon_{\Lambda} L_0^{-1} n(t_k)$, where
$L_0^{-1} n(t_k)$ is lower unipotent.
Thus, $X(t_k)$ is mapped into the flag manifold as
\begin{eqnarray}
j_{\Lambda}(X(t_k)) & = & L_0^{-1} n(t_k) \ \mbox{mod} \ B \  \\ \label{epsilon embedding n(t)}
& = & L_0^{-1} e^{t_k X^k} \ \mbox{mod} \ B \ . \label{epsilon embedding}
\end{eqnarray}
Notice that even at values of $t_k$ where (\ref{epsilon embedding n(t)}) is not defined because the factorization (\ref{general factorization}) is not
possible, the equivalent expression (\ref{epsilon embedding}) is defined.  In this way, the embedding of $X(t_k)$ into $G/B$ completes the flows
through the blow-up times.  This gives a compactification of ${\mathcal M}_{\Lambda}$ in $G/B$.
\cite{EFH} uses this embedding to study the nature of the blow-ups of $X(t_k)$.
  
To illustrate this in a simple case, consider Example 1.1 from the Section \ref{Extended Hessenberg Toda}.
The isospectral set of $2 \times 2$ Hessenberg matrices with both eigenvalues zero is embedded 
into the flag manifold $SL(2,{\mathbb C})/B$, which has the cell decomposition
\begin{equation}
{SL(2,{\mathbb C})}/{B} = \ {N B}/{B} 
\ \cup \
{N\left( \begin{array}{cc} 0 & -1 \\ 
 1 & 0 \end{array} \right) B}/{B} \ . \label{2x2 Bruhat}
\end{equation}
Here $N$ is the subgroup of lower unipotent matrices.
The big cell, ${NB}/{B}$, contains the image of the flow $X(t)$ whenever this flow is defined, that is, whenever
the factorization $e^{t X_0} = n(t) b(t)$ is possible.  
At $t = -1$, where $X(t)$ is undefined, the embedding $j_{\Lambda}$ completes the flow through the singularity.
The image $j_{\Lambda}(X(t))$ passes through
the flag $L_0^{-1} e^{-1 X(0)}$ at time $t = -1$, which is the cell on the right in (\ref{2x2 Bruhat}).

The cell decomposition (\ref{2x2 Bruhat}) is a special case of the cell stratification of the flag manifold $G/B$  known as the  
Bruhat decomposition.  This decomposition is defined in terms of the Weyl group, $W$, as
\begin{equation}
G/B = \bigcup_{w \in W} N w B / B\,. \label{Bruhat}
\end{equation}
In the present case of $G = SL(n, {\mathbb C})$, $W$ is essentially the group of permutation matrices.  
Thus, the Bruhat decomposition partitions flags according to which permutation matrix $w$ is needed to perform the factorization
$X = n w b$ for $X \in G$ with $n \in N$ and $b \in B$.  At all values of $t_k$ for which the flow $X(t_k)$ is defined, 
$j_{\Lambda}$ sends $X(t_k)$ into the big cell of the Bruhat
decomposition, since $w$ is the identity.  When the factorization (\ref{general factorization}) is not possible at time
$t_k = \bar{t}$, 
$e^{\bar{t} X^{k}(0)}$ can be factored as
\begin{equation}
e^{\bar{t} X^{k}(0)} = n(\bar{t}) w b(\bar{t}) \ , \label{w factorization}
\end{equation}
for some permutation matrix $w$.  In this case, the flow (\ref{epsilon embedding}) enters the Bruhat cell $N w B / B$ at time $t_k = \bar{t}$.
\cite{EFH} characterizes the Laurent expansion of each pole of $X(t_1)$ in terms of the Bruhat cell that the solution enters
at the blow-up time.

It is also seen in \cite{EFH} that for $k = 1, ..., n-1$, the flows (\ref{epsilon embedding}) generate a $({\mathbb C}^*)^{n-1}$ torus action on $G/B$ 
that has
trivial isotropy group at every $j_{\Lambda}(X)$ with $g_k(X) \neq 0$ for all $k$.  
The orbit through any such point is open and dense in 
$j_{\Lambda}({\mathcal M}_{\Lambda})$, and the closure of this orbit is the minimal compactification of $j_{\Lambda}({\mathcal M}_{\Lambda})$ in $G/B$.

\subsubsection{Compactification of iso-level set with arbitrary spectrum}

Here ${\mathcal M}$ and ${\mathcal M}_{\Lambda}$ are again defined as in Section \ref{Complex tridiagonal Hessenberg}.
Shipman \cite{Complex Tridiagonal} uses a different embedding, referred to as the Jordan embedding, of ${\mathcal M}_{\Lambda}$ into $G/B$
to describe the compactification of an isospectral set ${\mathcal M}_{\Lambda}$ with arbitrary spectrum.  The advantage of the Jordan embedding 
is that the maximal torus generated by the flows is diagonal if the eigenvalues are distinct and a product of a diagonal torus and a unipotent group
when eigenvalues coincide.  The orbits of these groups, specifically the torus component, are easily studied by taking their images under the moment
map, as explained below.  This leads to a simple description of the closure of 
${\mathcal M}_{\Lambda}$ in terms of faces of the moment polytope.
Recall that in the real tridiagonal Hessenberg form of the Toda lattice studied by Kodama and Ye \cite{Kodama Ye II} 
(see Section \ref{Extended Hessenberg Toda}), the flows
through an initial matrix $X$ preserve the sign of each $g_k$ that is not zero and preserve the vanishing of each $g_k$ that is zero.
The open subset of the isospectral set where no $g_k$ vanishes is partitioned into $2^{n-1}$ components, according to the signs of the $g_k$.
The compactification of the isospectral set is obtained by completing the flows through the blow-up times and pasting the $2^{n-1}$ components
together along the lower-dimensional pieces where one or more $g_k$ vanishes, producing a compact manifold. 
In contrast to this, when $X$ is complex, ${\mathcal M}_{\Lambda}$ is no longer partitioned by signs of the $g_k$; there is only one maximal component
where no $g_k$ vanishes.    The $n-1$ flows through any
initial $X$ with $g_k \neq 0$ for all $k$ generates the whole component, as was also observed in \cite{EFH}.

To define the {\it Jordan embedding},  let 
$C_{\Lambda}$ be the companion matrix of $X$,
\begin{equation}\label{companion matrix}
C_{\Lambda}=\begin{pmatrix}
0        &  1       &   0      &  \cdots  &  0   \\
0        &   0       &  1      &\cdots    & 0  \\
\vdots&\vdots&\ddots&\ddots    &\vdots \\
0        &  0        &    \cdots &  0     &   1   \\
s_{n} & s_{n-1}&\cdots & s_2& 0
\end{pmatrix} \,.
\end{equation}
Here the $s_j$'s are the symmetric polynomials of the eigevalues $\lambda_j$, i.e.
\[
{\rm det}(\lambda I-X)=\lambda^n-\sum_{j=2}^ns_j\lambda^{n-j}\,.
\]
Again, by \cite{Kostant}, there exists a unique lower unipotent
matrix $L$ such that $X = L C_{\Lambda} L^{-1}$.
In particular, all elements of ${\mathcal M}_{\Lambda}$ are conjugate.
Since the companion matrix has a single chain
of generalized eigenvectors for each eigenvalue, any matrix in Jordan canonical form 
that is conjugate to it contains one block for each eigenvalue.      

Following \cite{Nongeneric}, fix an ordering of
the eigenvalues, and let $J$ be the corresponding Jordan matrix.  Then
$C = W J W^{-1}$ where $W$ is a matrix whose columns are (generalized)
eigenvectors of $C$, where each eigenvector has a 1 in the first nonzero entry
and the vectors are ordered
according to the chosen ordering of eigenvalues, with generalized
eigenvectors ordered successively.  Once $W$ is fixed, for $X \in {\mathcal M}_{\Lambda}$, we can 
write $X = LW J W^{-1}L^{-1}$.  The Jordan embedding is the mapping
\begin{equation}\label{Jordan Embedding}
\begin{array}{ccccc}
 \gamma_{\Lambda}&: &{\mathcal M}_{\Lambda}& \longrightarrow &G/B \\[2.0ex]
  &{} & X &\longmapsto &W^{-1}L^{-1} \ \bmod B \end{array}
\end{equation}
That this is an embedding follows from the results in \cite{Kostant Whittaker}.    

Under this embedding, 
the flows $X(t_k) = n^{-1}(t_k) X(0) n(t_k)$ in (\ref{general factorization solution}) with
$e^{t_kX^{k-1}(0)} = n(t_k)b(t_k)$ as in (\ref{general factorization}) 
and $X(0) = LW J W^{-1}L^{-1}$ generate a group action as follows:
\begin{eqnarray*} \gamma_{\Lambda}(X(t_k)) & = & W^{-1} \ L^{-1} \ n_k(t_k) \ \mbox{mod} B \\
& = & W^{-1} \ L^{-1} \ \exp[ \ t_k X^k(0) \ ] \ \mbox{mod} B \\
& = & W^{-1} \ L^{-1} \ \exp[ \ t_k (L W  J_{\Lambda} W^{-1} L^{-1})^k \ ] \ \mbox{mod} B \\
& = & W^{-1} \ L^{-1} \ L \ W \exp[ \ t_k J_{\Lambda}^k \ ] \ W^{-1} L^{-1} \ \mbox{mod} B \\
& = & \exp[ \ t_k J_{\Lambda}^k \ ] \  W^{-1} \ L^{-1} \ \mbox{mod} B 
\end{eqnarray*}
The flows $\exp[ \ t_k J_{\Lambda}^k \ ]$ for $k = 1, ..., n-1$ generate the centralizer of $J$ in
$SL(n,{\mathbb C})$. We denote this subgroup as $A_J$.   

$A_J$ has $r$ blocks along the diagonal, 
\begin{equation}
A_i = \left\{  \left( \begin{array}{cccc} s & x_1 & \cdots & x_{d_i - 1}
\\ & \ddots & \ddots &  \vdots \\
& & \ddots & x_1 \\ & & & s \end{array} \right)  \ : \ s \in {\mathbb C}^*, \ 
x_1,...,x_{d_i-1} \in {\mathbb C} \right\} \ , \label{AJ}
\end{equation}
$i = 1, \ldots, r$, where $d_i$ is the multiplicity of the eigenvalue in
the $i$th block of $J$.   All the 
blocks together contain $n-r$ independent entries in 
${\mathbb C}$ above the diagonal and $r$ entries in ${\mathbb C}^*$ on the diagonal, where the
product of the diagonal entries is 1.   
$A_J$ is a semi-direct product of the diagonal torus $K_J$, obtained
by setting all the entries above the diagonal equal to zero, and the unipotent group $U_J$,
obtained by setting all the diagonal entries equal to 1.
The subgroup of $A_J$ that fixes every point in $G/B$ is the (discrete)
subgroup $D$ of all constant multiples of the identity.  The quotient $A_J/D$ 
has the manifold structure (but not the group structure) of 
$({\mathbb C}^*)^{r-1} \times {\mathbb C}^{n-r}$.    
When $r = n$ (distinct eigenvalues), 
$A_J$ is the maximal diagonal torus.  
The compactification, $\overline{{\mathcal M}_{\Lambda}}$,  of ${\mathcal M}_{\Lambda}$ in $G/B$ is the closure of one generic orbit of $A_J$.
Its boundary is a union of non-maximal orbits of $A_J$.  
\cite{Complex Tridiagonal} uses the moment map of the maximal torus action, which sends $SL(n,{\mathbb C})/B$ to a polytope in ${\mathbb{R}}^{n-1}$,
to identify each component of the boundary of $\overline{{\mathcal M}_{\Lambda}}$
with a specified face of the polytope, as described above. 

First we describe the boundary of ${\mathcal M}_{\Lambda}$ in ${\mathcal M}$.
Let $\alpha$ be a subset of $\{g_1, ..., g_{n-1}\}$, 
and denote by ${\mathcal M}_{\Lambda}^{\alpha}$ the subset of ${\mathcal M}_{\Lambda}$ on which 
exactly the $g_i$ in $\alpha$ are zero.  These subsets form a partition of
${\mathcal M}_{\Lambda}$ where the complex dimension of ${\mathcal M}_{\Lambda}^{\alpha}$ 
is equal to the number of $g_i$ that do not vanish.  There is one maximal component, 
on which no $g_k$ vanish, and one component consisting of the fixed points, 
where all the $g_k$ vanish.

Let $X \in {\mathcal M}_{\Lambda}^{\alpha}$.  
The blocks on the diagonal of $X$ where no $g_k$ vanish are full tridiagonal
Hessenberg matrices of a smaller dimension (all entries on their first subdiagonals 
are nonzero).   The union of the eigenvalues of these blocks,
counted with multiplicity, is the spectrum $\Lambda$. 
Let $P(\Lambda)$ be a partition of $\Lambda$ into subsets
$\Lambda = \Lambda_1 \cup ... \cup \Lambda_q$, where $\Lambda_k$ is the
spectrum of the $k$th block along the diagonal of $X$, and denote by
${\mathcal M}_{P(\Lambda)}^{\alpha}$ the component of ${\mathcal M}_{\Lambda}^{\alpha}$ 
where $\Lambda$ is partitioned among the blocks according to $P(\Lambda)$.
The Toda flows (\ref{general factorization solution}) through $X$ preserve the spectrum of each block
and therefore respect the partition
${\mathcal M}_{\Lambda} = \cup {\mathcal M}_{P(\Lambda)}^{\alpha}$.
The moment map, described above, gives a one-to-one correspondence between the components
${\mathcal M}_{P(\Lambda)}^{\alpha}$ and particular faces of a certain polytope \cite{Complex Tridiagonal}.

To see this, let 
$K_J$ be the torus that lies along the diagonal of $A_J$.  $K_J$ is a subtorus of the maximal diagonal torus $H$.
Its Lie algebra, ${\mathcal K}_J$, is the kernel of a subset $\Delta_J$ of the
simple roots; this determines the subset $S_J \subset S$ of reflections in the
hyperplanes perpendicular to the roots in $\Delta_J$. $S_J$ generates 
a subgroup $W_J$ of $W$.  
The elements in
$W^J = \{w \in W: l(sw) > l(w) \ \forall s \in S_J \}$
are the coset representatives of minimum length in the quotient $W_J \backslash W$.
The following result is proved in \cite{Complex Tridiagonal}:

\bigskip

\begin{Proposition} \cite{Complex Tridiagonal}
 The composition $\mu \circ \gamma_{\Lambda}: {\mathcal M}_{\Lambda} \rightarrow
{\mathcal H}^*_{\mathbb{R}}$ gives a one-to-one correspondence between the components
${\mathcal M}_{P(\Lambda)}^{\alpha}$ that partition ${\mathcal M}_{\Lambda}$
and the faces of the moment polytope with at
least one vertex in $W^J$.  The complex dimension of the component is equal to  
the real dimension of the face.  In particular,
the maximal orbit in $\overline{{\mathcal M}_{\Lambda}}$ corresponds to the full 
polytope, and the fixed points of $A_J$ in $\overline{{\mathcal M}_{\Lambda}}$ 
correspond to the vertices in $W^J$.  
\end{Proposition}

\bigskip

\subsection{The full Kostant-Toda lattice} \label{full Kostant-Toda}

Here we consider full complex Hessenberg matrices
\begin{equation}
X = \left( \begin{array}{ccccc} * & 1 & 0 & \cdots & 0 \\ 
* & * & 1 & \cdots & 0 \\
\vdots& \vdots & \vdots & \ddots & \vdots \\
* & * & * & \cdots & 1  \\
* & * & * &\cdots  & *
\end{array} \right) \ \label{full Hessenberg matrix}. 
\end{equation}
The set of all such $X$ is denoted $\epsilon + {\mathcal B}_-$, where $\epsilon$ is the matrix with 1's on the superdiagonal and zeros elsewhere
and ${\mathcal B}_-$ is the set of lower triangular complex matrices.

With respect to the symplectic structure on $\epsilon + {\mathcal B}_-$ defined below, the Toda hierarchy 
(\ref{Hessenberg hierarchy}) with $X$ as in (\ref{full Hessenberg matrix}) turns out to be completely integrable on the generic leaves. 
The complete integrability is observed in \cite{EFS}
by extending the results of \cite{DLNT} to $\epsilon + {\mathcal B}_-$.  For $n > 3$, the eigenvalues of the initial matrix do not constitute
enough integrals for complete integrability; a generic level set of the constants of motion is a subset of an isospectral set
that is cut out by additional integrals and Casimirs, which can be computed by a chopping construction given in Proposition (\ref{k-chops}).
 
Ikeda \cite{Ikeda} studies the level sets in $\epsilon + {\mathcal B}_-$ cut out by fixing only the eigenvalues.  He finds a compactification of
an isospectral set with distinct eigenvalues, showing that its cohomology ring is the same as that of the flag manifold $SL(n,{\mathbb C})/B$. 
This work differs from the previous works \cite{Tomei}, \cite{Complex Tridiagonal}, and \cite {Kodama Ye II} that compactify tridiagonal versions 
of the Toda lattice in that it does 
not use the Toda flows directly in producing the compactification. 

To describe the symplectic structure on $\epsilon + {\mathcal B}_-$, write
\begin{equation}
sl(n,{\mathbb C}) = {\mathcal N}_- \oplus {\mathcal B}_+   ,\label{decomp}
\end{equation}
where ${\mathcal N}_-$ and ${\mathcal B}_+$ are the strictly lower triangular and the upper triangular subalgebras, respectively.
With a non-degenerate inner product $\langle A,B\rangle={\rm tr}(AB)$ on $sl(n,{\mathbb C})$,
we have an isomorphism $sl(n,\mathbb{C})\cong sl^*(n,\mathbb{C})$ and
\[
sl^*(n,\mathbb{C})=\mathcal{N}_-^*\oplus \mathcal{B}_+^*=\mathcal{B}_+^{\perp}\oplus \mathcal{N}_-^{\perp} \,.
\]
 With the isomorphisms
\[
\mathcal{B}_+^*\cong \mathcal{N}_-^{\perp}=\mathcal{B}_- ,\qquad
\mathcal{N}_-^*\cong \mathcal{B}_+^{\perp}=\mathcal{N}_+\,
\]
we identify
\[
\epsilon+\mathcal{B}_-\cong \mathcal{B}_+^*\,,
\]
which defines the phase space of the full Kostant-Toda lattice. On the space $\mathcal{B}_+^*$,
we define the Lie-Poisson structure (Kostant-Kirillov form); that is, for any functions $f, h$ on $\mathcal{B}_+^*$, 
\[
\{f,h\}(X)=\langle \,L, [\Pi_{\mathcal{B}_+}\nabla f,\Pi_{\mathcal{B}_+}\nabla h]\, \rangle\,,\qquad {\rm for}\quad X\in \mathcal{B}_+^*\,,
\]
where $\langle Y, \nabla f\rangle=\lim_{\epsilon\to 0}\frac{d}{d\epsilon}f(X+\epsilon Y)$.
This Lie-Poisson structure gives a stratification of $\mathcal{B}_+^*\cong\epsilon+\mathcal{B}_-$.
The stratification of the Poisson manifold $\mathcal{B}_+^*$ with
this Lie-Poisson structure is complicated, having leaves of different types and different dimensions.

Denote by $B_+$ the upper-triangular subgroup of $SL(n,{\mathbb C})$ and by $g \cdot Y$ the adjoint action of $SL(n,{\mathbb C})$ on 
$sl(n,{\mathbb C})$.  Then, through the identification of $\epsilon+\mathcal{B}_-$ with $\mathcal{B}_+^*$, the abstract coadjoint action of $B_+$
on $\mathcal{B}_+^*$ becomes
\[
{\rm Ad}_b^*X = \epsilon + \Pi_{\mathcal{B}_-}b^{-1} \cdot (X - \epsilon)\,.
\]
The symplectic leaves in $\epsilon+\mathcal{B}_-$ are generated by the coadjoint orbits and additional Casimirs.

In general, the dimension of a generic leaf is
greater than $2(n-1)$, and more integrals are needed for complete integrability.
The chopping construction used in \cite{DLNT} to obtain a complete family of integrals for the full symmetric Toda lattice is adapted in \cite{EFS} 
to find a complete family of integrals for the full asymmetric Toda lattice.  

\bigskip

\begin{Proposition} \label{k-chops} \cite{EFS}
Choose $X \in \epsilon + {\mathcal B}_-$, and break it into blocks of the
indicated sizes as 
$$ X = \bordermatrix{&  k & n-2k & k \cr k & X_1 & X_2 & X_3 \cr
n-2k & X_4 & X_5 & X_6 \cr k & X_7 & X_8 & X_9 \cr}, $$
 where $k$ is an integer such that $0 \leq k \leq [\frac{(n-1)}{
2}]$.  If ${\rm det}( X_7) \neq 0$, define the matrix $\phi_k(X)$ by
\begin{eqnarray*} \phi_k(X) & = & X_5 - X_4 X_7^{-1} X_8 \in 
Gl(n-2k,{\mathbb C}), \quad k \neq 0, \\
\phi_0(X) & = & X. 
\end{eqnarray*}
The coefficients of the polynomial
${\rm det}(\lambda - \phi_k(X)) = \lambda^{n-2k} + I_{1k} \lambda^{n-2k-1} + \cdots + I_{n-2k,k}$
are constants of motion of the full Kostant-Toda lattice.  
The functions
$I_{1k}$ are  Casimirs on $\epsilon + {\mathcal
B}_-$, and the functions $I_{rk}$ for $r > 1$ constitute a complete involutive
family of integrals for the generic symplectic leaves of $\epsilon +
{\mathcal B}_-$ cut out by the Casimirs $I_{1k}$.   These integrals are known as the $k$-chop integrals.
\end{Proposition}

\noindent The $k$-chop integrals $I_{rk}$ are equivalent to the traces of the powers of $\phi_k(X)$.
The Hamiltonian system generated by an integral $I(X)$ is
\begin{equation}
\frac{d}{dt}X = [X,  \ \Pi_{{\mathcal N}_-}(\nabla I(X))]  \label{Toda equation for I} \ .
\end{equation}
When $I(X)$ is one of the original Toda invariants $H_k(X) = \frac{1}{k+1} \mbox{tr}(X^k+1)$ (a 0-chop integral), the flow is
\begin{equation}
\frac{d}{dt}X = [X,  \ \Pi_{{\mathcal N}_-}(X^k)]  \label{Toda equation for H_k} \ .
\end{equation} 

The solution may again be found via factorization \cite{EFS}.  Let
\begin{equation}
e^{t \nabla I(X_0)} = n(t)b(t) \ , \nonumber
\end{equation}
with $n(t)$ and $b(t)$ lower unipotent and upper-triangular, respectively.  Then
$$
X(t) = n^{-1}(t) X_0 n(t) \ .
$$

Let $X$ belong to $(\epsilon + {\mathcal B}_-)_{\Lambda}$.
Recall from Section \ref{Complex tridiagonal Hessenberg} that there exists a unique lower unipotent
matrix $L$ such that $X = L C_{\Lambda} L^{-1}$, where $C$ is the companion matrix (\ref{companion matrix}).
The mapping
\begin{equation}\label{companion embedding}\begin{array}{cccccc}
c_{\Lambda} &:&  {\mathcal M}_{\Lambda} &\longrightarrow& SL(n, {\mathbb C})/B\\[1.5ex]
 & &  X &\longmapsto & L^{-1} \ \bmod B \end{array}
\end{equation}
is an embedding \cite{Kostant Whittaker}, referred to as the {\it companion embedding}.
Its image is open and dense in the flag manifold. 
Under this embedding,  
the $n-1$ flows of the 0-chop integrals  $\frac{1}{k} {\rm tr}X^k$
generate the action of the
centralizer of $C_{\Lambda}$ in $SL(n,{\mathbb C})$ (the group acts by multiplication on the left).  

When the $\lambda_i$ are
distinct, $C_{\Lambda} = V \Lambda 
V^{-1}$, where $V$ is a Vandermonde matrix, and 
$$X = L V \Lambda V^{-1} L^{-1}. $$
The embedding
\begin{equation}\label{Torus Embedding}\begin{array}{ccccc}
\Psi_{\Lambda} &: &(\epsilon + {\mathcal B}_-)_{\Lambda}& \longrightarrow&
SL(n,{\mathbb C})/B\\[1.5ex]
& &  X  &\longmapsto &V^{-1} L^{-1} \bmod B \end{array}
\end{equation}
is a specific case of the Jordan embedding (\ref{Jordan Embedding}) when the eigenvalues are distinct.  
In this case, the group $A_J$ (see (\ref{AJ})) generated by Hamiltonian flows of $H_k = \frac{1}{k+1} \mbox{tr}(X^k)$
for $k = 1, ..., n-1$ is the maximal diagonal torus.
$\Psi_{\Lambda}$ is therefore referred to as
the {\it torus embedding}. 

When the values of the integrals are sufficiently generic (in particular, when the eigenvalues of each $k$-chop are distinct),
Ercolani, Flaschka, and Singer \cite{EFS} show how the flows of the $k$-chop integrals can be organized in the flag manifold by the torus embedding.  
(The companion embedding gives a similar structure, but the torus embedding is more convenient since the group action is diagonal.)
The guiding idea in \cite{EFS} is that 
the $k$-chop integrals for $SL(n, {\mathbb C})$ are equivalent to the 1-chop integrals for $SL(n-2(k-1), {\mathbb C})$.
Let $SL(m,{\mathbb C})/B$ denote the quotient of $SL(m,{\mathbb C})$ by its upper triangular subgroup, and let
$SL(m,{\mathbb C})/P$ denote the quotient of $SL(m,{\mathbb C})$ by the parabolic subgroup $P$ of $SL(m,{\mathbb C})$ whose 
entries below the diagonal in the first 
column and to the left of the diagonal in the last row are zero:
$$P \ =  \ 
\left( \begin{array}{ccccc} * & * & \cdots & * & * \\ 
0 & * & \cdots & * & * \\
\vdots & \vdots & \ddots & \vdots & \vdots \\
0 & * & \cdots & * & * \\
0 & 0 & \cdots & 0 & *
\end{array} \right)  \ .$$
The 1-chop integrals $I_{r1}$ depend only on the partial flag manifold
$SL(n,{\mathbb C})/P$.  In this partial flag manifold, 
a level set of the 1-chop integrals is generated by the flows of the 0-chop torus.  
The 1-chop flows generate a torus action along the fiber of the projection
\begin{eqnarray}
SL(n-2, {\mathbb C})/B & \rightarrow & SL(n, {\mathbb C})/B \nonumber \\
& & \hspace{1cm} \downarrow \nonumber \\
& & SL(n, {\mathbb C})/P \ . \nonumber
\end{eqnarray}
In this fiber, the 2-chop integrals depend only on the partial flag manifold 
$SL(n-2, {\mathbb C})/P$, where a level set of the 2-chop integrals is generated by the 1-chop torus.
This picture extends to all the $k$-chop flows.  \cite{EFS} builds a tower of fibrations 
\begin{eqnarray}
SL(n-2(k+1), {\mathbb C})/B & \rightarrow & SL(n-2k,{\mathbb C})/B   \nonumber \\
& & \hspace{1cm} \downarrow \nonumber \\
& & SL(n-2k,{\mathbb C})/P  \  \nonumber
\end{eqnarray}
where the $k$-chop flows generate a level set of the $(k+1)$-chop integrals
in the partial flag manifold $SL(n-2k,{\mathbb C})/P$ and the
$(k+1)$-flows act as a torus action along the fiber, $SL(n-2(k+1),{\mathbb C})/B$.
In the end, the closure of a level set of all the $k$-chop integrals in $SL(n, {\mathbb C})/B$ is realized as a product of closures of 
generic torus orbits in the product of partial flag manifolds
\begin{equation}
SL(n,{\mathbb C})/P \times SL(n-2, {\mathbb C})/P \times \cdots \times SL(n-2M, {\mathbb C})/P \label{product}
\end{equation}
where $M$ is largest $k$ for which there are $k$-chop integrals.

In \cite{GS:99}, Gekhtman and Shapiro generalize the full Kostant-Toda flows and
the $k$-chop construction of the integrals in Proposition \ref{k-chops} to arbitrary simple Lie algebras, showing that the Toda flows on a 
generic coadjoint orbit in a simple Lie algebra $g$
are completely integrable.  A key observation in making this extension is that the 1-chop matrix 
$\phi_1(X)$ can be obtained as the middle $(n-2) \times (n-2)$ block of
$Ad_{\Gamma(X)}(X)$, where $\Gamma(X)$ is a special element of the Borel subgroup of $G$.
This allows the authors to use the adjoint action of a Borel subgroup, followed by a projection onto a subalgebra, to define the appropriate analog 
of the 1-chop matrix.

Finally, we note that full Kostant-Toda lattice has a symmetry of order two induced by the nontrivial automorphism of the Dynkin diagram of the Lie 
algebra $sl(n, {\mathbb C})$.  In terms of the matrices in $\epsilon + {\mathcal B}_-$, the involution is reflection along the anti-diagonal.
It is shown in \cite{Symmetry} that this involution preserves all the $k$-chop integrals and thus defines an involution on each level set of the
constants of motion.  In the flag manifold, the symmetry interchanges the two fixed points of the torus action that correspond to antipodal 
vertices of the moment polytope under the moment map (\ref{moment map}).

\begin{Example}

In this example, we demonstrate the complexity of the Poisson stratification of $\epsilon + {\mathcal B}_-$ for $n = 3$ and $n = 4$.
The table of symplectic leaves of all dimensions has been calculated in notes by Stephanie Singer, a co-author of \cite{EFS}, as given below.
On the leaves of lower dimensions, the $k$-chop integrals are dependent.   

When $n = 3$, 
$$\epsilon + {\mathcal B}_- \ =  \ 
\left\{ \left( \begin{array}{ccc} f_1 & 1 & 0  \\ 
g_1 & f_2 & 1  \\
h &  g_2 & f_3 \end{array} \right) \ :  \ \ \sum_{i=1}^3 f_i = 0 \ \right\} \ . $$
Its symplectic leaves are listed in the following table.    The Casimirs are constants of motion that generate trivial Hamiltonian flows.
The value of each Casimir is fixed on a given symplectic leaf.

\bigskip

\begin{tabular}{cccc}
{Closed Conditions} & {Open Conditions} & {Casimirs} & {Dimension} \\[2.0ex]
--- & $h \neq 0$ & $f_2 - \frac{g_1 g_2}{h}$ & 4 \\[2.0ex]
$h = 0$ & $g_1 g_2 \neq 0$ & --- & 4 \\ [2.0ex]
$h, \ g_1 = 0$ & $g_2 \neq 0$ & $f_1$ & 2 \\ [2.0ex]
$h, \ g_2 = 0$ & $g_1 \neq 0$ & $f_3$ & 2 \\ [2.0ex]
$h,\  g_1,\ g_2 = 0$ & --- & $f_1,\ f_2,\ (f_3)$ & 0 
\end{tabular}

\bigskip
\noindent
On each 4-dimensional leaf, the functions $\frac{1}{2} \mbox{tr} (X^2)$ and $\frac{1}{3} \mbox{tr} (X^3)$ provide a complete family of integrals (Hamiltonians)
for the Toda hierarchy.  

When $n = 4$, the symplectic stratification is already much more complicated.  Here,
$$\epsilon + {\mathcal B}_- \ =  \ 
\left\{ \left( \begin{array}{cccc} f_1 & 1 & 0 & 0 \\ 
g_1 & f_2 & 1 & 0 \\
h_1 &  g_2 & f_3 & 1 \\
k & h_2 & g_3 & f_4
\end{array} \right) \ :  \ \ \sum_{i=1}^4 f_i = 0 \ \right\} \ .$$
The table of symplectic leaves is as follows:

\bigskip

\begin{tabular}{lccc}
{Closed Conditions} & {Open Conditions} & {Casimirs} & {Dimension} \\[3.0ex]
-- & $k(k g_2 - h_1 h_2) \neq 0$ & $I_{11} = -f_2 - f_3 + \frac{g_1 h_2 + g_3 h_1}{
k}$ & 8 \\
$k g_2 - h_1 h_2 = 0$ & $k \neq 0$ & ${\mathcal C}_1 = f_2 - \frac{g_1 h_2}{k}$, 
${\mathcal C}_2 = f_3
- \frac{g_3 h_1}{ k}$ & 6 \\[3.0ex]
$k = 0$ & $h_1 h_2 (g_1 h_2 + g_3 h_1) \neq 0$ & --- & 8 \\
$k,\ h_1 = 0$ & $h_2 g_1 \neq 0$ & ${\mathcal C}_3 = f_3 - \frac{g_2 g_3 }{h_2}$ & 6 \\
$k,\ h_2 = 0$ & $h_1 g_3 \neq 0$ & ${\mathcal C}_4 = f_2 - \frac{g_1 g_2 }{h_1}$ & 6 \\[3.0ex]
$k$, $g_1 h_2 + g_3 h_1 = 0$ & $h_1 h_2 \neq 0$ & ${\mathcal C}_5 = f_1 + f_3 - \frac{g_2 g_3
}{ h_2}$ & 6 \\
$k,\ h_1,\ h_2 = 0$ & $g_1 g_2 g_3 \neq 0$ & --- & 6 \\[3.0ex]
$k,\ h_1,\ g_1 = 0$ & $h_2 \neq 0$ & $f_1$, ${\mathcal C}_6 = f_3 - \frac{g_2 g_3}{
h_2}$ & 4 \\
$k,\ h_1,\ g_1,\ h_2 = 0$ & $g_2 g_3 \neq 0$ & $f_1$ & 4 \\
$k,\ h_2,\ g_3 = 0$ & $h_1 \neq 0$ & $f_4$, ${\mathcal C}_7 = f_2 - \frac{g_1 g_2}{
h_1}$ & 4 \\
$k,\ h_2,\ g_3,\ h_1 = 0$ & $g_1 g_2 \neq 0$ & $f_4$ & 4 \\
$k,\ h_1,\ h_2,\ g_2 = 0$ & $g_1 g_3 \neq 0$ & $f_1 + f_2 \ (= -(f_3 +
f_4))$ & 4 \\[2.0ex]
$k,\ h_1,\ h_2,\ g_1,\ g_2 = 0$ & $g_3 \neq 0$ & $f_1,\ f_2$ & 2 \\
$k,\ h_1,\ h_2,\ g_2,\ g_3 = 0$ & $g_1 \neq 0$ & $f_3,\ f_4$ & 2 \\ 
$k,\ h_1,\ h_2,\ g_1,\ g_3 = 0$ & $g_2 \neq 0$ & $f_1,\ f_4$ & 2 \\[2.0ex]
$k,\ h_1,\ h_2,\ g_1,\ g_2,\ g_3 = 0$ & --- & $f_1,\ f_2,\ f_3,\ (f_4)$
& 0 
\end{tabular}

\vspace{1cm}

On the maximal leaves, of dimension 8, the functions
$\frac{1}{k}  \ \mbox{tr} X^k$ for $k = 2, 3, 4$ provide three constants of motion.  
One 1-chop integral is needed to complete the family.

\end{Example}

\subsection{Nongeneric flows in the full Kostant-Toda lattice}

When eigenvalues of the initial matrix in $\epsilon + {\mathcal B}_-$ coincide, the torus embedding 
(\ref{Torus Embedding}) is not defined since any matrix 
in $\epsilon + {\mathcal B}_-$ has one Jordan block for each eigenvalue.  
In the most degenerate case of non-distinct eigenvalues, that is, when all eigenvalues are zero, the isospectral set can be embedded into 
the flag manifold by the companion embedding
(\ref{companion embedding}).  Under this embedding, the 0-chop integrals generate the action of the exponential of an abelian nilpotent algebra
\cite{PJM}.
The 1-chop integrals are again defined only in terms of the partial flag manifold $SL(n,{\mathbb C})/P$.  
Fixing the values of each 1-chop integral produces 
a variety in the flag manifold.  The common intersection of all
these varieties turns out to be invariant under the action of the diagonal torus and has a simple description in terms of the moment polytope \cite{PJM}.

\cite{Nongeneric} considers level sets where the eigenvalues of each $\phi_k(X)$ are distinct but one or more
eigenvalues of $\phi_j(X)$ and $\phi_{j+1}(X)$ coincide for one or more values of $j$. 
In this situation, the torus orbits generated by the $k$-chop integrals in the product (\ref{product})degenerate into unions of nongeneric orbits. The nature of this splitting
can be seen in terms of the moment polytopes of the partial flag manifolds in (\ref{product}).

\begin{figure}[t!]
\centering
\includegraphics[scale=0.31]{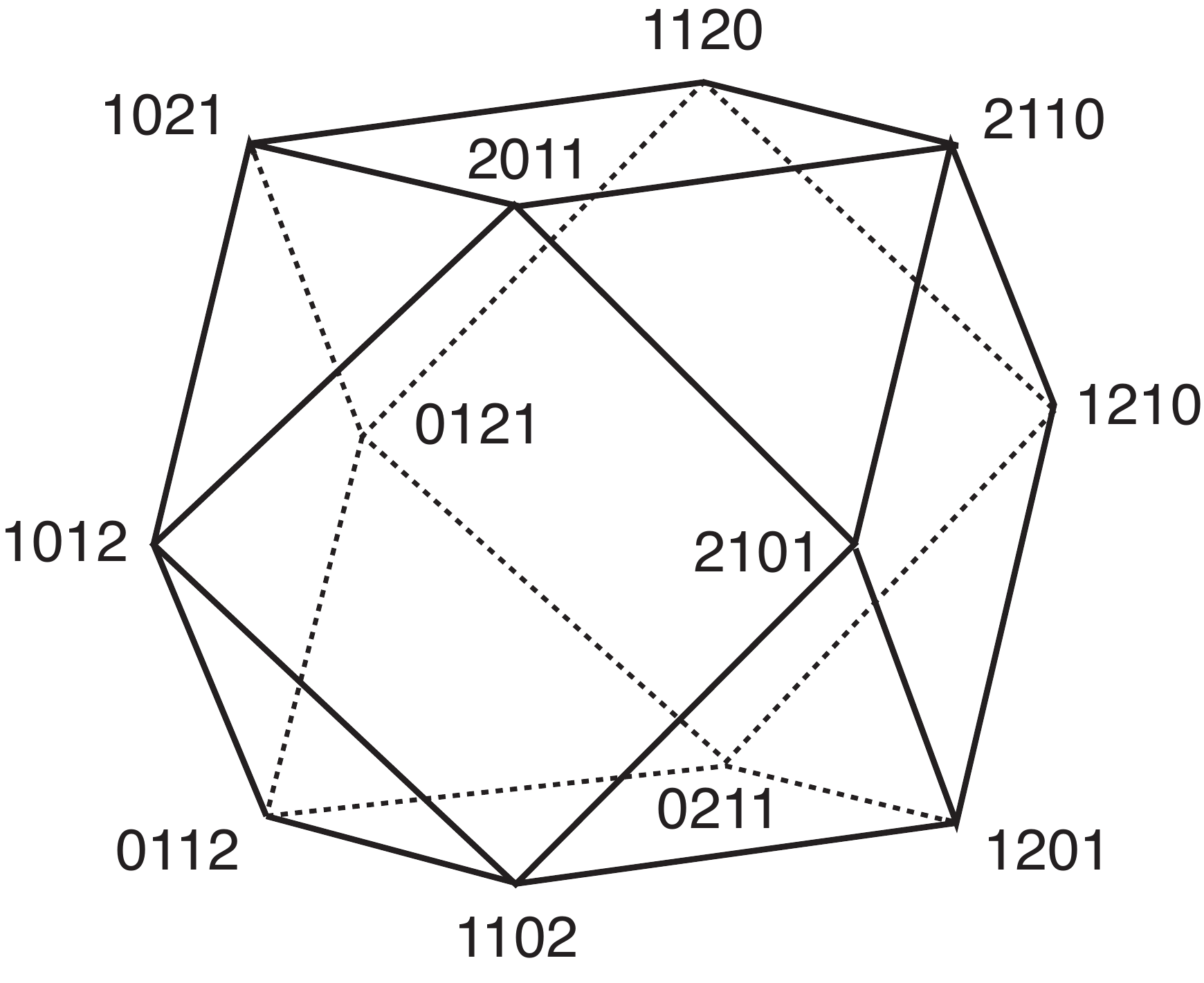}
\caption{The moment polytope $\Delta_4$ for $SL(4,\mathbb{C})/P$.
The vertices represent the weights $L_i-L_j$ which are expressed by
$i_1L_1+i_2L_2+i_3L_3+i_4L_4$ using $L_1+\cdots+L_4=0$,
e.g. $2110$ means $L_1-L_4$.}
\label{fig:partialpolytope}
\end{figure}

Recall the definition of the moment map $\mu$ in (\ref{moment map}).
Here $G$ is $SL(n,{\mathbb C})$ and $V$ is the adjoint representation.
$V$ may be realized as the subspace of ${\mathbb C}^n \otimes \wedge^{n-1}
{\mathbb C}^n$ with $\sum e_i \otimes e_i^* = 0$, where $\{e_i\}$ is
the standard basis of ${\mathbb C}$, and $e_i^* = (-1)^{i+1}e_1 \wedge
\ldots \wedge e_{i-1} \wedge e_{i+1} \wedge \ldots \wedge e_n$.
The partial flag manifold $SL(n,{\mathbb C})/P$ is the orbit of $G$ through $[e_1 \otimes e_n^*]$ in ${\mathbb P}(V)$.
The weight of $e_i \otimes e_j^*$ is $L_i - L_j$, where
$L_k$ is the linear function in ${\mathcal H}^*$ that sends an element
of ${\mathcal H}$ to its $k$th diagonal entry.   The weights
$L_i - L_j$ with $i \neq j$ are the vertices of the
weight polytope of $V$, which we denote by $\triangle_n$.   These vertices
are the images under the moment map
of the fixed points of the complex diagonal torus.
The image of the closure of a torus orbit under moment map is the convex hull of the weights corresponding to the 
fixed points of the torus in the closure of the orbit.
The real dimension of the image is equal to the complex dimension of the orbit \cite{Atiyah}. 
Figure \ref{fig:partialpolytope} shows the example of the moment polytope $\Delta_4$.

An element $gB$ in $SL(n,{\mathbb C})/P$ represents the partial flag
$V^1 \subset V^{n-1} \subset {\mathbb C}^n$ where $V^1$ is the 
span of the first column of $g$ and $V^{n-1}$ is the span of the first $n-1$ columns.
There are two natural
projections from $SL(n,{\mathbb C})/P$ to the projective space
${\mathbb{CP}}^{n-1}$ and its dual $({\mathbb{CP}}^{n-1})^*$ that send a partial flag
to the line $V^1$ and to  the hyperplane
$V^{n-1}$, respectively.   Let $\pi_i$ and $\pi_i^*$ be projective
coordinates on ${\mathbb{CP}}^{n-1}$ and $({\mathbb{CP}}^{n-1})^*$.

The coordinates $\pi_{L_i - L_j}$
that come from the embedding of $SL(n,{\mathbb C})/P$ into ${\mathbb P}(V)$ by the moment map (\ref{moment map})
are projectively equal to the products $\pi_i \pi_j^*$ for $i \neq j$:
$[\pi_{L_i - L_j}]_{i \neq j} = [\pi_i \pi_j^*]_{i \neq j}$.

\begin{figure}[t!]
\centering
\includegraphics[scale=0.29]{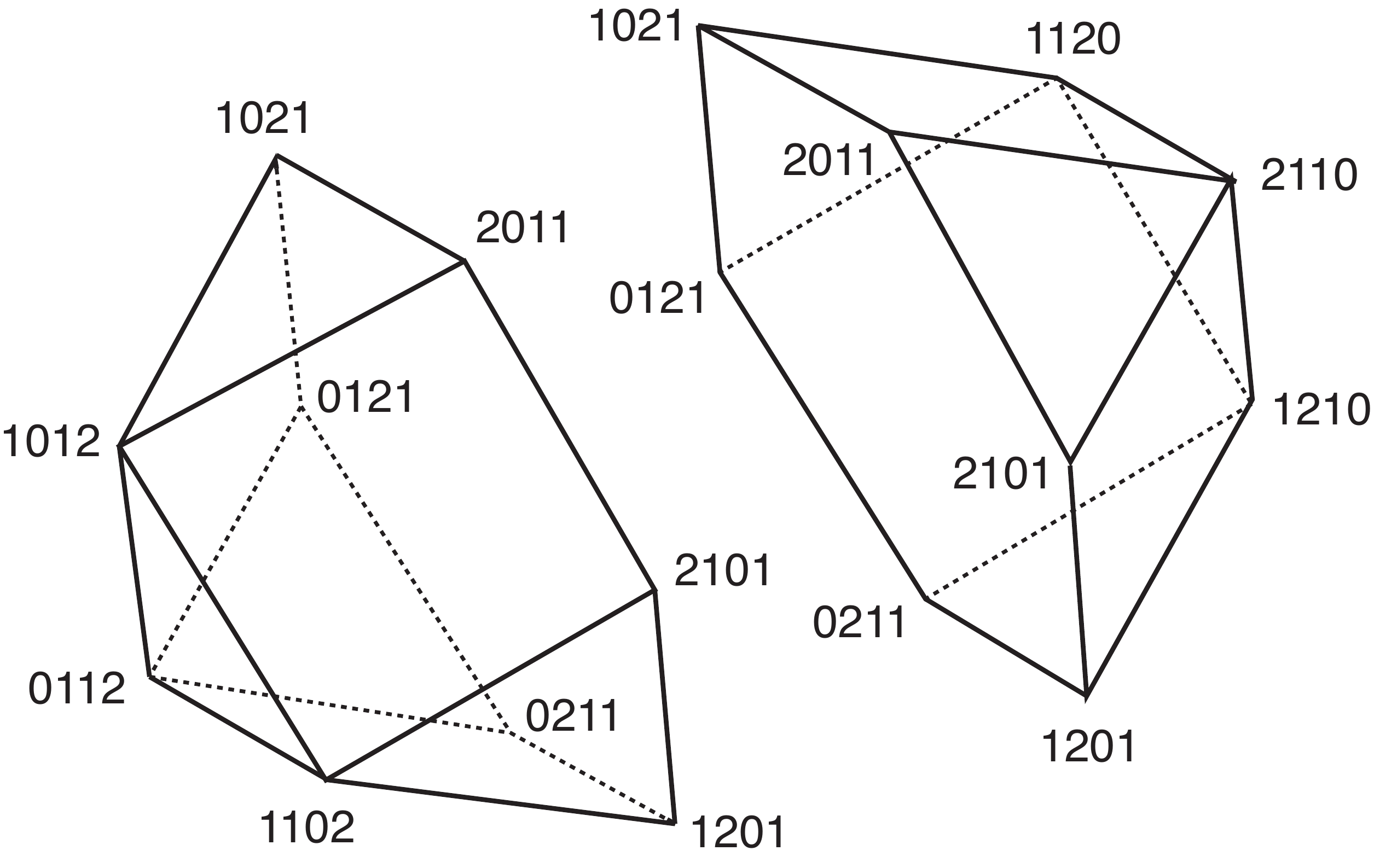}
\caption{The polytopes $\Delta_4(\backslash k)$ and $\Delta_4(\backslash k^*)$, obtained
by splitting $\Delta_4$ along an interior hexagon. Each is missing the vertices of one
triangular face.}
\label{fig:splitpolytope}
\end{figure}

At each fixed point of the diagonal torus in $SL(n,{\mathbb C})/P$, exactly one
$\pi_i$ and one $\pi_j^*$ does not vanish.  Those where $\pi_k \neq 0$
correspond to the vertices $L_k - L_i$ with $i \neq k$, whose
convex hull is an $(n-2)$-dimensional face of $\triangle_n$,
which we denote as $\triangle_n(k)$. 
The fixed points where $\pi_k^* \neq 0$
correspond to the vertices $L_i - L_k$ of the antipodal face, $\triangle_n(k^*)$.   
The polytope of an
$(n-1)$-dimensional torus orbit
where $\pi_k$ or $\pi_k^*$ is the only vanishing coordinate
is the convex hull of the vertices remaining
after the vertices of the face $\triangle_n(k)$, respectively
$\triangle_n(k^*)$ are removed.  These polytopes are denoted 
$\triangle_n(\backslash k)$ and $\triangle_n(\backslash k^*)$, respectively.  
They are congruent polytopes, obtained 
by splitting $\triangle_n$ along the hyperplane through
the vertices $L_i - L_j$ with $i,j \neq k$.  The convex
hull of these vertices
is an $(n-2)$-dimensional polytope in the interior of $\triangle_n$,
which we denote as $\triangle_n(\backslash k \backslash k^*)$.
We will refer to the pair $\triangle_n(\backslash k)$ and $\triangle_n(\backslash k^*)$ 
as a {\it split polytope}. In Figure \ref{fig:splitpolytope}, we illustrate the example of the split polytope
$\Delta_4(\backslash k)$ and $\Delta_4(\backslash k^*)$ \cite{Nongeneric}.
When two or more such splittings occur simultaneously, the 
collection of resulting polytopes will also be called a split polytope.

\begin{Proposition} \cite{Nongeneric} 
Let ${\mathcal F}$ be 
a variety in $SL(n,{\mathbb C})/P$ defined by fixing the values of the 1-chop integrals $I_{r1}$,
including the Casimir, where the values are chosen so that exactly one 
eigenvalue, say $\lambda_{i0}$, of $X$ is also an eigenvalue
of $\phi_1(X)$.
Then ${\mathcal F}$ is the union of the closures of two nongeneric
torus orbits, ${\mathcal O}^i$ and  ${\mathcal O}^{i^*}$, on which $\pi_i$, respectively
$\pi_i^*$, is the only coordinate that vanishes.  The images of their closures under the moment map,
$\triangle_n(\backslash i)$ and $\triangle_n(\backslash i^*)$,
are obtained by splitting $\triangle_n$
along the interior $(n-2)$-dimensional face $\triangle_n(\backslash i \backslash i^*)$.  
When exactly $p$ eigenvalues of $X$ are also
eigenvalues of $\phi_1(X)$ ($p \leq n-2$), then ${\mathcal F}$ is the
union of the closures of $2^p$ nongeneric $(n-1)$-dimensional orbits
whose images under the moment map are the polytopes obtained by splitting
$\triangle_n$ simultaneously 
along $p$ interior faces $\triangle_n(\backslash j \backslash j^*)$. 
\end{Proposition}
This result extends to the $k$-chop flows as follows:
\begin{Proposition} \cite{Nongeneric}
If $p$ eigenvalues of $\phi_{k}(X)$ and
$\phi_{k-1}(X)$ coincide, then the generic orbit of the diagonal
torus that generates the $(k-1)$-chop
flows in the component $SL(n-2(k-1),{\mathbb C})/P$ of (\ref{product})
becomes a union of $2^p$ nongeneric orbits.  Since the moment map on the product
(\ref{product}) is the product of the 
component moment maps, the moment map on the product of partial flag manifolds takes a level set in (\ref{product}) 
to a product of full and/or split polytopes, depending on where the coincidences of eigenvalues occur.  
\end{Proposition}

When a level set of the constants of motion is split into two or more nongeneric torus orbits, there
are separatrices in the Toda flows that generate the torus action.   The faces along which the polytope is split are the images under
the moment map of lower-dimensional torus orbits (the separatrices) that form the interface between the nongeneric orbits of maximum dimension.
The flow through an initial condition in one maximal orbit is confined to that orbit.  It is separated from the flows 
in the complementary nongeneric orbits by the separatrices.   \cite{Sl4} determines the monodromy around these singular level sets in the fiber
bundle of level sets where the spectrum of the initial matrix is fixed with distinct eigenvalues and the remaining constant of motion $I$
(equivalent to the determinant of the 1-chop matrix) is allowed to vary.  The flow generated by $I$ produces a ${\mathbb C}^*$-bundle
with singular fibers over the values of $I$.  The singularities occur both at values of $I$ where an eigenvalue of the 1-chop matrix coincides with an eigenvalue of the original matrix and at values of $I$ where the two eigenvalues of the 1-chop matrix coincide.   In a neighborhood of a singular
fiber of the first kind, the monodromy is characterized by a single twist of the noncompact cycle around the cylinder ${\mathbb C}^*$.  Near a singular fiber of the second kind, the monodromy produces two twists of the noncompact cycle.  This double twist is seen in the simplest case when $n = 2$ near the fiber where the two eigenvalues of the original matrix coincide; it as described in detail in \cite{Monodromy}.

When eigenvalues of $\phi_1(X)$ coincide, the torus embedding (\ref{Torus Embedding}) generalizes to the Jordan embedding
(\ref{Jordan Embedding}), under which the 1-chop flows generate the action of the group $A_J$ in (\ref{A}), a product of
a diagonal torus and a nilpotent group.  The general structure of a level set of the $1$-chop integrals with this type of singularity
is not known, in part because the orbit structure of $A_J$ in the flag manifold is not understood in sufficient detail.
When the eigenvalues are distinct, $A_J$ is a diagonal torus, and the closures of its orbits are toric 
varieties \cite{Oda}. The structure of torus orbits in flag manifolds is well-understood; see for example \cite{Atiyah}, \cite{FH G/P}, and \cite{GS}.  
The closures of orbits of $A_J$ in the flag manifold are generalizations of toric varieties, and much less is known about them.

The fixed points of the actions of the groups $A_J$ are studied in \cite{Fixed Points}, and the fixed point sets of the torus on the diagonal of
$A_J$ are characterized in \cite{Torus Actions}.   
If $A_J$ has $r$ blocks along the diagonal, where the dimension of the $i$th block is $d_i$, then the maximal diagonal subgroup of $A$ has
$\gcd(d_1, \ldots, d_r)$ connected components \cite{Connected}.  The subgroup of $A_J$ that fixes all
points in the flag manifold is the discrete group $D$ consisting of constant
multiples of the identity where the constants are the $n$th roots of unity; the group $A_J/D$ then acts effectively on the flag manifold.
\cite{Unipotent Fixed Points} describes the fixed-point set of the unipotent part
of $A_J$, giving an explicit way to express it in terms of canonical coordinates in each Bruhat cell. 
In the case where all eigenvalues coincide, $A_J$ is equal to its unipotent part.  \cite{Unipotent Actions} shows that the action of the 
group in this case preserves
each Bruhat cell and that its orbits in a given cell are characterized by the "gap sequence" of the permutation associated to the cell.

\section{Other Extensions of the Toda Lattice}\label{Other Extensions}

\subsection{Isospectral deformation of a general matrix} \label{Isospectral deformation}

In the full Hessenberg form of the Toda lattice, the matrix is diagonalizable if and only if the eigenvalues are distinct.
Kodama and Ye generalize this in \cite{general matrix}, where they consider an iso-spectral deformation of an arbitrary diagonalizable
matrix $L$.  The evolution equation is
\begin{equation}
\frac{d}{dt} L = [P, L] \ ; \label{general matrix}
\end{equation}
$P$ is defined by
\begin{equation}\label{P matrix}
P = \Pi(L) = (L)_{>0} - (L)_{<0} \ ,
\end{equation}
where $(L)_{>0 (<0)}$ is the strictly upper (lower) triangular part of $L$.
\cite{general matrix} establishes the complete integrability of (\ref{general matrix}) using inverse scattering, generalizing
the method used in \cite{Kodama McLaughlin} to solve the full symmetric real Toda lattice.  The method yields an explicit solution to the initial-value
problem.  The general context of the flow (\ref{general matrix})
includes as special cases the Toda lattices on other classical Lie algebras in addition to
$sl(n,{\mathbb{R}})$, which is most closely associated with Toda's original system.  
In this regard, Bogoyavlensky in \cite{B:76} formulated the Toda lattice on the real split semisimple
Lie algebras, which are defined as follows (the formulation below is in the Hessenberg (or Kostant)
form):
Let $\{h_{\alpha_i},e_{\pm\alpha_i}: i=1,\ldots,l\}$ be the Chevalley basis of the algebra $g$ of rank $l$,
i.e.
\[
[h_{\alpha_i},h_{\alpha_j}]=0,\quad [h_{\alpha_i},e_{\pm\alpha_j}]=\pm C_{ji}e_{\pm\alpha_j},\quad
[e_{\alpha_i},e_{-\alpha_j}]=\delta_{ij}h_{\alpha_j}\,,
\]
where $(C_{ij})_{1\le i,j\le l}$ is the Cartan matrix and $C_{ij}=\alpha_i(h_{\alpha_j})$.
Then the (non-periodic) Toda lattice associated with the Lie algebra $g$ is defined by
the Lax equation
\begin{equation}\label{generalLax}
\frac{dL}{dt}=[A,L]\,,
\end{equation}
where $L$ is a Jacobi element of $g$ and $A$ is the $\mathcal{N}_-$-projection of $L$,
\begin{align*}
L(t) &= \sum_{i=1}^l \, f_i(t)\,h_{\alpha_i}+\sum_{i=1}^l \,(g_i(t)\,e_{-\alpha_i}+e_{\alpha_i})\\
A(t)&=- \Pi_{\mathcal{N}_-}L(t)=-\sum_{i=1}^l\,g_i(t)\,e_{-\alpha_i}\,.
\end{align*}
The complete integrability is based on the existence of the Chevalley invariants of the algebra,
and the geometry of the isospectral variety has been discussed in terms of the representation
theory of Lie groups by Kostant in \cite{Kostant} for the cases
where $g_i$ are real positive, or complex. The general case for real $g_i$'s is studied
by Casian and Kodama \cite{CK:02, CK:02a}, which extends the results in
the $sl(n,\Real)$ Toda lattice in the Hessenberg form (see Section \ref{Extended Hessenberg Toda}) to the Toda lattice for any real split semisimple Lie algebra.
 
The Lax equation (\ref{generalLax}) then gives 
\begin{align*}
\frac{df_i}{dt}&=g_i\\
\frac{dg_i}{dt}&=-\left(\sum_{j=1}^lC_{ij}f_j\right)\,g_i\\
\end{align*}
 from which the $\tau$-functions are defined as
\begin{equation}\label{general tau}
f_k(t)=\frac{d}{dt}\ln\tau_k(t),\qquad g_k(t)=g_k(0)\prod_{j=1}^l (\tau_j(t))^{-C_{kj}}\,.
\end{equation}
In the case of $g=sl(n,\Real)$, those equations are (\ref{g indefinite}) and (\ref{f indefinite})
(note here that the superdiagonal of $L(t)$ is ${\rm diag}(f_1-f_2,f_2-f_3,\ldots, f_{l}-f_{l+1})$
with $n=l+1$). Those extensions have been discussed by many authors (see for example
\cite{Guest, Perelomov}).
One should note that Bogoyavlensky in \cite{B:76} also formulates those Toda lattices
for affine Kac-Moody Lie algebras, and they give the periodic Toda lattice. There has been much 
important progress on the periodic Toda lattices, but we will not cover the subject in this paper
(see for example \cite{AvM:80a, AvM:80b, Dub, RSTS, RS:94}).

From the viewpoint of Lie theory, the underlying structure of the integrable systems is
based on the Lie algebra splitting, e.g. $sl(n)=\mathcal{B}_-\oplus so(n)$ (the QR-decomposition) for the symmetric
Toda lattice, and $sl(n)=\mathcal{B}_+\oplus \mathcal{N}_-$ (the Gauss decomposition) for the Hessenberg form of
Toda lattice. Then one can also consider the following form of the evolution equation,
\begin{equation}
\frac{d}{dt}L=[Q_k,L]  \qquad {\rm with}\quad Q_k=\Pi_{\mathfrak{g}_1}(L^k)\,, \label{Q general Lax}
\end{equation}
where $\mathfrak{g}_1$ is a subalgebra in the Lie algebra splitting $sl(n)=\mathfrak{g}_1\oplus\mathfrak{g}_2$.
In this regard, we mention here the following two interesting systems directly connecting to the Toda lattice:

\medskip
\noindent
(a) The Kac-van Moerbeke system \cite{KM:75}:  We take $g_1=so(2n)$, and consider the equation for $L\in so(2n)$ (recall that $L$ is a symmetric matrix for the symmetric Toda lattice) .
Since $L^{2k-1}\in so(2n)$, the even flows are all trivial. Let $L$ be given by a tridiagonal form,
\[
L=\begin{pmatrix}
0   &   \alpha_1  &    0     &   \cdots    &    0    \\
-\alpha_1 &  0    &  \alpha_2 & \cdots  &   0  \\
\vdots  &  \vdots  &  \ddots   &   \cdots   & \cdots  \\
0          &    0        &    \cdots  &      0   &  \alpha_{2n-1} \\
0     &     0     &  \cdots   &  -\alpha_{2n-1}  &   0   \\
\end{pmatrix} \, \in \, so(2n,\Real)
\]
Then the even flows are the Kac-van Moerbeke hierarchy, $\displaystyle{\frac{\partial L}{\partial t_{2j}}=[\Pi_{{so}}(L^{2j}),L]}$ 
(recall that $\Pi_{{so}}(L^{2j}) = \mbox{Skew}(L^{2j})$),
where the first member of $t_2$-flow gives
\[
\frac{\partial \alpha_{k}}{\partial t_{2}}=\alpha_{k}(\alpha_{k-1}^2-\alpha_{k+1}^2)\,,\qquad k=1,\ldots,2n-1\,,
\]
with $\alpha_0=\alpha_{2n}=0$.
This system is equivalent to the symmetric Toda lattice which can be written as (\ref{Q general Lax})
for the square $L^2$. Note here that $L^2$ is a symmetric matrix given by
\[
L^2=T^{(1)}\otimes \begin{pmatrix}1 & 0\\0&0\end{pmatrix}\,+\, T^{(2)}\otimes \begin{pmatrix}0&0\\0&1\end{pmatrix}\,,
\]
where $T^{(i)}$, for $i=1,2$, are $n\times n$ symmetric tridiagonal matrices given by
\[
T^{(i)}=\begin{pmatrix}
b^{(i)}_1    & a^{(i)}_1 & 0 & \cdots & 0  \\
a^{(i)}_1 &  b_2^{(i)} & a_2^{(i)} & \cdots  & 0\\
\vdots    &   \vdots   &  \ddots   &  \ddots   &   \vdots  \\
0     &           0        &             \cdots  &     b_{n-1}^{(i)}    &   a_{n-1}^{(i)}\\
0     &           0       &          \cdots  &     a_{n-1}^{(i)} &b_{n}^{(i)}
\end{pmatrix}\,,
\] 
with $a^{(1)}_k = \alpha_{2k-1}
\alpha_{2k} $, $b^{(1)}_k = - \alpha_{2k-2}^2 - \alpha_{2k-1}^2 $, 
$a^{(2)}_k=\alpha_{2k}\alpha_{2k+1}$, and
$b^{(2)}_k=-\alpha_{2k-1}^2-\alpha_{2k}^2$ (see \cite{GHSZ:93}). Then
one can show that each $T^{(i)}$ gives the symmetric Toda lattice, that is,
the Kac-van Moerbeke hierarchy for $L^2$ matrix
splits into two Toda lattices,
\[ 
\frac{\partial T^{(i)}}{\partial t_{2j}} = [\Pi_{{so}}(T^{(i)})^{j},
  T^{(i)}]\,\quad i=1,2\,.
\]
The equations for $T^{(i)}$ are connected by the Miura-type transformation, with the functions
 $(a^{(i)}_k,b^{(i)}_k)$, through the Kac-van Moerbeke variables $\alpha_k$ (see \cite{GHSZ:93}).
 
\medskip
\noindent
(b) The Pfaff lattice for a symplectic matrix \cite{AM:02, KP:07}: The Pfaff lattice is defined in the same form
with $g_1=sp(2n)$ and $L$ in the Hessenberg form with $2\times 2$ block structure.
In particular, we consider the case $L\in sp(2n)$ having the form,
\[
L=
\begin{pmatrix}
\begin{matrix} 0 & s_1 \\ b_1 & 0 \end{matrix} &\vline&
\begin{matrix} 0  & 0 \\ a_1 & 0 \end{matrix} &\vline& \cdots &\vline& 0_2
\\ \hline\bigxstrut
\begin{matrix} 0 & 0 \\ a_1 & 0 \end{matrix} &\vline&
\begin{matrix} 0 & s_2 \\ b_2 & 0 \end{matrix} &\vline&\cdots &\vline& 0_2
\\ \hline\bigxstrut
\vdots &\vline& \vdots &\vline& \ddots &\vline& \vdots
\\ \hline\bigxstrut
0_2 & \vline& 0_2 &\vline& \cdots &\vline &\begin{matrix} 0 & s_n \\ b_n & 0\end{matrix}
\end{pmatrix} \, \in \,sp(2n,\Real) \,,
\]
where $0_2$ is the $2\times 2$ zero matrix. The variables $(a_k,b_k)$ and $s_k=\pm1$
are those in the indefinite Toda lattice. It should be noted again that the odd members
are trivial (since $L^{2k-1}\in sp(2n)$), and the even members give the indefinite Toda
lattice hierarchy \cite{KP:08}. Here one should note that $L^2$ can be written as
\[
L^2=\tilde{L}^T\otimes\begin{pmatrix}1 & 0\\0&0\end{pmatrix} \, +\, 
\tilde{L}\otimes\begin{pmatrix}0&0\\0&1\end{pmatrix}\,,
\]
where $\tilde L$ is given by (\ref{Indefinite Symmetric1}). Then one can show that 
the generator $Q_{2j}$ of the Lax equation is given by
\[
Q_{2j}=\Pi_{{sp}}(L^{2j})=-\tilde{B}^T_{j}\otimes\begin{pmatrix}
1&0\\0&0\end{pmatrix}\,+\,\tilde{B}_j\otimes\begin{pmatrix}0&0\\0&1\end{pmatrix}\,,
\]
where $\tilde{B}_j=\frac{1}{2}[(\tilde{L}^j)_{>0}-(\tilde{L}^j)_{<0}]$.
Then the hierarchy $\frac{d}{dt}L=[Q_{2j},L]$ gives the indefinite Toda lattice hierarchy.

\subsection{Gradient formulation of Toda flows} \label{gradient formulation}

In \cite{Bloch:90}, Bloch observed that the symmetric tridiagonal Toda equations (\ref{Symmetric Lax Equation}) can also be written in the 
double-bracket form
\begin{equation}\label{double bracket}
\frac{d}{dt}L(t) = [L(t), \ [L(t), \ N]]  \,,
\end{equation}
where $N$ is the constant matrix $\mbox{diag}(1, 2, ..., n)$ and $L$ is as in (\ref{Symmetric}). 
He showed that this double-bracket equation is the gradient flow of the function $f(L) = \mbox{tr}(LN)$ with respect to the normal metric on an adjoint orbit of $SO(n)$.
The normal metric is defined as follows: Let $\kappa(\,,\,)=-\langle\,,\,\rangle$ be the Killing
form of a semisimple Lie algebra $g$, and decompose $g$ orthogonally relative to $\langle\,,\,\rangle$ into $g=g^L\oplus g_L$ where $g_L$ is the centralizer of $L$ and $g^L={\rm Im~ad}(L)$.
For $X\in g$, denote by $X^L$ the projection of $X$ onto $g^L$. 
Then given two tangent vectors to the orbit at $L$, $[L,X]$ and $[L,Y]$, the normal metric is defined by
$\langle [L,X],[L,Y]\rangle_N=\langle X^L,Y^L\rangle$. Then the right hand side of (\ref{double bracket})
can be written as 
grad$\,H=[L,[L,N]]$ for the Hamiltonian function $H(L)=\kappa(L,N)$ (Proposition 1.4 in \cite{BBR:92}).
Thus the Toda lattice (\ref{Symmetric Lax Equation}) is both Hamiltonian and a gradient flow on the isospectral set. 
Brockett shows in \cite{Br:88} that any symmetric matrix $L(0)$ can be diagonalized by the flow (\ref{double bracket}), and the flow can be used to solve various combinatorial optimization problems
such an linear programming problems (see \cite{HM:94} for the connections of the Toda lattice with
several optimization problems).

The flow (\ref{double bracket}) is extended in \cite{BBR:90} and \cite{BBR:92} to show that the generalized tridiagonal symmetric Toda lattice can also be expressed
as a gradient flow.   
In Section \ref{Isospectral deformation}, we give the equations of the generalized tridiagonal Toda lattice in the Hessenberg form on a real split
semisimple Lie algebra.  The symmetric version of this is as follows (see \cite{BBR:92}).
Let $g$ be a complex semisimple Lie algebra of rank $l$ with normal real form $g_n$.  Choose a Chevalley basis 
$\{h_{\alpha_k},e_{\pm\alpha_k}: k=1,\ldots,l\}$ as in Section \ref{Isospectral deformation}.  The generalized tridiagonal symmetric Toda lattice is defined by 
the Lax equation
\[
\frac{d}{dt}L(t)=[A(t),L(t)]\,, \label{general tridiagonal symmetric}
\]
where 
\begin{align*}
L(t) &= \sum_{k=1}^l \, b_k(t)\,h_{\alpha_k}+\sum_{k=1}^l \,a_k(t)(e_{\alpha_k}+e_{-\alpha_k})\\
A(t)&= \sum_{k=1}^l\,a_k(t)\,(e_{\alpha_k}-e_{-\alpha_k}) \,.
\end{align*}
This flow defines a completely integrable Hamiltonian system on the coadjoint orbit of the lower Borel subalgebra of $g_n$ through 
$\sum_{k=1}^l(e_{\alpha_k}-e_{-\alpha_k})$.  The Hamiltonian is $H(L) = \frac{1}{2} K(L,L)$, where $K$ is the Killing form.

\cite{BBR:92} shows that (\ref{general tridiagonal symmetric}) is a gradient flow with respect to the normal metric on the orbit. 
The gradient formulation in \cite{BBR:92} is given in the context of the compact form $g_u$ of $g$ (see also the survey in \cite{BG:07}).
Their key result is the following.
\begin{Proposition} \label{gradient vector field}
The gradient vector field of the function $f(L) = K(LN)$ on the adjoint orbit in $g_u$ containing the initial condition $L_0$, with respect to 
the normal metric, is  
\begin{equation}
\frac{d}{dt}L(t) = [L(t), \ [L(t), \ N]] \label{gradient flow} \,.
\end{equation}
\end{Proposition}
Now let $H_u$ be a maximal abelian subalgebra of $g_u$, and take $H = H_u \oplus iH_u$ as the Cartan subalgebra of $g$. 
Choose a Chevalley basis for $g$ as above.   Bloch, Brockett, and Ratiu \cite{BBR:92} show the following.
\begin{Theorem} \label{Toda flow is gradient}
Let $N$ be $i$ times the sum of the simple coweights of $g$, and let 
\[
L(t) = \sum_{k=1}^l \,i\,b_k(t)\,h_{\alpha_k}+\sum_{k=1}^l \,i\,a_k(t)(e_{\alpha_k}+e_{-\alpha_k})\,.
\]
Then the gradient vector field (\ref{gradient flow}) gives the flow of the generalized tridiagonal symmetric Toda lattice on the adjoint orbit in $g_u$ containing the initial condition $L_0$.
Explicitly,
\[
N = \sum_{k=1}^l \,i\,x_k\,h_{\alpha_k}\,,
\]
where $(x_1, ..., x_l)$ is the unique solution of the system
\[
\sum_{k=1}^l \,x_k\,\alpha_p(h_{\alpha_k})\, = \, -1 \,, p = 1, ..., l \,.
\]
\end{Theorem}
A list of the coefficients $(x_1, ..., x_l)$ for all the semi-simple Lie algebras is given on p.~62 of \cite{BBR:92}.
Proposition (\ref{gradient vector field}) and Theorem (\ref{Toda flow is gradient}) are extended in \cite{dMP:99} to the generalized full
symmetric Toda lattice and in \cite{BG:98, BG:07} to the generalized signed Toda lattice to show that these extensions of the Toda flows are also
gradient flows.

\section{Connections with the KP equation}  \label{KP connection}
Here we give a brief review of the paper \cite{BK:03} whose main result is to show that the $\tau$-functions of the Toda hierarchy (\ref{Symmetric Toda Hierarchy})
with a symmetric tridiagonal matrix provide a new class of solutions of the Kadomtsev-Petviashvili (KP) equation. We also provide a geometric 
description of the $\tau$-functions in terms of the Grassmann
manifolds (see \cite{K:04}).

\subsection{The $\tau$-functions for the symmetric Toda lattice hierarchy}

We return to the Toda lattice hierarchy (\ref{Symmetric Toda Hierarchy}) with symmetric tridiagonal
matrix. The solution $L({\bf t})$ can be explicitly expressed in terms of the $\tau$-functions (\ref{Todatau}): Let us summarize the process of solution method based on the Gram-Schmidt orthogonalization.
First we consider
\[
g({\bf t})=\exp\left(\frac{1}{2}\theta(L(0),\mathbf{t})\right)\,,
\]
where $\theta(\lambda,\mathbf{t}):=\sum_{k=1}^{n-1}t_k\lambda^k$ with the $k$-th flow parameter
$t_k$ of the Toda hierarchy, i.e.
\[
\frac{\partial L}{\partial t_k}=[B_k,L] \qquad{\rm with}\quad B_k=\frac{1}{2} {\rm Skew}(L^k)\,.
\]
(Note here that we rescale the time $t_k\to t_k/2$.) Then define the matrix
\begin{align*}
M(\mathbf{t})&:=g^T(\mathbf{t})g(\mathbf{t})= e^{\theta(L(0),\mathbf{t})} \\
     &= \Phi(0)\,e^{\theta(\Lambda,\mathbf{t})}\Phi^T(0)=\left(\langle \phi_i^0\phi_j^0e^{\theta(\lambda,\mathbf{t})}\rangle\right)_{1\le i,j\le n}
\end{align*}
where $\Phi(0)=(\phi_i^0(\lambda_j))_{1\le i,j\le n}$ is the eigenmatrix of $L(0)$, i.e. $L(0)\Phi(0)=\Phi(0)\Lambda$, and $\Phi(0)\in SO(n)$. Since $L(0)$ is a tridiagonal matrix, the entries $m_{i,j}(\mathbf{t}):=\langle\phi_i^0\phi_j^0e^{\theta(\lambda,\mathbf{t})}\rangle$ can be written in terms of the moment by the Gram-Schmidt
orthogonalization process (see \cite{Kodama McLaughlin} for the details), 
\[
m_{i,j}(\mathbf{t})=\langle \lambda^{i+j-2}\rho(\lambda)\, e^{\theta(\lambda,\mathbf{t})}\rangle=\sum_{k=1}^n
\lambda_k^{i+j-2}\,\rho_k \,e^{\theta_k(\mathbf{t})}\,,
\]
where $\rho(\lambda)=\phi_1^0(\lambda)^2$ with $\rho_k=\rho(\lambda_k)$, and
$\theta_k(\mathbf{t})=\theta(\lambda_k,\mathbf{t})$.
In particular, we have
\begin{equation}\label{tau1}
\tau_1(\mathbf{t})=\langle \rho(\lambda)e^{\theta(\lambda,\mathbf{t})}\rangle=\sum_{k=1}^n\, \rho_k\,e^{\theta_k(\mathbf{t})}\,.
\end{equation}
Then the $\tau$-functions are given by the Wronskian of the set of functions 
of $\tau_1(\mathbf{t})$ and its $x$-derivatives,
\[
\tau_k(\mathbf{t})={\rm Wr}\,(\tau_1(\mathbf{t}),\tau_1'(\mathbf{t}),\ldots,\tau_1^{(k-1)})\qquad {\rm for}\quad  k=1,2,\ldots,n-1\,.
\]
Using the Binet-Cauchy theorem, one can write $\tau_k$ in the form
\begin{equation}\label{tauk}
\tau_k(\mathbf{t})=\sum_{1\le i_1<\cdots<i_k\le n}\xi(i_1,\ldots,i_k)\,E(i_1,\ldots,i_k)(\mathbf{t})\,,
\end{equation}
where 
\begin{equation}\label{plucker xi}
\left\{\begin{array}{llll}
\displaystyle{\xi(i_1,\ldots,i_k)=\left[\prod_{1\le l<j\le k} (\lambda_{i_j}-\lambda_{i_l})\right]\prod_{j=1}^k\rho_{i_j}},\\
{}\\
\displaystyle{ E(i_1,\ldots,i_k)={\rm Wr}\,(e^{\theta_{i_1}},e^{\theta_{i_2}},\ldots,e^{\theta_{i_k}})\,.}
\end{array}\right.
\end{equation}
Here we assume the ordering in $\lambda_j$ as
\begin{equation}\label{ordering}
\lambda_1<\lambda_2<\cdots <\lambda_n\,,
\end{equation}
so that all $\xi(i_1,\ldots,i_k)$ are positive. As we show below that this form of $\tau_k$
has interesting geometric interpretation, which makes a connection with the KP equation.

\begin{Example}\label{sl4Toda}
In Figure \ref{fig:sl4Toda}, we illustrate the solutions $(a_1(\mathbf{t}),a_2(\mathbf{t}),a_3(\mathbf{t}))$ for the $sl(4,\Real)$ Toda lattice.
The $\tau$-functions are given by
\begin{align*}
\tau_1(\mathbf{t})&=e^{\theta_1(\mathbf{t})}+\cdots +e^{\theta_4(\mathbf{t})},\\
\tau_2(\mathbf{t})&=\displaystyle{\sum_{i<j}(\lambda_j-\lambda_i)^2e^{\theta_i(\mathbf{t})+\theta_j(\mathbf{t})}},\\
\tau_3(\mathbf{t})&=\displaystyle{\sum_{i<j<k}[(\lambda_i-\lambda_j)(\lambda_j-\lambda_k)(\lambda_i-\lambda_k)]^2e^{\theta_i(\mathbf{t})+\theta_j(\mathbf{t})+\theta_k(\mathbf{t})}},
\end{align*}
with $\theta_k(\mathbf{t})=\lambda_kt_1+\lambda_k^2t_2+\lambda_k^3t_3$.
The solutions $a_k(\mathbf{t})$ are then given by
\[
a^2_k(\mathbf{t})=\frac{\tau_{k-1}(\mathbf{t})\,\tau_{k+1}(\mathbf{t})}{\tau^2_k(\mathbf{t})},
\]
which can be also written as $a_k^2=\partial^2\ln\tau_k/\partial t_1^2$.  

Each line in Figure \ref{fig:sl4Toda} is given by the equation $\theta_k(\mathbf{t})=\theta_j(\mathbf{t})$
for some $k$ and $j$, for example, in the middle graphs, two lines are
\begin{itemize}
\item[(i)]  $\theta_1(\mathbf{t})=\theta_3(\mathbf{t})$ which gives
\[
t_1+(\lambda_1+\lambda_3)\,t_2+(\lambda_1^2+\lambda_1\lambda_3+\lambda_3^2)\,t_3=0\,,
\]
\item[(ii)] $\theta_2(\mathbf{t})=\theta_4(\mathbf{t})$ which gives
\[
t_1+(\lambda_2+\lambda_4)\,t_2+(\lambda_2^2+\lambda_2\lambda_4+\lambda_4^2)\,t_3=0\,.
\]
\end{itemize}
Each line indicates a balance between two exponential terms in the $\tau$-function, for example,
$\theta_j=\theta_k$ shows the balance between $e^{\theta_j}$ and $e^{\theta_k}$ in $\tau_1$, or
$e^{\theta_j+\theta_l}$ and $e^{\theta_k+\theta_l}$ in $\tau_2$ for some $l$.
Then near each line for $\theta_j=\theta_k$, we have, from $\tau_1\sim e^{\theta_j}+e^{\theta_k}$,
\[
a_k^2(\mathbf{t})=\frac{1}{4}(\lambda_k-\lambda_j)^2\,{\rm sech}^2\,\frac{1}{2}(\theta_j(\mathbf{t})-\theta_k
(\mathbf{t}))\,.
\]
This is a soliton solution of the KP equation, and we will discuss some details of the connection
to the KP equation in the next section. In this regard, the graphs in Figure \ref{fig:sl4Toda} shows
several solutions of the KP equation, and in particular those solutions indicate the soliton
resonances (see \cite{BK:03, K:04, ChK:07} for more details).

\begin{figure}[t!]
\centering
\includegraphics[scale=0.45]{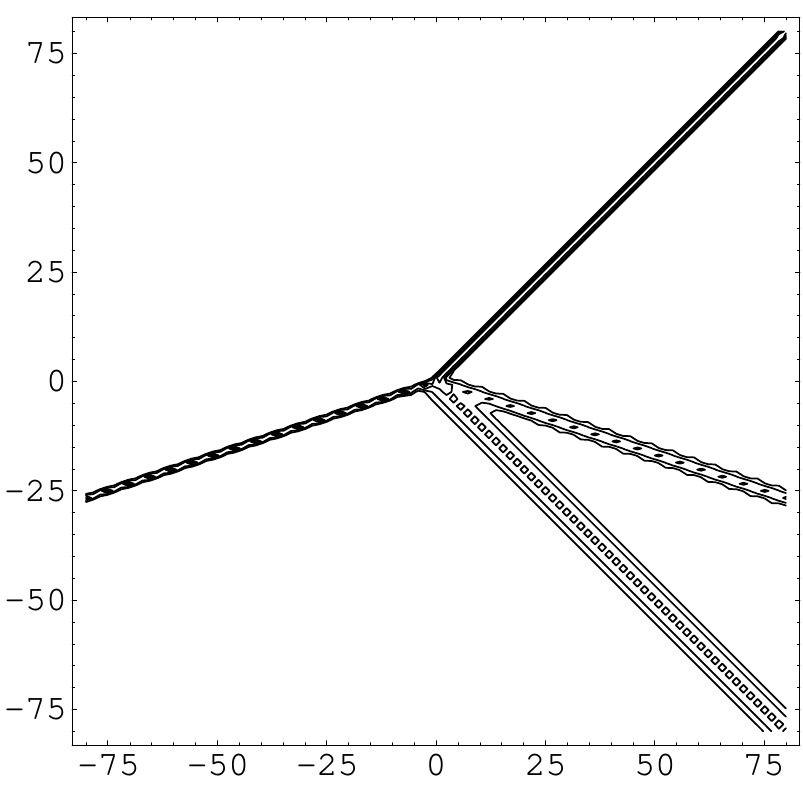} \hskip 1cm 
\includegraphics[scale=0.45]{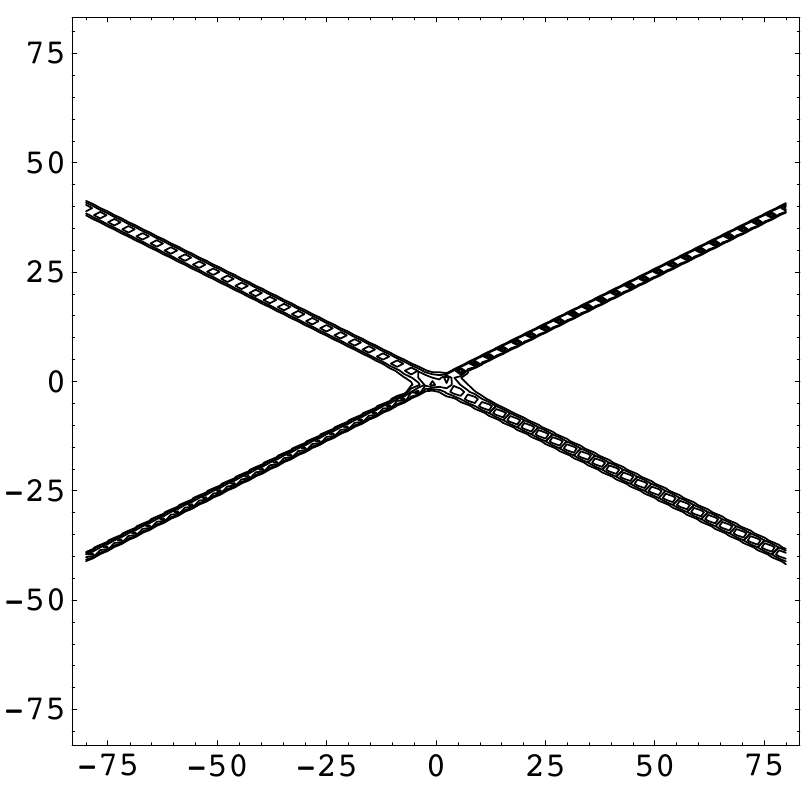}\hskip 1cm
\includegraphics[scale=0.45]{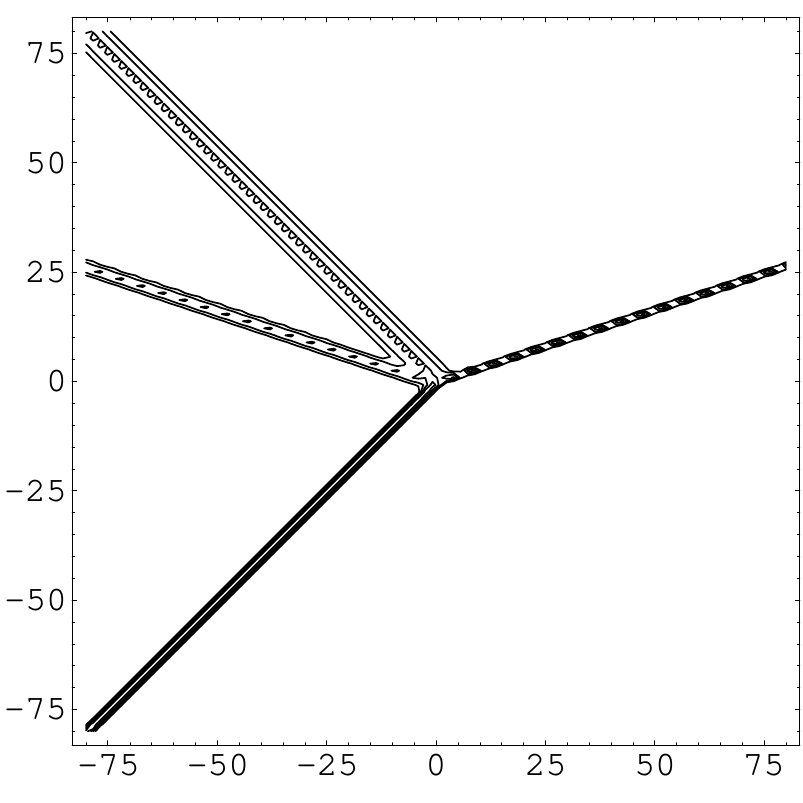} 
\caption{The solutions $(a_1,a_2,a_3)$ for $sl(4,\Real)$ Toda equation.
The graphs show the contour lines of the solutions $a_k(t_1,t_2,t_3)$
for $t_1$-$t_2$ plane with $t_3=0$. 
The left graph is  for $a_1$, the middle one for $a_2$ and the right one for $a_3$.
Each line is given by the balance $\theta_j(\mathbf{t})=\theta_k(\mathbf{t})$ where $\theta_j(\mathbf{t})
=\lambda_jt_1+\lambda_j^2t_2+\lambda_j^3t_3$. Here the eigenvalues $(\lambda_1,\ldots,\lambda_4)$ are given by
$(-3,0,1,2)$. For example, the lines in the middle graph are given by $t_1+2t_2=0$ and $t_1-2t_2=0$.
Those graphs also show soliton solutions of the KP equation (see Section \ref{KPsolitons}). }
\label{fig:sl4Toda}
\end{figure}

\end{Example}


\subsection{The KP equation and the $\tau$-function}\label{KPsolitons}
The KP equation is given by the following partial differential equation,
\begin{equation}\label{KP}
\frac{\partial }{\partial x}\left(-4\frac{\partial u}{\partial t}+\frac{\partial^3u}{\partial x^3}+12u\frac{\partial u}{\partial x}\right)+3\frac{\partial^2 u}{\partial y^2}=0\,,
\end{equation}
where $u=u(x,y,t)$ with $(x,y)$ represents a coordinate of two space dimensions and $t$ the time.
The KP equation is one of the prototypical (2+1)-dimensional integrable 
 equations, originally derived in \cite{KP:70}
as a model for small-amplitude, quasi two-dimensional 
 waves in a weakly dispersive medium. It arises in many different physical
applications including shallow water waves and ion acoustic waves in plasmas (for a review, 
see e.g. \cite{IR:00}). 

The goal of this section is to show that a large class of the solutions of the KP equation can be
obtained in the Wronskian forms which contain the $\tau$-functions of the symmetric Toda lattice
(this is based on \cite{S:81}, and see also \cite{MJD:00} and Appendix in \cite{BK:03}).
We start with a brief explanation of the inverse scattering scheme for the KP equation:
 Let $\mathcal{L}$ be a pseudo-differential operator
defined by
\[
{\mathcal L}=\partial +u_2\partial^{-1}+u_3\partial^{-2}+\cdots,
\]
where $\partial$ is a derivation satisfying
$\partial\,\partial^{-1}=\partial^{-1}\,\partial=1$ and
the generalized Leibnitz rule,
\[
\partial^{\nu}(fg)=\sum_{k=0}^{\infty}\bigg(\!\!
\begin{array}{cc}\nu\\k\end{array}\!\!\bigg)\,
\frac{\partial^kf}{\partial x^k}\,\,\partial^{\nu-k}g\,,
\qquad\mathrm{for}\quad\nu\in{\mathbb Z}\,.
\]
(Note that the series terminates if $\nu$ is a positive integer.)
Then the following infinite set of equations is called the KP hierarchy:
\begin{equation}
\label{e:kpl}
\frac{\partial {\mathcal L}}{\partial t_j}=[{\mathcal B}_n,{\mathcal L}],
\quad\mathrm{with}\quad
{\mathcal B}_j:=({\mathcal L}^j)_{+}, \quad j=1,2,\ldots.
\end{equation}
Here $({\mathcal L}^n)_{+}$ represents the projection of $\mathcal{L}^n$ onto the polynomial
(differential) part in $\partial$. 
For examples, the first three members of $\mathcal{B}_j$ are given by
\begin{align*}
\mathcal{B}_1&=\partial,\qquad  \mathcal{B}_2=\partial^2+2u_2 \\
\mathcal{B}_3&=\partial^3+3u_2\partial+3(u_{2,x}+u_3)
\end{align*}
where $u_{2,x}=\partial u_2/\partial x$.  The equation $\mathcal{B}_1=\partial$ implies
\[
\frac{\partial \mathcal{L}}{\partial t_1}=\partial (\mathcal{L})=\frac{\partial\mathcal{L}}{\partial x}\,,
\]
 from which we identify $t_1=x$.
The compatibility among the equations in (\ref{e:kpl})
is given by the Zakharov-Shabat (ZS) equations,
\[
\frac{\partial \mathcal{B}_i}{\partial t_j}-\frac{\partial\mathcal{B}_j}{\partial t_i}+[\mathcal{B}_i,\mathcal{B}_j]=0\,,
\]
which are a direct consequence of the definition of $\mathcal{B}_j$. In particular, the ZS equation with $i=2$ and $j=3$ gives
\begin{align*}
&\displaystyle{2\frac{\partial u_3}{\partial x}-\frac{\partial u_2}{\partial t_2}+\frac{\partial^2u_2}{\partial x^2}  =0 },\\  
&\displaystyle{3\left(\frac{\partial u_3}{\partial t_2}-\frac{\partial^2u_3}{\partial x^2}\right)
-2\frac{\partial u_2}{\partial t_3}+3\frac{\partial^2u_2}{\partial t_2\partial x}-\frac{\partial^3u_2}{\partial x^3}+6u_2\frac{\partial u_2}{\partial x}=0\,.}
\end{align*}
Eliminating $u_3$ from those equations, we obtain the KP equation (\ref{KP}) with the identifications
$u=u_2$ and $y=t_2, t=t_3$.  This means that any solution of the KP hierarchy is also a solution of the
KP equation.

Now writing ${\mathcal L}$ in the dressing form,
\begin{equation}\label{LW}
{\mathcal L}={\mathcal W}{\partial}{\mathcal W}^{-1}, \quad {\rm with}\quad
{\mathcal W}=1+w_1\partial^{-1}+w_2\partial^{-2}+\cdots,
\end{equation}
the KP hierarchy becomes
\begin{equation}
\label{e:weq}
\frac{\partial{\mathcal W}}{\partial t_j}={\mathcal B}_j{\mathcal
W}-{\mathcal W}\,\partial^j.
\end{equation}
Using~\eqref{LW}, the variables $u_i$ can be expressed in terms of the
$w_j$'s; for example,
\[ 
u_2=-\frac{\partial w_1}{\partial x}, \qquad u_3=-\frac{\partial w_2}{\partial x}+w_1\frac{\partial w_1}{\partial x}\,,
\]
and so on.
The equations for $w_j$ are, for example,
\begin{align*}
&\frac{\partial w_1}{\partial t_2}=-2w_1\frac{\partial w_1}{\partial x}+\frac{\partial^2 w_1}{\partial x^2}+2\frac{\partial w_2}{\partial x}\\
&\frac{\partial w_2}{\partial t_2}=-2w_2\frac{\partial w_1}{\partial x}+\frac{\partial ^2 w_2}{\partial x^2}+2\frac{\partial w_3}{\partial x}\,,
\end{align*}
and so on.

To find some exact solutions, we consider a finite truncation of ${\mathcal W}$, for some $k\ge1$,
\[
{\mathcal W}_k:=1+w_1\partial^{-1}+\cdots +w_k\partial^{-k}\,.
\]
 For example,
the ${\mathcal W}$-equation (\ref{e:weq}) for the $k=1$ truncation, i.e.
${\mathcal W}_1=1+w_1\partial^{-1}$, is just the Burgers equation,
\begin{equation}
\label{e:burgers}
\frac{\partial w_1}{\partial t_2}=-2w_1\frac{\partial w_1}{\partial x}+\frac{\partial w_1}{\partial x^2}\,,
\end{equation}
which can be solved by a Cole-Hopf transformation, $w_1=-\partial \ln f/\partial x$, leading to
a linear diffusion equation, $f_{t_2}=f_{xx}$. Note here that the Cole-Hopf transformation is just
$\mathcal{W}_1\partial f=\partial f+w_1 f=0$.
For the $k$-truncation,  we consider the generalization,
\begin{equation}
\label{e:fm}
{\mathcal W}_k\partial^kf=f^{(k)}+w_1f^{(k-1)}+\cdots+w_kf=0.
\end{equation}
The invariance of this equation under the evolution (\ref{e:weq}) can be shown as follows:
\begin{align*}
\frac{\partial}{\partial t_j}(\mathcal{W}_k\partial^kf)&=\frac{\partial\mathcal{W}_k}{\partial t_j}\partial^k f
+\mathcal{W}_k\partial^k\frac{\partial f}{\partial t_j} \\
&= \left(\mathcal{B}_j\mathcal{W}_k-\mathcal{W}_k\partial^j\right)\partial^kf+\mathcal{W}_k\partial^k\frac{\partial f}{\partial t_j} \\
&=\mathcal{W}_k\partial^k\left(\frac{\partial f}{\partial t_j}-\frac{\partial^j f}{\partial x^j}\right)\,.
\end{align*}
This implies that the equation $\mathcal{W}_k\partial^kf=0$ is invariant, if $f$ satisfies the {\it linear}
equations,
\begin{equation}\label{linear}
\frac{\partial f}{\partial t_j}=\frac{\partial^jf}{\partial x^j}\qquad {\rm for}\quad j=1,2,\ldots.
\end{equation}

Now we construct the solutions of the KP equation from the linear equations (\ref{linear}):
Let $\{f_j\,|~j=1,\dots,k\}$ be a fundamental set of solutions of
(\ref{e:fm}), i.e.
\[
f_j^{(k)}+w_1f_j^{(k-1)}+\cdots + w_{k-1}f_j'+w_kf_j=0,\qquad j=1,\ldots,k\,.
\]
Using the Cramer's rule, the function $w_1$ is expressed in terms of the
Wronskian with those $f_j$'s, i.e.,
\[
w_1=-\frac{\partial}{\partial x}\ln {\rm Wr}(f_1,\dots,f_k).
\]
We then obtain a solution of the KP equation,
\[
\displaystyle{u=u_2=-\frac{\partial}{\partial 
x}w_1=\frac{\partial^2}{\partial x^2}\ln {\rm Wr}(f_1,\ldots,f_k)\,.}
\]
Here the Wronskian is called the $\tau$-function of the KP equation (see also \cite{F:83}). 
Thus the solution $u$ in this form is characterized by the kernel of the differential operator,
$\mathcal{W}_k\partial^k$, and the Wronskian structure of
the $\tau$-function leads to the notion of the Grassmannian
as explained below.  

It is now clear that 
this $\tau$-function agrees with $\tau_k$ in (\ref{tauk}) of the Toda lattice,
if we take $f_j$ to be
\[
f_j=\tau_1^{(j-1)} =\frac{\partial^{j-1}\tau_1}{\partial x^{j-1}}\qquad {\rm for}\quad j=1,2,\ldots,k\,.
\]
One can also show that the square of $a_k$ in the $L$ matrix 
of the symmetric Toda lattice is a solution of the KP equation, i.e.
\[
a_k^2= u=\frac{\partial^2}{\partial x^2}\ln\tau_k\qquad {\rm with}\quad \tau_k={\rm Wr}(\tau_1,\tau_1',\ldots,\tau_1^{(k-1)})\,.
\]


\begin{Example}
In Figure \ref{fig:2soliton}, we show the 2-soliton solution given by
$\tau_2(\mathbf{t})$ in Example \ref{sl4Toda}, that is, $u(x,y,t)=a_2^2(t_1,t_2,t_3)$
with $t_1=x,t_2=y$ and $t_3=t$. This solution illustrates a resonant interaction
of two solitons, which was first found in \cite{BK:03}.
 Notice that there are six distinct regions in the $t_1$-$t_2$ plane,
four unbounded regions and two bounded regions at the interaction point.
Those six regions correspond to the dominant exponentials in $\tau_2$ function,
and those are identified as the vertices of the moment polytope of $Gr(2,4)$
(see Section \ref{grassmannian} and also \cite{KP:07}).

\begin{figure}[t!]
\centering
\includegraphics[scale=0.34]{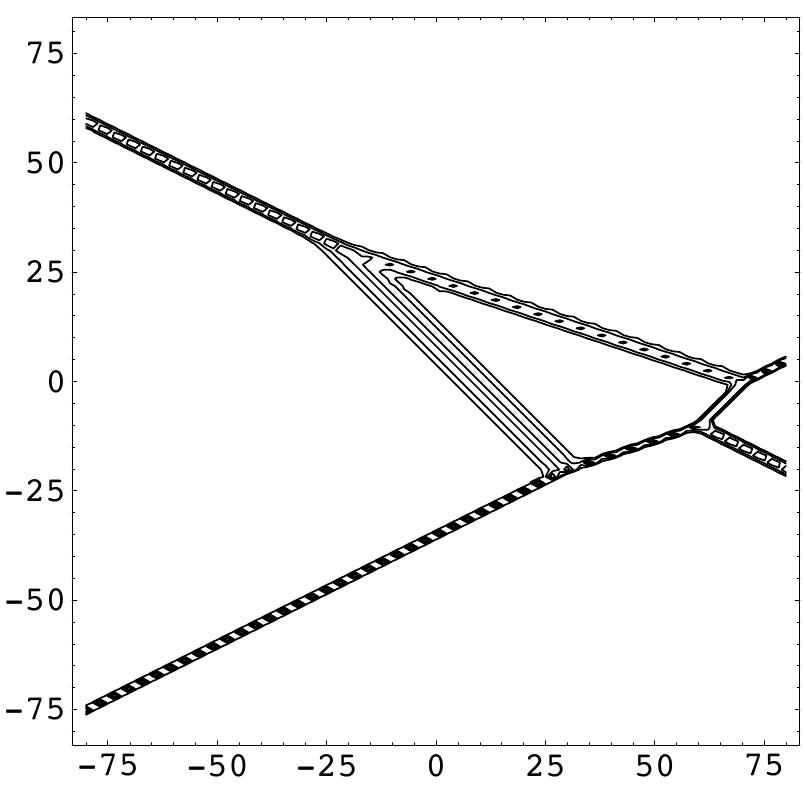} \hskip 0.2cm 
\includegraphics[scale=0.34]{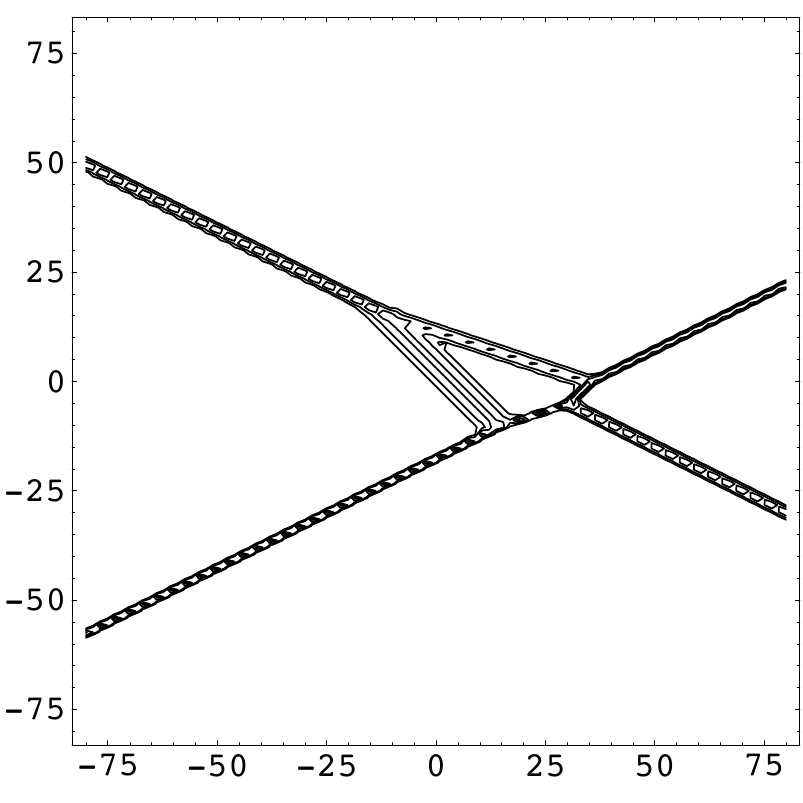}\hskip 0.2cm
\includegraphics[scale=0.34]{F2sola0.pdf} \hskip 0.2cm
\includegraphics[scale=0.34]{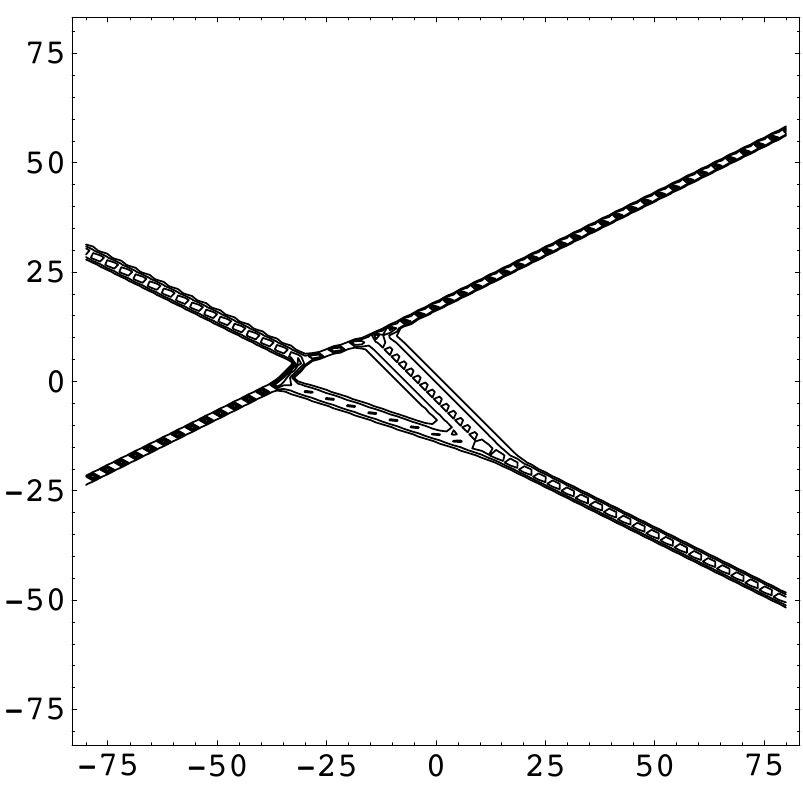}\hskip 0.2cm
\includegraphics[scale=0.34]{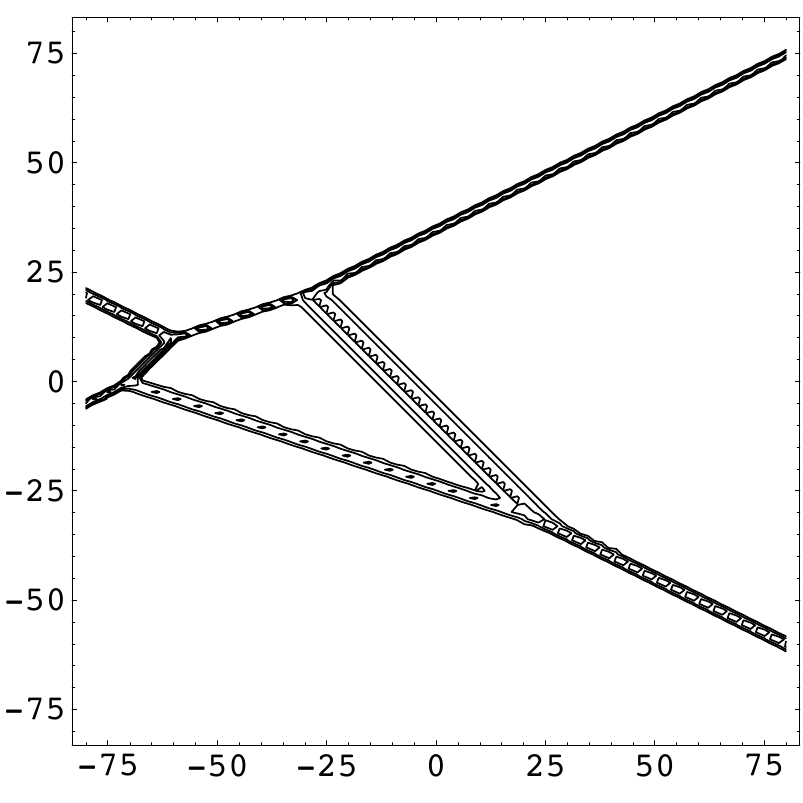}
\caption{A 2-soliton solution of the KP equation. The figures show the contour plots
of the solution $u(x,y,t)$ in the $x$-$y$ plane with the times for $t=-10$ (left) $-5,~0,5$ and
$10$ (right). The $\tau$-function is given by $\tau_2(\mathbf{t})$ in Example \ref{sl4Toda} with $t_1=x,t_2=y$ and $t_3=t$. The parameters $(\lambda_1,\ldots,\lambda_4)$ are the same as
in Figure \ref{fig:sl4Toda}.}
\label{fig:2soliton}
\end{figure}
\end{Example}


\subsection{Grassmannian $Gr(k,n)$}\label{grassmannian}
There exists a natural identification between the space of the $\tau$-functions 
of the Wronskian form ${\rm Wr}(f_1,\ldots,f_k)$ and the Grassmannian
$Gr(k,n)$, the set of $k$-dimensional subspaces in $\mathbb{R}^n$. To explain this, we
take the following functions 
as a fundamental set of the solutions (\ref{e:fm}), i.e. finite Fourier series solutions of (\ref{linear}),
\begin{equation}
\label{f-function}
f_i=\sum_{j=1}^n a_{ij}\,e^{\theta_j}\,,\quad {\rm for}\quad i=1,\ldots,k\le n\,,
\end{equation}
with some constants $a_{ij}$ which define the $k\times n$ matrix $A_{(k,n)}:=(a_{ij})$, and the phase functions ${\theta_j}$ are given by
\begin{equation}
\label{theta}
\theta_j(\mathbf{t})=\sum_{i=1}^{n-1}\lambda_j^i\,t_i\,\quad {\rm for}\quad j=1\ldots,n\,.
\end{equation}
Here $\lambda_j$ are arbitrary constants which can be identified as the eigenvalues
of the $L$ matrix for the Toda lattice. We assume $\lambda_j$ being ordered as (\ref{ordering}),
i.e. $\lambda_1<\cdots<\lambda_n$.

Since $\{e^{\theta_j}: j=1,\ldots,n\}$ and $\{f_i:i=1,\ldots,k\}$ are
linearly independent sets of functions, it follows that   
${\rm Span}_{\Real}\{e^{\theta_j}:j=1,\ldots,n\} \cong \Real^n$, and also
that ${\rm Span}_{\Real}\{f_i:i=1,\ldots,k\}$ defines a $k$-dimensional
subspace of $\Real^n$. We then consider the identifications for the generic set $\{\lambda_1,\ldots,\lambda_n\}$,
\begin{subequations}
\begin{equation*}
e^{\theta_j} \longleftrightarrow E_j := e^{\theta_j}\,(1,\lambda_j, \ldots, \lambda_j^{n-1})^{T} \in \Real^n\,,
\quad \forall\, j=1,2, \ldots n\,,  
\end{equation*}
and
\begin{equation*}
f_i \longleftrightarrow F_i := \sum_{j=1}^M a_{ij}E_j \,, \quad \forall \, i=1,2,\ldots,k .
\end{equation*}
\end{subequations}
Note here that 
\[
f^{(j-1)}_i=\langle F_i,e_j\rangle, \qquad j=1,\ldots,n,
\]
where $e_j$ is the $j$-th vector of the standard basis $\{e_i\}_{i=1}^n$, and $\langle \cdot,\cdot\rangle$
is the scalar product on $\mathbb{R}^n$.
Then a point $F \in Gr(k,n)$ is defined by $F={\rm Span}_{\mathbb R}\{F_i:i=1,\ldots,k\}$. 
 Equivalently, $F \in Gr(k,n)$
is represented by the $k \times n$, full rank coefficient matrix $A = (a_{ij})$ whose
rows are the coordinates of the basis vectors $\{F_i:i=1,\ldots,k\}$ with respect to
the fixed basis $\{E_j:j=1,\ldots,n\}$ of $ \Real^n$. Since $A$ depends on
the choice of basis, the matrix representation for the point $F \in Gr(k,n)$ is 
unique up to a left multiplication: $A \to HA$ by any $H \in GL(k)$. So, 
the Grassmannian can be considered as the factor space of $M_{k\times n}(\Real)$,
the set of all $k\times n$ matrices of rank$=k$, with $GL(k)$, i.e.
\[
 Gr(k,n) \cong M_{k \times n}(\Real)/GL(k) \,.
\]
Note dim $Gr(k,n) = k(n-k)$. The matrix $H\in GL(k)$ is chosen so that the coefficient matrix $A$ is put in the reduced row-echelon 
form (RREF) via Gaussian elimination.
(Recall that, for an $k\times n$ matrix in RREF,
the leftmost nonvanishing entry in each nonzero row is called a pivot,
which is normalized to 1. The submatrix of the pivot columns of the full rank
matrix $A$ is the $k \times k$ identity matrix). A convenient parametrization of $Gr(k,n)$
is given by the {\em Pl\"ucker embedding},
\[\begin{array}{ccc}
Gr(k,n) &\longrightarrow & {\mathbb P}(\bigwedge^{k}{\Real}^{n})\\[1.0ex]
F = {\rm Span}_{\Real}\{F_i: i=1,\ldots,k\} &\longmapsto& F_1\wedge\cdots\wedge F_{k}\,. 
\end{array}\]
With respect to the basis $\{E_{i_1}\wedge\cdots\wedge E_{i_k}:i_1<\cdots<i_k\}$ for $\wedge^k\Real^n$,
\begin{equation}\label{wedgeF}
F_1\wedge\cdots\wedge F_{k}=\sum_{1\le i_1<\ldots<i_k\le n}
A{(i_1,\cdots,i_k)}\,E_{i_1}\wedge\cdots\wedge E_{i_k}\,,
\end{equation}
where the coefficients $A{(i_1,\ldots,i_k)}$ of $A$ are the $k\times k$ maximal minors of $A$,
called the Pl\"ucker coordinates. These are the homogeneous coordinates
of the image of $F$ in ${\mathbb P}(\bigwedge^{k}{\Real}^{n})$, and they satisfy the {\it Pl\"ucker relations},
\[
\sum_{r=1}^{k+1}(-1)^r\,A(j_r,i_1,\ldots,i_{k-1})\,A(j_1,\ldots,\hat{j}_r,\ldots,j_{k+1})=0\,,
\]
for any set of numbers $\{i_1,\ldots,i_{k-1},j_1,\ldots,j_{k+1}\}\in \{1,2,\ldots,2k\}$.
Here $\hat{j}_r$ implies the deletion of the index $j_n$.
Then $Gr(k,n)$ is also defined as the set of vectors in the form (\ref{wedgeF})
with the coordinates satisfying the Pl\"ucker relations. Notice that the functions $\xi(i_1,\ldots,i_k)$
are the Pl\"ucker coordinates, that is, they satisfy the Pl\"ucker relations (this is the general fact for the
$\tau$-functions, see below).

Let $\langle\cdot,\cdot\rangle$ be the usual inner product on $\wedge^k\Real^n$, i.e.
\[
\langle v_1\wedge\cdots\wedge v_k,\,w_1\wedge\cdots\wedge w_k\rangle:={\rm det}(\langle
v_i,w_j\rangle)_{1\le i,j\le n}\,.
\]
Then the Wronskian $E(i_1,\ldots,i_k)$ is given by
\[
E(i_1,\ldots,i_k)=\langle E_{i_1}\wedge\cdots\wedge E_{i_k}, v_{[1\ldots,k]}\rangle\,,
\]
with $v_{[1,\ldots,k]}=e_1\wedge\cdots\wedge e_k$. Note here that the independence of the exponential functions $E(i_1,\ldots,i_k)$ can be
shown for the generic choice of the parameters $\{\lambda_1,\ldots,\lambda_n\}$.

Now let us define the set of $\tau$-functions,
\[
\mathcal{T}(k,n):=\left\{\tau_A={\rm Wr}(f_1,\ldots,f_k):   (f_1,\ldots,f_k)=(e^{\theta_1},\ldots,e^{\theta_n})A^T\right\}
\]
Then comparing the expression of a point on $Gr(k,n)$ and the $\tau$-function
of the Wronskian determinant ${\rm Wr}(f_1,\ldots,f_k)$, we obtain the following Proposition:
\begin{Proposition}\label{bijection}
Let $\lambda(i_1,\ldots,i_k)=\sum_{j=1}^k\lambda_{j_i}$, and assume that those are all distinct. Then
there exists a bijection,
\[
\begin{array}{cccc}
\mathcal{T}(k,n) &~ \overset{\cong}{ \longrightarrow } &~Gr(k,n) \\[1.0ex]
\displaystyle{\tau_k={\rm Wr}(f_1,\ldots,f_k)}&~\longmapsto &~F=F_1\wedge\cdots \wedge F_k
\end{array}
\]
\end{Proposition}
\begin{Proof}
Using the Binet-Cauchy theorem, we have
\begin{align}\label{BC tau}
\tau_k=\sum_{1\le i_1<\cdots<i_k\le n} A(i_1,\ldots,i_k)\,E(i_1,\ldots,i_k)
 \end{align}
where $A(i_1,\ldots,i_k)$ are the Pl\"ucker coordinates given in (\ref{wedgeF}), and
\[
E(i_1,\ldots,i_k)={\rm Wr}(e^{\theta_{i_1}},\ldots,e^{\theta_{i_k}})=\prod_{j<l}
(\lambda_{i_l}-\lambda_{i_j})\, \exp \theta(i_1,\ldots,i_k)\,.
\]
Here $\theta(i_1,\ldots,i_k)=\sum_{j=1}^k\theta_{i_j}$.
Since $\{\lambda(i_1,\ldots,i_k)\}$ are distinct, the set $\{E(i_1,\ldots,i_k)\}$ is  linearly independent as the functions of $x$. This implies that one can identify
the basis $\{E_{i_1}\wedge \cdots\wedge E_{i_k}\}$ of $\wedge^k\mathbb{R}^n$
with $\{E(i_1,\ldots,i_k)\}$ of $\mathcal{T}(k,n)$, and we have $\tau_k=\langle F_1\wedge\cdots\wedge F_k,\, v_{[1,\ldots,k]}\rangle$.
\end{Proof}
We thus identify each $\tau$ function in the form (\ref{BC tau}) as a point of $Gr(k,n)$, and
the solution of the KP equation given by $\tau_k(\mathbf{t})$ defines a torus orbit on $Gr(k,n)$,
\[
[F_1(\mathbf{t}),\ldots,F_k(\mathbf{t})]=[E_1(\mathbf{t}),\ldots,E_k(\mathbf{t})]\,A^T
=K\,{\rm diag}(e^{\theta_1(\mathbf{t})},\ldots,e^{\theta_n(\mathbf{t})})\,A^T\,,
\]
where $K=(k_{ij})_{1\le i,j\le n}$ with $k_{ij}=\lambda_j^{i-1}$. Note in particular that 
\[
\frac{\partial E_j(\mathbf{t})}{\partial t_k}=C_{\Lambda}^k\,E_j(\mathbf{t})=\lambda_j^k\,E_j(\mathbf{t})\,,
\]
where $C_{\Lambda}$ is the companion matrix (\ref{companion matrix}).

\begin{figure}[t!]
\centering
\includegraphics[scale=0.3]{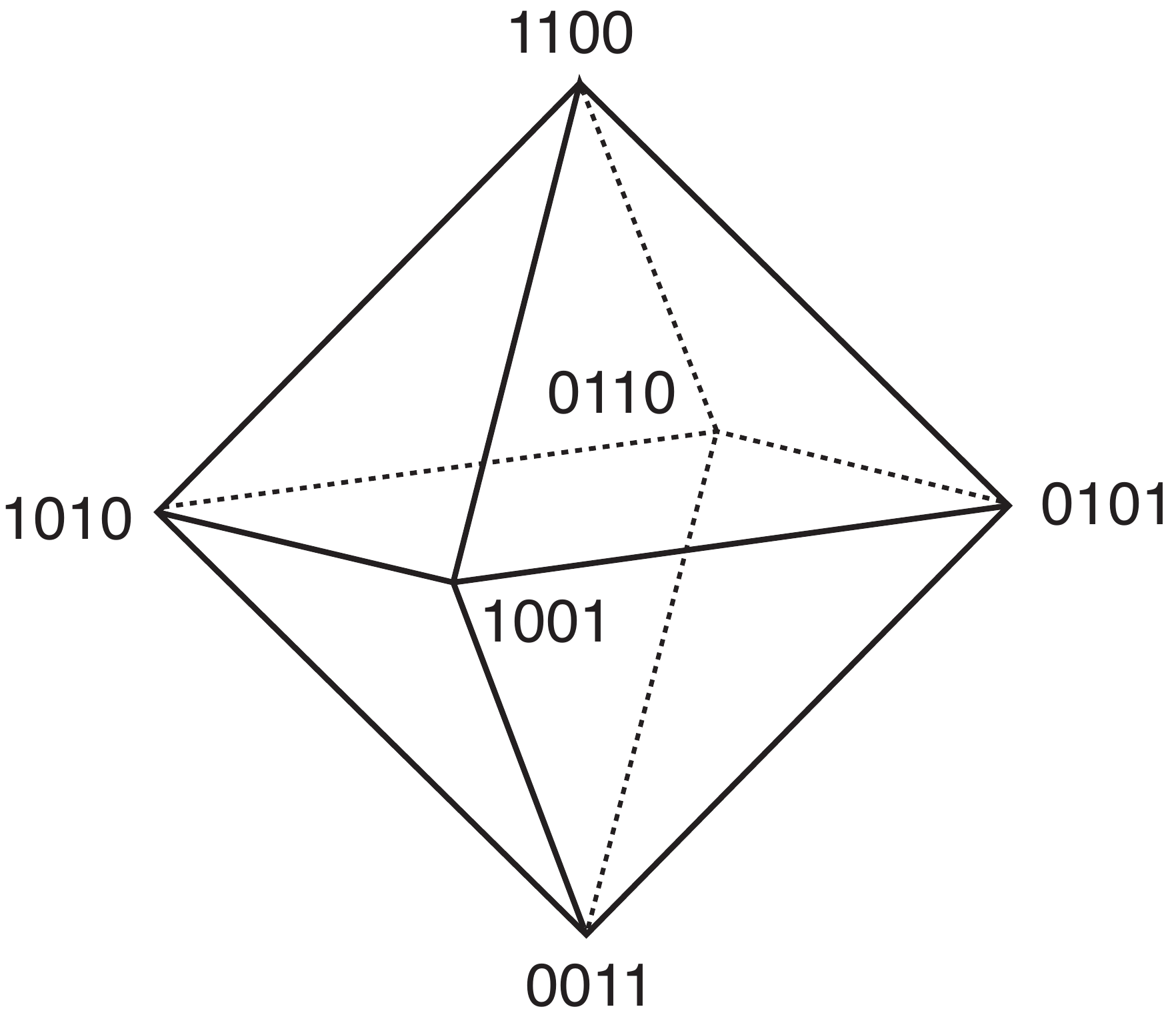}
\caption{The moment polytope of $Gr(2,4)$.
Each vertex corresponds to a fixed point of the KP flow given by
the $\tau_2$ function on $\Real^4$, and it represents the dominant
exponent in $\tau_2$, e.g. the weight  $1100=L_1+L_2$ corresponds to
the term $e^{\theta_1+\theta_2}$, which is dominant as $t_1\to -\infty$ for
$\lambda_1<\cdots<\lambda_4$.}
\label{fig:Grassmann24}
\end{figure}

As discussed in Section \ref{moment}, let us define the moment map $\mu:Gr(k,n)\to \mathcal{H}_{\Real}^*,~
\tau_k\mapsto \mu(\tau_k)$ with
\begin{equation}\label{moment tau}
\mu(\tau_k)=\frac{\sum_{1\le i_1<\cdots<i_k\le n}|A(i_1,\ldots,i_k)e^{\theta(i_1,\ldots,i_k)}|^2\,(L_{i_1}+\cdots+L_{i_k})}
{\sum_{1\le i_1<\cdots<i_k\le n}|A(i_1,\ldots,i_k)e^{\theta(i_1,\ldots,i_k)}|^2}
\end{equation}
where $\theta(i_1,\ldots,i_k)=\sum_{j=1}^k\theta(\lambda_{i_j},\mathbf{t})$, and $L_j$ are the weights
of the standard representation of $SL(n)$ (see (\ref{moment map})), and $\mathcal{H}^*_{\Real}$ is
defined by
\[
\mathcal{H}_{\Real}^*={\rm Span}_{\Real}\left\{L_1,\ldots, L_n:\sum_{j=1}^nL_j=0\right\}\cong{\Real}^{n-1}\,.
\]
Then from (\ref{moment tau}), we can see that the image under the moment map of the toric variety 
generated by the solutions of the KP equation with (\ref{BC tau}) is a convex polytope whose vertices
are the fixed points of the orbit. In the representation theory, this polytope is a weight polytope
of the fundamental representation of $sl(n)$ on $\wedge^kV$
with the standard representation $V$. In Figure \ref{fig:Grassmann24},
we illustrate the moment polytope of $Gr(2,4)$. The orbit given by a KP solution can be
realized as a curve inside of the polytope.

\section{The Toda lattice and integral cohomology of real flag manifolds} \label{Toda-cohomology connections}
Here we briefly explain how one can get the integral cohomology of real flag variety
from the isospectral variety of the Toda lattice (this is an introduction of the papers \cite{CK:06,CK:07}):
We consider the Toda lattice hierarchy (\ref{Hessenberg hierarchy}) on the {\it real} split semi-simple Lie algebra
$sl(n,\Real)$, and assume $X\in sl(n,\Real)$ to be a generic element in the tridiagonal
Hessenberg form, that is, it has all real and distinct eigenvalues (see \cite{CK:06,CK:07}, for the general case associated with real split semisimple Lie algebra).  

\subsection{The moment polytope and Weyl group action}
Let $G=SL(n,\Real)$, $B$ be the Borel subgroup of upper triangular matrices of
$G$, and $N$ be the subgoup of lower unipotent matrices of $G$. 
As in (\ref{companion embedding}), we consider the companion embedding \cite{CK:02a, FH:91},
\begin{equation}\label{companion embedding c}
\begin{array}{ccccccc}
c_{\Lambda} &:&  \mathcal{M}_{\Lambda} & \longrightarrow & G/B \\[1.0ex]
                         &   &    X(0)   &\longmapsto &  n_0^{-1}~{\rm mod}\,B
\end{array}
\end{equation}
where $n_0\in N$ is the unique element such that $X(0)=n_0C_{\Lambda}n_0^{-1}$ with
the companion matrix (\ref{companion matrix}) \cite{Kostant}.
With the factorization (\ref{general factorization}) for the hierarchy (\ref{Hessenberg hierarchy}), i.e.
\[
e^{\theta(X(0),\mathbf{t})}=n(\mathbf{t})b(\mathbf{t})\qquad {\rm with}\quad \theta(X(0),\mathbf{t})=\sum_{j=1}^{n-1}t_jX(0)^j\,,
\]
we have 
\[X(\mathbf{t})=n^{-1}(\mathbf{t})\,X(0)\,n(\mathbf{t})\,.
\]
The Toda lattice hierarchy can be linearized on the flag variety $G/B$, that is, we have a commutative diagram \cite{FH:91},
\[
\begin{CD}
X(0)  @>c_{\Lambda}>> n_0^{-1}\,{\rm mod}\,B\\
@V Ad(n^{-1}(\mathbf{t}))VV @VVV \\
X(\mathbf{t}) @>c_{\Lambda}>>  
       e^{\theta(C_{\Lambda},\mathbf{t})}n_0^{-1}~{\rm mod}~B
\end{CD}
\]
where we have used $n_0^{-1}n(\mathbf{t})\equiv n_0^{-1}e^{\theta(X(0),\mathbf{t})}~{\rm 
mod}\,B$ and $X(0)n_0=n_0C_{\Lambda}$.    
With this diagram, the flows of the Toda lattice hierarchy form a toric variety, denoted by $\widetilde{\mathcal{M}}_{\Lambda}$,  in the flag  
manifold $G/B$, that is,
\begin{equation}\label{toric}
\widetilde{\mathcal{M}}_{\Lambda}=\overline{G^{C_\Lambda}n_0^{-1}B}\qquad {\rm with}\quad G^{C_{\Lambda}}=\left\{e^{\theta(C_{\Lambda},\mathbf{t})}:\mathbf{t}\in\Real^{n-1}\right\}\,.
\end{equation}
In order to characterize this toric variety, let us first consider a diagonal embedding of $G/B$,
\begin{equation}\label{diagonal embedding}
\phi:G/B \longrightarrow Gr(1,n)\times Gr(2,n)\times \cdots \times Gr(n-1,n)\,,
\end{equation}
which defines the Bruhat decomposition, that is, each point in $G/B$ determines a set of
points $(V_1,\ldots,V_{n-1})$ with $V_k\in Gr(k,n)$ such that
\[
\{0\}\subset V_1\subset \cdots\subset V_{n-1}\subset \Real^{n}\,,
\]
where $V_k$ can be expressed in terms of a frame $V_k=[F_1,\ldots,F_k]$, where the $\{F_j:j=1,\ldots,k\}$ is a basis of $V_k$. 
As shown in Proposition \ref{bijection}, each $\tau_k$ can be identified with a point on $Gr(k,n)$ so that one may write
\[
\phi\circ c_{\Lambda}(X(\mathbf{t}))=\phi(e^{\theta(C_{\Lambda},\mathbf{t})}n_0^{-1}B)=(\tau_1(\mathbf{t}),\ldots,\tau_{n-1}(\mathbf{t}))\,.
\]
Now we define the moment map $\mu:G/B\to \mathcal{H}^*_{\Real}$, which is expressed in
terms of the map (\ref{moment tau}),
\begin{equation}\label{Toda moment map}
\begin{array}{ccccccc}
\mu &: & G/B &\longrightarrow & \mathcal{H}_{\Real}^*  \\[0.5ex]
   &  &   e^{\theta(C_{\Lambda},\mathbf{t})}n_0^{-1}B & \longmapsto  &  \sum_{k=1}^{n-1}\mu(\tau_k(\mathbf{t}))
\end{array}
\end{equation}
The fixed points of the flows in $G/B$ are the images of the matrices $X$ with all 
$g_k=0$. There are $n!$ fixed points, which is the order of the symmetric group $S_n$, and those are the orbit of the Weyl group
$W=S_n$. 
As we explained in Section \ref{moment}, we also note that  the image under $\mu$ of the toric
variety $\widetilde{\mathcal{M}}_{\Lambda}$ in (\ref{toric}) is the convex hull of the vertices corresponding to
the fixed points, which are the weights $L=\sum_{k=1}^{n} i_kL_k$ with $i_k\in\{0,1,\ldots,n-1\}$
and $i_k\ne i_j$ if $k\ne j$. The highest weight is given by $L_*=\sum_{k=1}^n(n-k)L_k$, which corresponds to the matrix $X(\mathbf{t})$ as $t_1\to -\infty$ with the ordering
$\lambda_1<\cdots<\lambda_n$, i.e. ${\rm diag}(X)={\rm diag}(\lambda_1,\ldots,\lambda_n)$ (see, for example, Figure \ref{fig:A3momentpolytope}).

\begin{figure}[t!]
\centering
\includegraphics[scale=0.32]{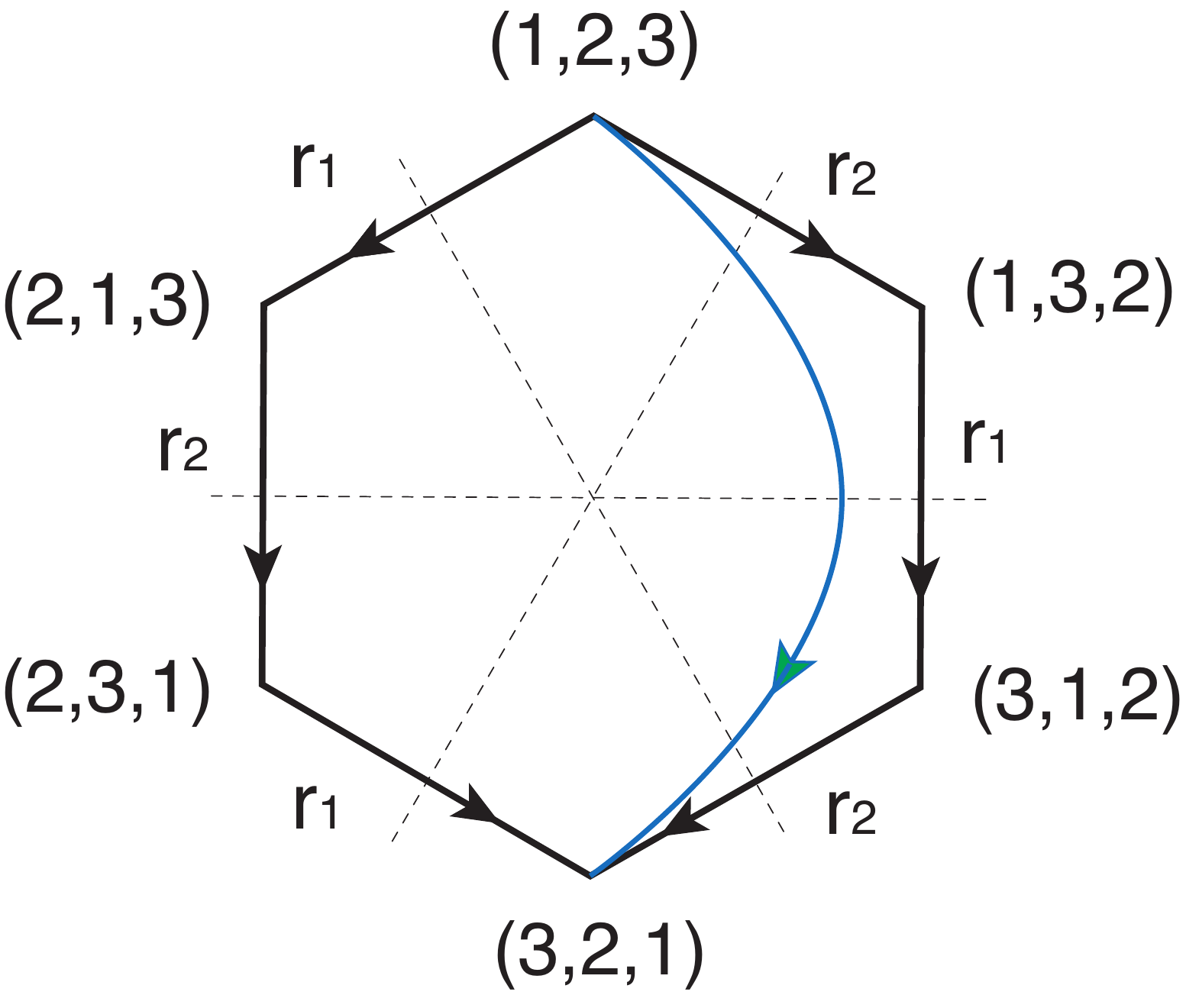}
\caption{The moment polytope for the $sl(3,\Real)$ Toda lattice.  
The vertices are marked by the ordered sets of the eigenvalues, i.e. $(i,j,k)=(\lambda_i, \lambda_j, \lambda_k)$. Each edge corresponds to a simple reflection $r_j$ exchanging the eigenvalues at the
$j$th and $(j+1)$th places. Then the vertices are also marked by
the elements of the Weyl group, e.g. $(1,2,3)=e, (2,1,3)=r_1$ etc. Each directed edge corresponds to an $sl(2,\Real)$ Toda flow, and
the directed curve in the middle shows a generic flow of $sl(3,\Real)$ Toda lattice.}
\label{fig:moment}
\end{figure}

This can be seen from the dominant exponential term in $\tau_k(t_1)$ for $t_1\to -\infty$, i.e.
\[
\tau_k(t_1)\approx A(1,\ldots,k)E(1,\ldots,k) \quad {\rm as}\quad t_1\to -\infty\,,
\]
which implies $\mu(\tau_k)=L_1+\cdots+L_k$. Since the vertices are the orbit of the Weyl group $W$,
those can be also parametrized by the elements of $W=S_n$. Let $r_i\in W$ be a simple reflection, that is,
$r_i$ exchanges the entries in positions $i$ and $i+1$ on the diagonal of the fixed point matrix. Each vertex of the polytope can be also represented by the $n$-tuple of the eigenvalues, that is, 
$(\lambda_{w^{-1}(1)},\lambda_{w^{-1}(2)},\ldots,\lambda_{w^{-1}(n)})$ is  the diagonal of the fixed
point matrix marked by $w\in W$. In Figure \ref{fig:moment}, we illustrate the weight
polytope associated with the $sl(3,\Real)$ Toda lattice where $(i,j,k)=(\lambda_i,\lambda_j,\lambda_k)$. Each edge given by an arrow corresponds to a simple reflection, and it gives the flow associated with $sl(2,\Real)$ Toda lattice.

\subsection{Integral cohomology of $G/B$}
Here we give a brief summary of the cohomology of $G/B$ as a background for
the next section where we explain how one gets the cohomology of $G/B$ from
the isospectral variety of the Toda lattice associated with real split semisimple Lie group $G$.

Let us first recall the Bruhat decomposition of $G/B$,
\[
G/B=\bigsqcup_{w\in W}X_w \qquad {\rm with}\quad X_w=NwB/B\,.
\]
Each Bruhat cell $X_w$ is labeled by the element $w\in W$ and ${\rm codim}\,(X_w)=l(w)$
where $l(w)$ represents the length of $w$. Then we can define the chain complex,
\[
\mathcal{C}^*=\bigoplus_{k=0}^{l(w_*)}\mathcal{C}^k\qquad{\rm with}\quad  \mathcal{C}^k=\sum_{l(w)=k}\mathbb{Z}\,X_w\,,
\]
where $w_*$ is the longest element of $W$, and the coboundary operators $\delta_k$ on $X_w$ with $l(w)=k$ is given by
\[
\delta_k(X_w)=\sum_{l(w')=k+1}[w:w']\,X_{w'}\,,
\]
where $[w:w']$ is the incidence number associated with $X_w\overset{\delta_k}{\longrightarrow}X_{w'}$.
It has been known (see \cite{K:95, CS:99}) that the incidence number is either $0$ or $\pm2$ for the real flag manifold
$G/B$ of real split semi-simple Lie group $G$. The Bruhat order defined on $W$
is given as follows: For two elements $w$ and $w'$ in $W$, we have the order
\[
w\le w' \qquad {\rm iff}\qquad \overline{X}_w\supset\overline{X}_{w'}\,,
\]
where the closure $\overline{X}_w$ is called the Schubert manifold. Then the cohomology
of $G/B$ can be calculated from the incidence graph $\mathcal{G}_{G/B}$ defined as follows:
\begin{Definition}\label{incidence graph}
The incidence graph $\mathcal{G}_{G/B}$ consists of
the vertices labeled by $w\in W$ and the edges $\Rightarrow$ defined by
\[
w\,\Rightarrow\, w' \quad {\rm iff}\quad \left\{
\begin{array}{lllll}
{\rm (i)}~~ w\le w' \\[0.4ex]
{\rm (ii)}~~ l(w')=l(w)+1 \\[0.4ex]
{\rm (iii)}~~ [w:w']\ne 0
\end{array}
\right.
\]
The incidence number for each edge is either $0$ or $\pm 2$ (see \cite{CS:99}). The integral cohomology is then calculated from the graph.
\end{Definition}
\begin{Example}
In the case of $G=SL(3,\Real)$, the incidence graph is given by
\[
\begin{matrix} 
{}                     &{e}& {}                \\[1.5ex]
[1]                     &{} & [2]                   \\
 \Downarrow&{}&\Downarrow  \\
  [12]               &{}& [21]                  \\[1.5ex]
   {}&{[121]}&{ }                      \\
    \end{matrix} 
\] 
where the vertices of the hexagon are marked by the elements of the Weyl group denoted as $r_{i}\cdots r_j=[i\cdots j]$. Since the nonzero incidence numbers are $\pm 2$, the integral cohomology of $G/B$ in this case is give by
\[
\left\{ \begin{array}{lllll}
H^0(G/B,\mathbb Z )&= &\mathbb Z   \\[0.4ex]
H^1(G/B,\mathbb Z )&=& 0               \\[0.4ex]
H^2(G/B,\mathbb Z )&=& \mathbb{Z}_2\oplus\mathbb{Z}_2  \\[0.4ex]
H^3(G/B,\mathbb Z)&=&\mathbb Z \\
\end{array}\right. 
\]
\begin{figure}[t!]
\centering
\includegraphics[scale=0.46]{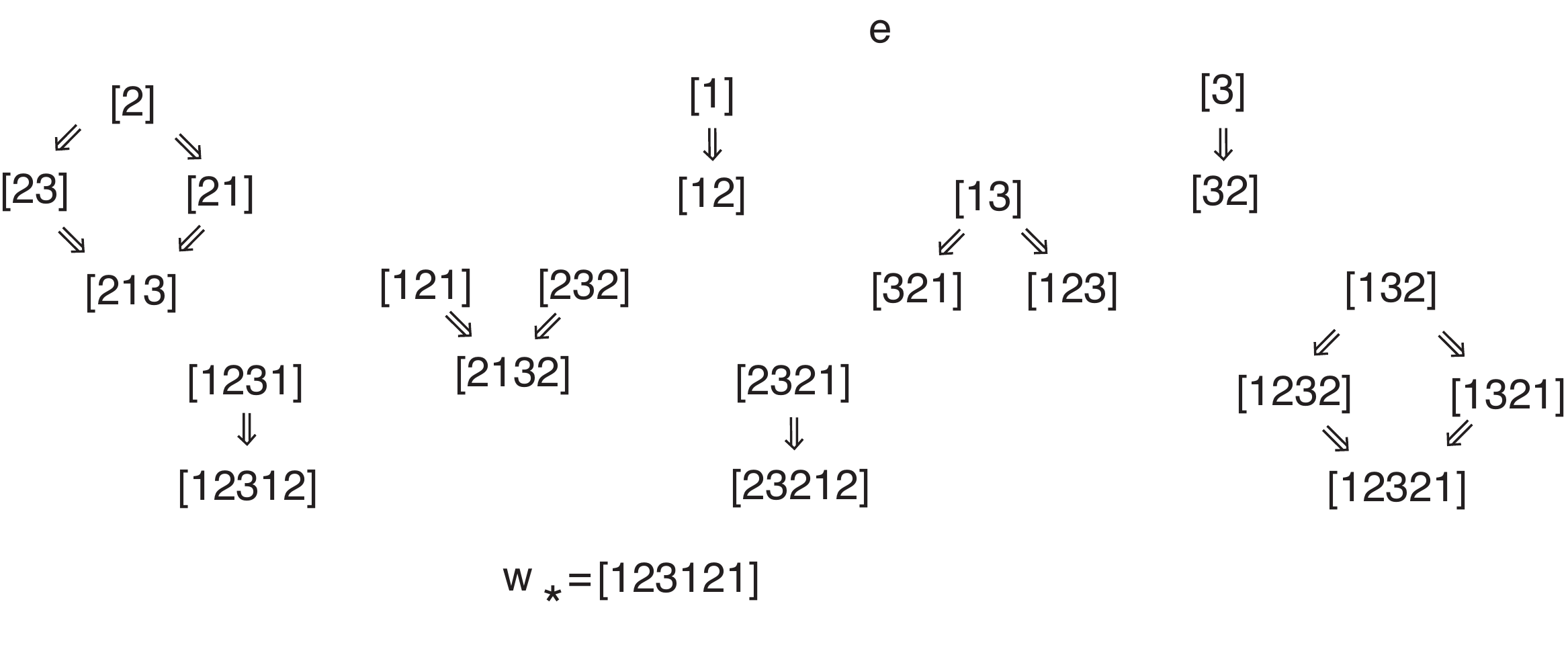}
\caption{The incidence graph $\mathcal{G}_{G/B}$ for the real flag manifold $SL(4,\Real)/B$.
The Bruhat cells $X_w=NwB/B$ are denoted by $[ij\ldots k]$ for $w=r_ir_j\ldots r_k$.
$w_*=[123121]$ is the longest element of $W=S_4$.
The incidence numbers associated with the edges $\Rightarrow$ are $\pm 2$ (see also Example 8.1
in \cite{CS:99}).}
\label{fig:A3incidence}
\end{figure}
In Figure \ref{fig:A3incidence}, we show the incidence graph for $G/B$ with $G=SL(4,\Real)$,
  from which one can compute the integral cohomology \cite{CS:99}.
We then obtain
\[
\left\{ \begin{array}{lllll}
H^0(G/B,\mathbb Z )&= &\mathbb Z   \\[0.4ex]
H^1(G/B,\mathbb Z )&=& 0               \\[0.4ex]
H^2(G/B,\mathbb Z )&=& \mathbb{Z}_2\oplus\mathbb{Z}_2\oplus\mathbb{Z}_2  \\[0.4ex]
H^3(G/B,\mathbb Z)&=&\mathbb{Z}\oplus\mathbb{Z}\oplus\mathbb{Z}_2\oplus\mathbb{Z}_2 \\[0.4ex]
H^4(G/B,\mathbb Z)&=&\mathbb{Z}_2\oplus\mathbb{Z}_2\\[0.4ex]
H^5(G/B,\mathbb{Z})&=&\mathbb{Z}_2\oplus\mathbb{Z}_2\oplus\mathbb{Z}_2\\[0.4ex]
H^6(G/B,\mathbb{Z})&=&\mathbb{Z}
\end{array}\right. 
\]

\end{Example}

The incidence graph for the general case of real split semisimple $G$ can be found in \cite{CS:99}.
Then the integral cohomology of $G/B$ can be computed from the incidence graph with the incidence
numbers $[w:w']$ being $0$ or $\pm2$.

For the rational cohomology, we have 
\begin{equation}\label{KG relation}
H^*(G/B,\mathbb{Q})=H^*(K/T,\mathbb{Q})=H^*(K,\mathbb{Q})\,,
\end{equation}
where $K$ is the maximal compact subgroup of $G$, and $T$ is the maximal torus of $G$. e.g. for $G=SL(n,\Real)$, $K=SO(n)$, and 
$T={\rm diag}(\pm1,\ldots,\pm1)$ (see Proposition 6.3 in \cite{CK:07}).
It is well known that the cohomology ring $H^*(K,\mathbb{Q})$ of compact connected group $K$ of rank $l$ is given by the exterior product algebra,
\[
H^*(K,\mathbb{Q})=\bigwedge{}_{\mathbb{Q}}\{x_{m_1},x_{m_2},\ldots, x_{m_l}\}\,,
\]
where $\{x_{m_1},\ldots,x_{m_l}\}$ are the generators of the exterior product representation with
${\rm deg}(x_{m_i})=m_i$ (odd) for $i=1,\ldots,l$ and $m_1+\cdots+m_l={\rm dim}(K)$ (see for example \cite{C:89}).
In the case of $K=SO(n)$, we have,
\begin{itemize}
\item[(a)] for $n=2m+1$,
\[H^*(SO(2m+1),\mathbb{Q})=\bigwedge{}_{\mathbb{Q}}(x_3,x_7,\ldots,x_{4m-1})\]
\item[(b)] for $n=2m$,
\[H^*(SO(2m),\mathbb{Q})=\bigwedge{}_{\mathbb{Q}}(x_3,x_7,\ldots, x_{4m-5},y_{2m-1})\,.
\]
Note here that the generators include the additional $y_{2m-1}$. For example, $H^*(SO(4),\mathbb{Q})$ is generated by two elements $x_3,y_3$ of the same degree, $V=\mathbb{Q} x_3+\mathbb{Q} y_3$ and $\wedge^2V=\mathbb{Q} x_3\wedge y_3$.
\end{itemize}

We also note that the number of points on the finite Chevalley group $K(\mathbb{F}_q)$ of the
compact connected group $K$ is given by certain polynomial of $q$. Here $\mathbb{F}_q$ is a finite
field with $q$ elements. Although this polynomial can be computed by using the Lefschetz fixed point theorem for the Frobenius map $\Phi:K(\overline{\mathbb{F}}_q)\to K(\overline{\mathbb{F}}_q),\, 
x\mapsto x^q$, 
we here give an elementary calculation to find those polynomials for $K=SO(n)$
(see also \cite{CK:06}). As we show in the next section, those polynomials are also related to
the indefinite Toda lattice.

Let us first assume that $q$ is a power of a prime number $p\ne 2$, such that in $\mathbb{F}_q$
the polynomial $x^2+1$ is not irreducible, i.e. $\sqrt{-1}\in\mathbb{F}_q$. Then we have
the following results for $|S^n(\mathbb{F}_q)|$, the number of $\mathbb{F}_q$ points on $S^n$:
\begin{Lemma}\label{FqpointS}
We have
\[
|S^n(\mathbb{F}_q)|=\left\{\begin{array}{lllll}
q^{m-1}(q^m-1)\quad & {\rm if}\quad & n=2m-1, \\[1.0ex]
q^m(q^m+1) & {\rm if}  &  n=2m.
\end{array}\right.
\]
\end{Lemma}
\begin{Proof}
Let us first consider the case $n=1$, i.e.
 \[
 S^1(\mathbb{F}_q)=\{(x,y)\in\mathbb{F}_q^2:x^2+y^2=1\}\,.
 \]
 Then using the formulae for the stereographic projection; $x=\frac{2u}{u^2+1},
 y=\frac{u^2-1}{u^2+1}$ with $y\ne 1$ and $\{u\in \mathbb{F}_q:u^2+1\ne0\}$. 
 Since $\sqrt{-1}\in\mathbb{F}_q$,
 we have $2$ points in $\{u^2+1=0\}$. Counting the point $(0,1)$, the north pole, we have
 $|S^{1}(\mathbb{F}_q)|=q-2+1=q-1$. Now consider the case $n=2$, we have
 $x=\frac{2u_1}{u_1^2+u_2^2+1}, y=\frac{2u_2}{u_1^2+u_2^2+1}, z=\frac{u_1^2+u_2^2-1}{u_1^2+u_2^2+1}$ with $z\ne 1$ and $\{(u_1,u_2)\in \mathbb{F}_q^2: u_1^2+u_2^2+1\ne 0\}$.
 This gives $q^2-(q-1)$ points (note $(q-1)$ is the number of points in $u_1^2+u_2^2+1=0$).
 We now add the points of the north pole $(x,y,1)$ with $x^2+y^2=0$. This gives $2(q-1)+1$,
 where $2(q-1)$ for $x=\pm\sqrt{-1}y\ne 0$ and $1$ for $(0,0,1)$. Then we have
 $|S^2(\mathbb{F}_q)|=q^2-(q-1)+2(q-1)+1=q(q+1)$. Using the induction, one can show
 that the number of points in the north pole is given by
 $|\{(x_1,\ldots, x_{2m-1})\in \mathbb{F}_q^{2m-1}:x_1^2+\cdots+x_{2m-1}^2=0\}|=q^{2m-2}$ and
 $|\{x_1,\ldots,x_{2m})\in \mathbb{F}_q^{2m}:x_1^2+\cdots+x_{2m}^2=0\}|=q^{2m-1}+q^m-q^{m-1}$.
 Then one can obtain the above formulae for $|S^n(\mathbb{F}_q)|$.
\end{Proof}
We can now find the number of $\mathbb{F}_q$ points
of finite Chevalley group $SO(n,\mathbb{F}_q)$:
First recall that $SO(n+1)/SO(n)\cong S^n$. Then we obtain
\[
|SO(n,\mathbb{F}_q)|=\prod_{k=1}^n |S^{n-k}(\mathbb{F}_q)|\,,
\]
which leads to the results \cite{C:89}:
\begin{itemize}
\item[(a)] For $n=2m$,
\[
|SO(2m,\mathbb{F}_q)|=2q^{m(m-1)}(q^2-1)(q^4-1)\cdots (q^{2m-2}-1)(q^m-1)\,.
\]
\item[(b)] For $n=2m+1$,
\[
|SO(2m+1,\mathbb{F}_q)|=2q^{m^2}(q^2-1)(q^4-1)\cdots (q^{2m}-1)\,.
\]
\end{itemize}

In general,  the number of $\mathbb{F}_q$ points on the compact group $K$ can be
expressed by (see e.g. \cite{C:89})
\begin{equation}\label{Fq points} 
\left|K(\mathbb{F}_q)\right|=q^r\,p(q)\qquad {\rm with}\quad p(q)=\prod_{i=1}^l\,(q^{d_i}-1)\,,
\end{equation}
where $d_i$'s are degree of basic Weyl group 
invariant polynomials for $K$ given by $d_i=(m_i+1)/2$,
and $r={\rm dim}(K)-{\rm deg}(p(q))$. 
In the next section, we show that those polynomials can be reproduced by
counting the blow-ups in the solution of the indefinite Toda lattice 
(see \cite{CK:06} for the general case).

\subsection{Blow-ups of the indefinite Toda lattice on $G$ and the cohomology of $G/B$}\label{blow-ups} 
Now we explain how we get the cohomology of $G/B$ from the moment polytope of
the indefinite Toda lattice of Section \ref{Extended Hessenberg Toda}.

First we note that the $\tau$-functions can change their signs if some (but not all) of $s_i$'s are negative.
This can be seen from (\ref{indefinite tau}), and implies that the solution blows up for some time $t_1=\bar{t}_1$, (see also (\ref{g indefinite}) and (\ref{f indefinite})). The explicit form of the $\tau$-functions
can be obtained from (\ref{indefinite tauk}), and they are expressed by (see (\ref{BC tau}) and also Proposition 3.1 in \cite{Kodama Ye II}),
\begin{equation}\label{sign tau}
\tau_k(t)=\sum_{1\le j_1<\cdots<j_k\le n}s_{j_1}\cdots s_{j_k}\,K(j_1,\ldots,j_k)\exp\left(\sum_{i=1}^k\lambda_{j_i}t\right)\,,
\end{equation}
where $K(j_1,\ldots,j_k)$ are positive and given by
\[
K(j_1,\ldots,j_k)=\left(\varphi^0(\lambda_{j_1})\cdots\varphi^0(\lambda_{j_k})\right)^2\left|
\begin{matrix}
1          & \cdots    & 1 \\
\vdots & \ddots    & \vdots \\
\lambda_{j_1}^{k-1} & \cdots & \lambda_{j_k}^{k-1}\\
\end{matrix}\right|^2\,>0\,.
\]
As the simplest case, let us consider the $sl(2,\Real)$ Toda lattice: We have one $\tau$-function, 
\[
\tau_1(t)=s_1\rho_1e^{\lambda_1t}+s_2\rho_2e^{\lambda_2t}\,.
\]
If $s_1s_2<0$, $\tau_1(t)$ has zero at a time $t=\frac{1}{\lambda_2-\lambda_1}\ln(\frac{\rho_1}{\rho_2})$, that is, we have a blow-up in the solution.
The image of the moment map $\mu(\tau_1)$ is given by a line segment whose end points correspond 
to the weights $L_1$ and $L_2=-L_1$. Although the dynamics are so different in the cases $s_1s_2 > 0$ and $s_1s_2 < 0$,
the moment polytope (a line segment) is independent of the signs of the $s_i$'s.  
Notice that $s_1s_2={\rm sgn}(g_1)$,
and in general, if ${\rm sgn}(g_k)<0$ for some $k$, then the solution blows up sometime in $\Real$.

In order to find the general pattern of the sign changes in $(g_1(t),\ldots,g_{n-1}(t))$ of the matrix $X$ in
(\ref{X indefinite}),
we first recall
that the isospectral variety is characterized by the moment polytope $\mathcal{M}_{\epsilon}$ whose vertices are given by
the orbit of Weyl group action.  Here the set of signs $\epsilon=(\epsilon_1,\ldots,\epsilon_{n-1})$
is defined by the signs of $g_i$ for $t\to -\infty$.
From the ordering $\lambda_1<\cdots<\lambda_n$, 
we first see that  $\tau_k(t)\approx s_1\cdots s_k K(1,\ldots,k)\exp((\lambda_1+\cdots+\lambda_k)t)$.
Then from the definition of $g_k(t)$ in (\ref{g indefinite}), i.e. $g_k=\tau_{k-1}\tau_{k+1}/\tau_k^2$,
the sign of
$g_k(t)$ for $t\to-\infty$ is given by
\[
\epsilon_k={\rm sgn}(g_k)=s_ks_{k+1}\, \quad {\rm for}\quad k=1,\ldots, n-1\,.
\]  
Then from the moment map (\ref{moment tau}), one notes that the moment polytope given as the image of the moment map $\mu(\mathcal{M}_{\epsilon})$
in (\ref{Toda moment map}) is independent of the sign set $\epsilon$.
However the dynamics of the Toda lattice with a different $\epsilon$ is quite different, and
the solution with at least one $\epsilon_k<0$ has a blow-up at some $t\in\Real$.

\begin{figure}[t!]
\centering
\includegraphics[scale=0.42]{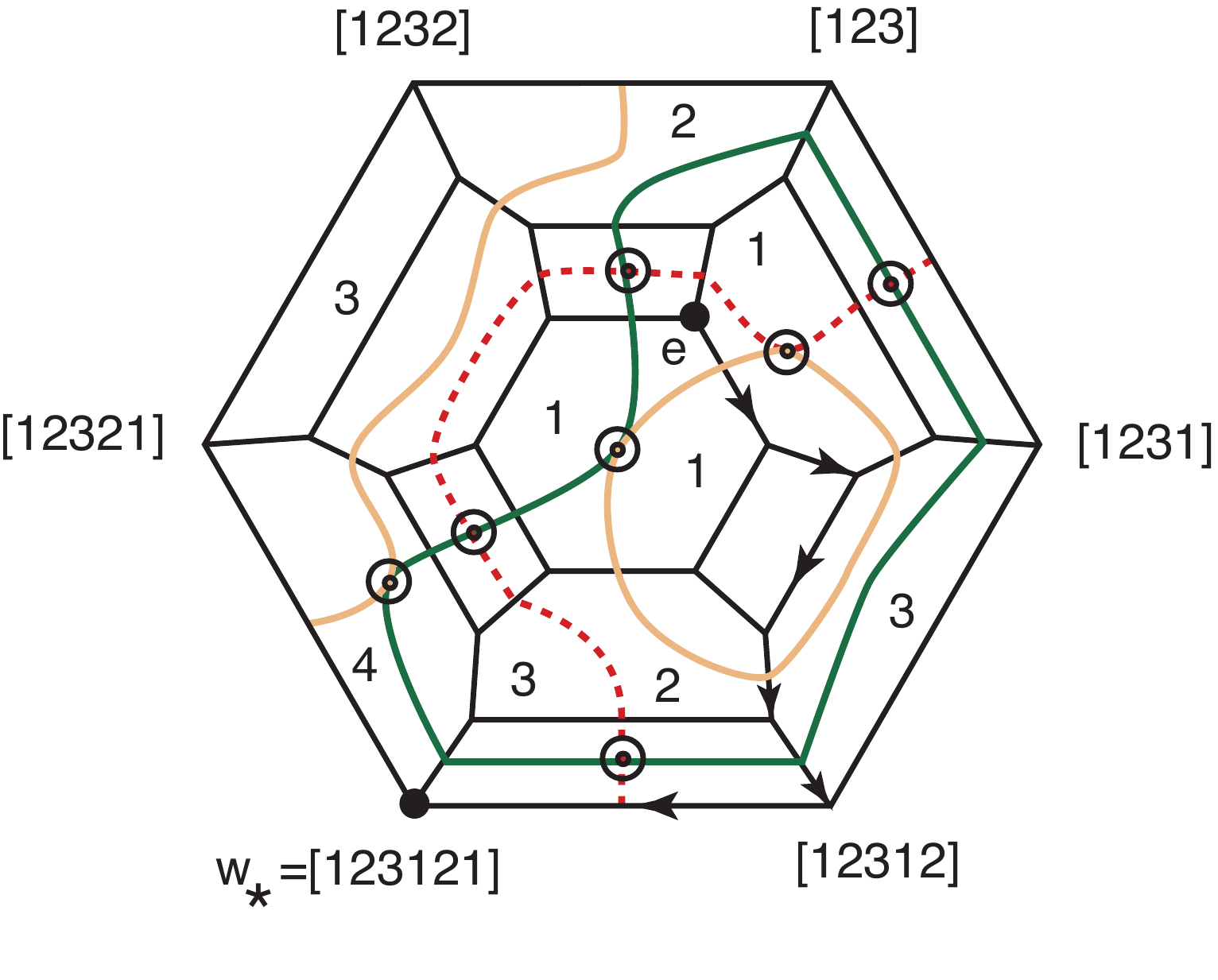}
\caption{The moment polytope $\mathcal{M}_{---}$ for the $sl(4,\Real)$ indefinite Toda lattice.
The divisors defined by the set of zero points for the $\tau$-functions are shown by 
the dotted curve for $\{\tau_1=0\}$, by the light color one for $\{\tau_2=0\}$ and the dark one for 
$\{\tau_3=0\}$.
The double circles indicate the divisors with $\{\tau_i=0\}\cap\{\tau_j=0\}$ which are all connected
at the center of the polytope corresponding to the point with $\{\tau_1=\tau_2=\tau_3=0\}$. The numbers in the polytope indicate the number of blow-ups
along the flow. An example of a path from the top vertex $e$ to the bottom vertex $w_*$, the longest
element of $S_4$, is shown by directed edges.}
\label{fig:A3polytope}
\end{figure}

We now consider each edge of the polytope which corresponds to an $sl(2,\Real)$ indefinite Toda
lattice, that is, where $g_j\ne 0$ for only one $j$. This edge can be also expressed by
a simple reflection $r_j\in W$.
Since the simple reflection $r_j$ exchanges $s_j$ and $s_{j+1}$, we have an action of $r_j$
on all the signs $\epsilon_k$, $r_j:\epsilon_k\to\epsilon'_k$,
\[
\epsilon'_k=r_j(\epsilon_k)=\left\{ \begin{array}{llll}
\epsilon_k\epsilon_{k-1} \quad &{\rm if} ~~& j=k-1\\[0.4ex]
\epsilon_k\epsilon_{k+1}\quad &{\rm if} ~~& j=k+1\\[0.4ex]
\epsilon_k    \quad &{\rm if}  ~~ &j=k, ~{\rm or}~|j-k|>1
\end{array}\right.
\]
which can be also shown directly from the form of $\tau_k(t)$ in (\ref{sign tau}). This formula can be extended to the indefinite Toda lattice on any real split semisimple Lie algebras, and we have
(see (\ref{general tau}) and Proposition 3.16 in \cite{CK:02}):
\begin{Proposition}\label{sign change}
Let $\epsilon_j={\rm sgn}(g_j)$ for $j=1,\ldots,n-1$. Then the Weyl group action on the signs is given by
\[
r_j : \epsilon_k \mapsto \epsilon_k\epsilon_j^{-C_{kj}}\,,
\]
where $(C_{ij})_{1\le i,j\le n-1}$ is the Cartan matrix of $sl(n,\Real)$.
\end{Proposition}
With this $W$-action on the signs $\epsilon=(\epsilon_1,\ldots,\epsilon_{n-1})$
with $\epsilon_k={\rm sgn}(g_k)$ at each vertex of the polytope, we now define the relation between the
vertices labeled by $w$ and $w'=wr_i$ as follows: Notice that if $\epsilon_i=+$, then $
(\epsilon_1,\cdots,\epsilon_{n-1})$ remains the same under $r_i$-action. Then we write
\[
w\Longrightarrow w' \qquad {\rm with}\quad w'=wr_i\,.
\]
Now the following definition gives the number of blow-ups in the Toda orbit from the top vertex $e$
to the vertex labeled by $w\in W$:  Choose a reduced expression $w=r_{j_1}\cdots r_{j_k}$.
Then consider the sequence of the signs at the orbit given by $w$-action,
\[
\epsilon\, \to\, r_{j_1}\epsilon\,\to \,r_{j_2}r_{j_1} \epsilon\,\to\, \cdots\,\to \,w^{-1}\epsilon\,.
\]
We then define the function $\eta(w,\epsilon)$ as the number of $\to$ which are not of the form
$\Rightarrow$. The number $\eta(w_*,\epsilon)$ for the longest element $w_*$ gives the total
number of blow-ups along the Toda flow in the polytope of $\mathcal{M}_{\epsilon}$.
Whenever $\epsilon=(-,\ldots,-)$, we just denote $\eta(w,\epsilon)=\eta(w)$.
This number $\eta(w,\epsilon)$ does not depend on the choice of the reduced expression
of $w$ (see Corollary 5.2 in \cite{CK:06}). Hence the number of blow-up points along the 
trajectories in the edges of the polytope is independent of the trajectory parametrized by the
reduced expression. In Figure \ref{fig:A2indefinite}, we illustrate the numbers $\eta(w,\epsilon)$
for the $sl(3,\Real)$ indefinite Toda lattice.
For example, on $\mathcal{M}_{--}$, we have $\eta(e)=0, \eta(r_1)=\eta(r_2)=\eta(r_1r_2)=\eta(r_2r_1)=1$ and $\eta(r_1r_2r_1)=2$, i.e the total number of blow-ups is 2. We also illustrate this for the $sl(4,\Real)$
Toda lattice in Figure \ref{fig:A3polytope}. Along the path shown in this Figure, we have $\eta(e)=0, \eta([2])=\eta([21])=\eta([213])=1,
\eta([2132])=2,\eta([21323])=3$ and $\eta(w_*)=4$, where $[ij\cdots k]=r_ir_j\cdots r_k$, and
note $[21323]=[12312]$.

In general, the total number of blow-ups $\eta(w_*,\epsilon)$ depends only
the initial signs $\epsilon=(\epsilon_1,\ldots,\epsilon_{n-1})$ with $\epsilon_i={\rm sgn}(g_i(t))$
for $t\to-\infty$, which is given by $\epsilon_i=s_is_{i+1}$. Then in the case of $sl(n,\Real)$ indefinite
Toda lattice, the number $\eta(w_*,\epsilon)=m(n-m)$ where $m$ is the total number of negative $s_i$'s
(Proposition 3.3 in \cite{Kodama Ye II}).
In particular, the maximum number of blow-ups occurs the case with $\epsilon=(-,\ldots,-)$, and it is given by $[(n+1)/2](n-[(n+1)/2])$. Those numbers $\eta(w_*,\epsilon)$ are related to the polynomials
appearing in $\mathbb{F}_q$ points on certain compact groups defined in (\ref{Fq points}).

We now introduce polynomials in terms of the numbers $\eta(w,\epsilon)$, which play a key
role for counting the number of blow-ups and give a surprising connection to the rational
cohomology of the maximal compact subgroup $SO(n)$ (Definition 3.1 in \cite{CK:06}).
\begin{Definition}\label{p(q) polynomial}
We define a monic polynomial associated to the polytope $\mathcal{M}_{\epsilon}$,
\[
p(q,\epsilon)=(-1)^{l(w_*)}\sum_{w\in W}(-1)^{l(w)}q^{\eta(w,\epsilon)}\,,
\]
where $l(w)$ indicates the length of $w$. Notice that the degree of $p(q,\epsilon)$, denoted by ${\rm deg}(p(q,\epsilon))$,
is the total number of blow-ups, i.e. $\eta(w_*,\epsilon)={\rm deg}(p(q,\epsilon))$. For the case $\epsilon=(-,\ldots,-)$, we simply denote it by $p(q)$.
\end{Definition}    
\begin{Example}
In the case of the $sl(2,\Real)$ Toda lattice, 
\begin{itemize}
\item[(a)] for $\epsilon=(+)$, we have $e\Rightarrow s_1$ which gives $p(q,+)=0$,
\item[(b)] for $\epsilon=(-)$, we have a blow-up between $e$ and $s_1$, hence $p(q,-)=q-1$.
\end{itemize}
Recall from the previous section that the polynomial $p(q)=p(q,-)$ appears in $|SO(2,\mathbb{F}_q)|=q-1$.

In the case of the $sl(3,\Real)$ Toda lattice, from Figure \ref{fig:A2indefinite},
\begin{itemize}
\item[(a)] for all the cases of $\epsilon=(\epsilon_1,\epsilon_2)$ except $(-,-)$, we have
$p(q,\epsilon)=0$.
\item[(b)] for $\epsilon=(-,-)$, we have $p(q)=q^2-1$.
\end{itemize}
Note again that the polynomial $p(q)$ appears in $|SO(3,\mathbb{F}_q)|=q(q^2-1)$.

In the case of $sl(4,\Real)$, we have, from Figure \ref{fig:A3polytope},
\begin{itemize}
\item[(a)] for all $\epsilon=(\epsilon_1,\epsilon_2,\epsilon_3)$ except $(-,-,-)$, $p(q,\epsilon)=0$.
\item[(b)] for $\epsilon=(-,-,-)$, $p(q)=q^4-2q^2+1=(q^2-1)^2$.
\end{itemize}
Again note that  $|SO(4,\mathbb{F}_q)|=q^2(q^2-1)^2$.
\end{Example}
Casian and Kodama then prove that the polynomial $p(q)$ for $\mathcal{M}_{\epsilon}$ with
$\epsilon=(-,\ldots,-)$ in Definition \ref{p(q) polynomial} agrees with the polynomial $p(q)$ in 
$|K(\mathbb{F}_q)|$ in (\ref{Fq points}) where $K$ is the maximal compact subgroup
of real split semisimple Lie group $G$ for the Toda lattice (Theorem 6.5 in \cite{CK:06}).

Thus the polynomial $p(q)$ contains all the information on the $\mathbb{F}_q$ points on the
compact subgroup $K$ of $G$, which is also related to the rational cohomology, i.e.
$H^*(K,\mathbb{Q})=H^*(G/B,\mathbb{Q})$ (see (\ref{KG relation})).
Now recall that the integral cohomology of the real flag variety $G/B$ is obtained
by the incidence graph $\mathcal{G}_{G/B}$ in Definition \ref{incidence graph}.
In \cite{CK:06}, Casian and Kodama show that the graph $\mathcal{G}_{G/B}$ can be
obtained from the blow-ups of the Toda flow. They define a graph $\mathcal{G}_{\epsilon}$ associated to the blow-ups as follows:
\begin{Definition} 
The graph $\mathcal{G}_{\epsilon}$ consists of vertices labeled by the elements of the Weyl
group $W$ and oriented edges $\Rightarrow$. The edges are defined as follows:
\[
w_1\Rightarrow w_2\quad{\rm iff}\quad \left\{\begin{array}{lllll}
{\rm (a)}~ w_1\le w_2~ ({\rm Bruhat~order})\\[0.5ex]
{\rm (b)}~ l(w_1)=l(w_2)+1 \\[0.5ex]
{\rm (c)}~ \eta(w_1,\epsilon)=\eta(w_2,\epsilon)\\[0.5ex]
{\rm (d)}~w_1^{-1}\epsilon=w_2^{-1}\epsilon
\end{array}\right.
\]
When $\epsilon=(-,\ldots,-)$, we simply denote $\mathcal{G}=\mathcal{G}_{\epsilon}$.
\end{Definition}
Then they prove that $\mathcal{G}_{\epsilon}$ with $\epsilon=(-,\ldots,-)$ is equivalent to $\mathcal{G}_{G/B}$ (Theorem 3.5 in \cite{CK:06} which is the main theorem in the paper).
For example, the graph $\mathcal{G}$ associated with Figure \ref{fig:A3polytope}
agrees with the incidence graph $\mathcal{G}_{G/B}$ given in Figure \ref{fig:A3incidence}.
The proof of the equivalence $\mathcal{G}_{G/B}=\mathcal{G}$ contains several technical steps,
which are beyond the scope of this review.



\medskip

\end{document}